\documentclass[aps,prx,twocolumn,superscriptaddress]{revtex4-2}
\pdfoutput=1

\usepackage{amssymb}
\usepackage[export]{adjustbox}
\usepackage{multirow}
\usepackage{bm}
\usepackage{amsmath}
\usepackage{graphicx}
\usepackage{epstopdf}
\usepackage{subfigure}
\usepackage{natbib}
\usepackage{epsfig}
\usepackage{amsfonts}
\usepackage{mathrsfs}
\usepackage[toc,page,title,titletoc,header]{appendix}
\usepackage[colorlinks,linkcolor=blue,citecolor=blue,anchorcolor=blue]{hyperref}
\usepackage{comment}
\usepackage{physics}
\usepackage{xcolor}

\usepackage{braket}

\def\Pf{{\rm Pf}}

\usepackage{dsfont,amsthm,amsbsy}

\def\Tr{{\rm Tr}}

\usepackage{fancyhdr}

\usepackage{ulem}

\usepackage{bbm}

\newcommand{\bea}{\begin{equation} \begin{aligned}}
\newcommand{\eea}{\end{aligned} \end{equation} }

\newcommand{\bpm}{\begin{pmatrix}}
\newcommand{\epm}{\end{pmatrix}}

\newcommand{\bs}{\boldsymbol}

\def\spinc{spin$_\mathbb{C}$ }

\usepackage{tikz}
\usetikzlibrary{calc}

\newcommand{\BraidDX}{0.95}    
\newcommand{\BraidDY}{1.10}    
\newcommand{\BraidLW}{0.9pt}   
\newcommand{\BraidGap}{3.2pt}  
\newcommand{\BraidDot}{1.7pt}  

\newcommand{\JoinEps}{0.03}       
\newcommand{\WhiteShorten}{1.2pt} 

\providecommand{\Bar}[1]{\overline{#1}}

\newcount\BraidN
\newcount\BraidT

\tikzset{
  branchline/.style={line width=\BraidLW, line cap=round, line join=round},
  strandline/.style={line width=\BraidLW, line cap=round, line join=round},
  treelabel/.style={font=\small, inner sep=1pt}
}

\newcommand{\Xof}[1]{(#1-1)*\BraidDX}

\newcommand{\SplitOne}[5]{%
  \pgfmathsetmacro{\XP}{#1}%
  \pgfmathsetmacro{\XL}{#2}%
  \pgfmathsetmacro{\XR}{#3}%
  \pgfmathsetmacro{\Yzero}{#4}%
  \pgfmathsetmacro{\Yone}{#5}%
  \pgfmathsetmacro{\Ymid}{(\Yzero+\Yone)/2}%
  \pgfmathsetmacro{\CTRL}{0.55*(\Yone-\Ymid)}%

  \draw[branchline] (\XP,\Yzero) -- (\XP,\Ymid+\JoinEps);
  \draw[branchline] (\XP,\Ymid) .. controls (\XP,\Ymid+\CTRL) and (\XL,\Yone-\CTRL) .. (\XL,\Yone+\JoinEps);
  \draw[branchline] (\XP,\Ymid) .. controls (\XP,\Ymid+\CTRL) and (\XR,\Yone-\CTRL) .. (\XR,\Yone+\JoinEps);

  \fill (\XP,\Ymid) circle[radius=\BraidDot];
}

\newcommand{\TreeOneTwoFour}[4]{%
  \pgfmathsetmacro{\Yroot}{#2}%
  \pgfmathsetmacro{\YA}{#3}%
  \pgfmathsetmacro{\YB}{#4}%

  \pgfmathsetmacro{\Xone}{\Xof{#1}}%
  \pgfmathsetmacro{\Xtwo}{\Xof{\numexpr#1+1\relax}}%
  \pgfmathsetmacro{\Xthree}{\Xof{\numexpr#1+2\relax}}%
  \pgfmathsetmacro{\Xfour}{\Xof{\numexpr#1+3\relax}}%

  \pgfmathsetmacro{\Xcenter}{(\Xone+\Xfour)/2}%
  \pgfmathsetmacro{\XleftMid}{(\Xone+\Xtwo)/2}%
  \pgfmathsetmacro{\XrightMid}{(\Xthree+\Xfour)/2}%

  \SplitOne{\Xcenter}{\XleftMid}{\XrightMid}{\Yroot}{\YA}
  \SplitOne{\XleftMid}{\Xone}{\Xtwo}{\YA}{\YB}
  \SplitOne{\XrightMid}{\Xthree}{\Xfour}{\YA}{\YB}

  \fill (\Xcenter,\Yroot) circle[radius=\BraidDot];
}

\newcommand{\TreeOneTwoFourPositions}[7]{%
  \pgfmathsetmacro{\Yroot}{#5}%
  \pgfmathsetmacro{\YA}{#6}%
  \pgfmathsetmacro{\YB}{#7}%

  \pgfmathsetmacro{\Xone}{(\numexpr#1-1\relax)*\BraidDX}%
  \pgfmathsetmacro{\Xtwo}{(\numexpr#2-1\relax)*\BraidDX}%
  \pgfmathsetmacro{\Xthree}{(\numexpr#3-1\relax)*\BraidDX}%
  \pgfmathsetmacro{\Xfour}{(\numexpr#4-1\relax)*\BraidDX}%

  \pgfmathsetmacro{\XleftMid}{(\Xone+\Xtwo)/2}%
  \pgfmathsetmacro{\XrightMid}{(\Xthree+\Xfour)/2}%
  \pgfmathsetmacro{\Xcenter}{(\XleftMid+\XrightMid)/2}%

  \SplitOne{\Xcenter}{\XleftMid}{\XrightMid}{\Yroot}{\YA}
  \SplitOne{\XleftMid}{\Xone}{\Xtwo}{\YA}{\YB}
  \SplitOne{\XrightMid}{\Xthree}{\Xfour}{\YA}{\YB}

  \fill (\Xcenter,\Yroot) circle[radius=\BraidDot];
}

\newcommand{\TreeOuterLabels}[4]{%
  \pgfmathsetmacro{\Ylab}{#2}%
  \pgfmathsetmacro{\Xone}{\Xof{#1}}%
  \pgfmathsetmacro{\Xfour}{\Xof{\numexpr#1+3\relax}}%
  \node[treelabel, left=-8pt]  at (\Xone,\Ylab) {$#3$};
  \node[treelabel, right=-8pt] at (\Xfour,\Ylab) {$#4$};
}

\newcommand{\TopDotsLabels}[2]{%
  \foreach \i in {1,...,#1}{%
    \pgfmathsetmacro{\XI}{(\i-1)*\BraidDX}%
    \fill (\XI,#2) circle[radius=\BraidDot];
    \node[above=2pt] at (\XI,#2) {$\sigma_{\i}$};
  }%
}

\newcommand{\BraidFromOffset}[3]{%
  \BraidN=#2\relax
  \BraidT=0\relax
  \foreach \GEN in {#3}{%
    \begingroup
    \pgfmathtruncatemacro{\K}{abs(\GEN)}%
    \pgfmathtruncatemacro{\SGN}{ifthenelse(\GEN>0,1,-1)}%

    \pgfmathsetmacro{\YA}{#1 + \BraidT*\BraidDY}%
    \pgfmathsetmacro{\YB}{#1 + (\BraidT+1)*\BraidDY}%
    \pgfmathsetmacro{\YAo}{\YA - \JoinEps}%
    \pgfmathsetmacro{\YBo}{\YB + \JoinEps}%

    \pgfmathsetmacro{\XL}{(\K-1)*\BraidDX}%
    \pgfmathsetmacro{\XR}{(\K)*\BraidDX}%

    \foreach \I in {1,...,\the\BraidN}{%
      \ifnum\I=\K\relax\else
        \ifnum\I=\numexpr\K+1\relax\else
          \pgfmathsetmacro{\XI}{(\I-1)*\BraidDX}%
          \draw[strandline] (\XI,\YAo) -- (\XI,\YBo);
        \fi
      \fi
    }%

    \def\CTRL{0.45*\BraidDY}%

    \ifnum\SGN=1\relax
      \draw[strandline]
        (\XR,\YAo) .. controls +(0,\CTRL) and +(0,-\CTRL) .. (\XL,\YBo);
      \draw[strandline,
            preaction={
              draw,white,
              line width=\BraidGap,
              line cap=butt,
              line join=round,
              shorten <=\WhiteShorten,
              shorten >=\WhiteShorten
            }]
        (\XL,\YAo) .. controls +(0,\CTRL) and +(0,-\CTRL) .. (\XR,\YBo);
    \else
      \draw[strandline]
        (\XL,\YAo) .. controls +(0,\CTRL) and +(0,-\CTRL) .. (\XR,\YBo);
      \draw[strandline,
            preaction={
              draw,white,
              line width=\BraidGap,
              line cap=butt,
              line join=round,
              shorten <=\WhiteShorten,
              shorten >=\WhiteShorten
            }]
        (\XR,\YAo) .. controls +(0,\CTRL) and +(0,-\CTRL) .. (\XL,\YBo);
    \fi

    \global\advance\BraidT by 1\relax
    \endgroup
  }%
}

\newcommand{\BraidParStep}[2]{%
  \begingroup
  \pgfmathsetmacro{\YA}{#1 + \BraidT*\BraidDY}%
  \pgfmathsetmacro{\YB}{#1 + (\BraidT+1)*\BraidDY}%
  \pgfmathsetmacro{\YAo}{\YA - \JoinEps}%
  \pgfmathsetmacro{\YBo}{\YB + \JoinEps}%
  \def\CTRL{0.45*\BraidDY}%

  \foreach \I in {1,...,\the\BraidN}{%
    \pgfmathsetmacro{\XI}{(\I-1)*\BraidDX}%
    \draw[strandline] (\XI,\YAo) -- (\XI,\YBo);
  }%

  \foreach \GEN in {#2}{%
    \pgfmathtruncatemacro{\K}{abs(\GEN)}%
    \pgfmathsetmacro{\XL}{(\K-1)*\BraidDX}%
    \pgfmathsetmacro{\XR}{(\K)*\BraidDX}%
    \draw[white, line width={\BraidGap+1.2pt}, line cap=butt, line join=round]
      (\XL,\YAo) -- (\XL,\YBo);
    \draw[white, line width={\BraidGap+1.2pt}, line cap=butt, line join=round]
      (\XR,\YAo) -- (\XR,\YBo);
  }%

  \foreach \GEN in {#2}{%
    \pgfmathtruncatemacro{\K}{abs(\GEN)}%
    \pgfmathtruncatemacro{\SGN}{ifthenelse(\GEN>0,1,-1)}%
    \pgfmathsetmacro{\XL}{(\K-1)*\BraidDX}%
    \pgfmathsetmacro{\XR}{(\K)*\BraidDX}%

    \ifnum\SGN=1\relax
      \draw[strandline]
        (\XR,\YAo) .. controls +(0,\CTRL) and +(0,-\CTRL) .. (\XL,\YBo);
      \draw[strandline,
            preaction={
              draw,white,
              line width=\BraidGap,
              line cap=butt,
              line join=round,
              shorten <=\WhiteShorten,
              shorten >=\WhiteShorten
            }]
        (\XL,\YAo) .. controls +(0,\CTRL) and +(0,-\CTRL) .. (\XR,\YBo);
    \else
      \draw[strandline]
        (\XL,\YAo) .. controls +(0,\CTRL) and +(0,-\CTRL) .. (\XR,\YBo);
      \draw[strandline,
            preaction={
              draw,white,
              line width=\BraidGap,
              line cap=butt,
              line join=round,
              shorten <=\WhiteShorten,
              shorten >=\WhiteShorten
            }]
        (\XR,\YAo) .. controls +(0,\CTRL) and +(0,-\CTRL) .. (\XL,\YBo);
    \fi
  }%

  \global\advance\BraidT by 1\relax
  \endgroup
}

\def\PermAt#1{\csname perm@#1\endcsname}
\def\SetPerm#1#2{\expandafter\xdef\csname perm@#1\endcsname{#2}}
\def\InitPerm#1{%
  \foreach \p in {1,...,#1}{\SetPerm{\p}{\p}}%
}
\def\SwapPerm#1{%
  \edef\tmpA{\PermAt{#1}}%
  \edef\tmpB{\PermAt{\number\numexpr#1+1\relax}}%
  \SetPerm{#1}{\tmpB}%
  \SetPerm{\number\numexpr#1+1\relax}{\tmpA}%
}
\def\ColorOfOrig#1{%
  \ifnum#1=1 red\else black\fi
}

\newcommand{\BraidParStepTracked}[2]{%
  \begingroup
  \pgfmathsetmacro{\YAstep}{#1 + \BraidT*\BraidDY}%
  \pgfmathsetmacro{\YBstep}{#1 + (\BraidT+1)*\BraidDY}%
  \pgfmathsetmacro{\YAo}{\YAstep - \JoinEps}%
  \pgfmathsetmacro{\YBo}{\YBstep + \JoinEps}%
  \def\CTRL{0.45*\BraidDY}%

  \foreach \I in {1,...,\the\BraidN}{%
    \pgfmathsetmacro{\XI}{(\I-1)*\BraidDX}%
    \edef\orig{\PermAt{\I}}%
    \edef\col{\ColorOfOrig{\orig}}%
    \draw[strandline, draw=\col] (\XI,\YAo) -- (\XI,\YBo);
  }%

  \foreach \GEN in {#2}{%
    \pgfmathtruncatemacro{\K}{abs(\GEN)}%
    \pgfmathsetmacro{\XL}{(\K-1)*\BraidDX}%
    \pgfmathsetmacro{\XR}{(\K)*\BraidDX}%
    \draw[white, line width={\BraidGap+1.2pt}, line cap=butt, line join=round]
      (\XL,\YAo) -- (\XL,\YBo);
    \draw[white, line width={\BraidGap+1.2pt}, line cap=butt, line join=round]
      (\XR,\YAo) -- (\XR,\YBo);
  }%

  \foreach \GEN in {#2}{%
    \pgfmathtruncatemacro{\K}{abs(\GEN)}%
    \pgfmathtruncatemacro{\SGN}{ifthenelse(\GEN>0,1,-1)}%
    \pgfmathsetmacro{\XL}{(\K-1)*\BraidDX}%
    \pgfmathsetmacro{\XR}{(\K)*\BraidDX}%

    \edef\origK{\PermAt{\K}}%
    \edef\origKp{\PermAt{\number\numexpr\K+1\relax}}%
    \edef\colK{\ColorOfOrig{\origK}}%
    \edef\colKp{\ColorOfOrig{\origKp}}%

    \ifnum\SGN=1\relax
      \draw[strandline, draw=\colKp]
        (\XR,\YAo) .. controls +(0,\CTRL) and +(0,-\CTRL) .. (\XL,\YBo);
      \draw[strandline, draw=\colK,
            preaction={
              draw,white,
              line width=\BraidGap,
              line cap=butt,
              line join=round,
              shorten <=\WhiteShorten,
              shorten >=\WhiteShorten
            }]
        (\XL,\YAo) .. controls +(0,\CTRL) and +(0,-\CTRL) .. (\XR,\YBo);
    \else
      \draw[strandline, draw=\colK]
        (\XL,\YAo) .. controls +(0,\CTRL) and +(0,-\CTRL) .. (\XR,\YBo);
      \draw[strandline, draw=\colKp,
            preaction={
              draw,white,
              line width=\BraidGap,
              line cap=butt,
              line join=round,
              shorten <=\WhiteShorten,
              shorten >=\WhiteShorten
            }]
        (\XR,\YAo) .. controls +(0,\CTRL) and +(0,-\CTRL) .. (\XL,\YBo);
    \fi
  }%

  \foreach \GEN in {#2}{%
    \pgfmathtruncatemacro{\K}{abs(\GEN)}%
    \SwapPerm{\K}%
  }%

  \global\advance\BraidT by 1\relax
  \endgroup
}

\begin{document}

\title{Charge-$4e$ superconductor with parafermionic vortices: \\
A path to universal topological quantum computation}

\author{Zhengyan~Darius~Shi}
\thanks{These authors contributed equally to this work.}
\affiliation{Leinweber Institute for Theoretical Physics, Stanford University, Stanford, California 94305, USA}

\author{Zhaoyu~Han}
\thanks{These authors contributed equally to this work.}
\email{zhan@fas.harvard.edu}
\affiliation{Department of Physics, Harvard University, Cambridge, Massachusetts 02138, USA}

\author{Srinivas~Raghu}
\affiliation{Leinweber Institute for Theoretical Physics, Stanford University, Stanford, California 94305, USA}

\author{Ashvin~Vishwanath}
\email{avishwanath@g.harvard.edu}
\affiliation{Department of Physics, Harvard University, Cambridge, Massachusetts 02138, USA}

\begin{abstract}

Topological superconductors (TSCs) provide a promising route to fault-tolerant quantum information processing. However, the canonical Majorana platform based on $2e$ TSCs remains computationally constrained. In this work, we find a $4e$ TSC that overcomes these constraints by combining a charge-$4e$ condensate with an Abelian chiral $\mathbb{Z}_3$ topological order in an intertwined fashion. Remarkably, this $4e$ TSC can be obtained by proliferating vortex-antivortex pairs in a stack of two $2e$ $p+ip$ TSCs, or by melting a $\nu=2/3$ quantum Hall state. Specific to this TSC, the $hc/(4e)$ fluxes act as charge-conjugation defects in the topological order, whose braiding with anyons transmutes anyons into their antiparticles. This symmetry enrichment leads to $\mathbb{Z}_3$ parafermion zero modes trapped in the elementary vortex cores, which naturally encode qutrits. Braiding the parafermion defects alone generates the full many-qutrit Clifford group. We further show that a single-probe interferometric measurement enables topologically protected magic-state preparation, promoting Clifford operations to a universal gate set. Because the non-Abelian modes are bound to flux defects, they can, in principle, be externally controlled using superconducting circuit-based technology.
More broadly, our results highlight hierarchical electron aggregation, the formation and condensation of higher-charge electron clusters, as a design principle for topological quantum matter with increased computational capability.


\end{abstract}

\maketitle

\section{Introduction}

Topological superconductivity (TSC) has long been pursued as a platform for topological quantum computation (TQC)~\cite{Kitaev1997_TQCmanifesto,Nayak2008_TQC_review}. In two dimensions, the paradigmatic example is the spinless $2e$ $p+ip$ TSC, whose vortices trap Majorana zero modes~\cite{volovik2003_universe,Read1999_pairfermion,Ivanov2000_nonabelian_pwave}. These zero modes can also be realized at the boundary of a one-dimensional $p$-wave superconductor known as the Kitaev chain~\cite{kitaev2001unpaired}. 
The resulting non-locally encoded Hilbert space and non-Abelian statistics offer a robust route to storing and manipulating quantum information with hardware-level protection against local decoherence. Yet a central bottleneck remains: in the standard four-vortex qubit encoding, vortex-braiding only generates single-qubit Clifford gates~\cite{BRAVYI2002_fermionicQC}, while realizing the full multi-qubit Clifford group typically requires measurements or redundant encodings~\cite{Karzig2016_Majorana_architect,Calzona2020_majoranaCNOT,frey2025_MajoaranaEntanglingGate}. Computational universality further relies on resource intensive magic state distillation protocols~\cite{Bravyi2005_Cliffordmagic} or rather complicated ``twisted"\cite{bonderson2013_twistedinterferometry} or ``tilted"\cite{Freedman2006_tilted,bonderson2010_blueprinttopologicallyfaulttolerantquantum} interferometry schemes involving multiple probe anyons or time-dependent, topology-changing deformations of the medium,  which are difficult to realize. 

Recent progress in Van der Waals stacked 2D materials has underscored a broader lesson: assembling well-understood building blocks can unlock qualitatively new phases that are absent in any single component. Motivated by this ``materials-by-design" viewpoint, we ask: Can we use an analogous strategy to elevate the computational power of TSCs by using $2e$ TSC not as an endpoint, but as an ingredient?

Here we answer this question affirmatively by showing that a $4e$ TSC can emerge as a quantum ``vestigial order"~\cite{Demler2002_vortex_frac,berg2009_4e,Radzihovsky2009_FractionalVortices,Herland2010_metallicSF,Fernandes2019_vestigial,Ge2024_4e,volovik2024_fermionic} from a stack of two copies of $2e$ TSCs with identical electron density (we will also discuss an alternative route to this state and its higher-charge counterparts via melting lattice Jain states towards the end). Concretely, we argue that a strong inter-component current-current coupling can drive the condensation of vortex-antivortex pairs across the two components and give rise to a $4e$ TSC. 

This $4e$ superconductor (SC) coexists with intrinsic topological order - specifically a  chiral bosonic topological order with $\mathbb{Z}_3$ anyons. Furthermore, the charge $4e$ condensate leaves behind an unbroken $\mathbb{Z}_4$ symmetry, which acts nontrivially on the anyons by permuting their types.
This symmetry enrichment leads to two important consequences for vortices nucleated by an externally applied $hc/(4e)$ flux. Since vortices act as defects of the $\mathbb{Z}_4$ symmetry, they change the type of any anyon that circles them. Moreover, they trap $\mathbb{Z}_3$ parafermion zero modes.

Since $\mathbb{Z}_3$ parafermions have a larger quantum dimension than their $\mathbb{Z}_2$ Majorana counterparts, they naturally encode qu{\it trits} rather than qubits. In the standard encoding using four vortices, braiding alone already generates the full multi-qutrit Clifford gates~\cite{Hutter2015_parafermion_QC}. To reach universality, we further introduce an  interferometric measurement requiring only a single probe vortex, which enables an efficient, topologically protected projection onto non-stabilizer states. This operation promotes Clifford gates to a universal gate set via magic-state injection~\cite{Bravyi2005_Cliffordmagic}.

An advantage of our approach is that the non-Abelian computational objects are externally defined. In proposals based on intrinsic non-Abelian topological orders, identifying and reliably pinning specific non-Abelian anyons can be challenging; here by contrast, the non-Abelian objects are uniquely trapped by $hc/(4e)$ fluxes. Also, unlike schemes based on discrete symmetry defects such as dislocations~\cite{Bombin2010_dislocation,Yoder2017_surfacecodetwist,Barkeshli2012_nematic_disloc} or genons~\cite{Barkeshli2012_genon,Barkeshli2013_abelian_defect}, the discrete symmetry in our system emerges from spontaneously breaking a continuous one, rendering the defects (SC vortices) intrinsically mobile. Consequently an external flux—implementable in principle by bringing a fluxonium~\cite{Manucharyan2009_Fluxonium} adjacent to the thin-film TSC—provides a natural handle for fusing and braiding these computational elements. Since all the operations are defined by point defects within the bulk, this platform offers a native route to TQC, that does not rely on engineered interfaces in heterostructures~\cite{LIndner2012_fractionalizing,Cheng2012_SC_FCI_proximity,clarke2013_exotic,clarke2014_hybrid,Mong2014_FQHSChybrid,Lian2018_chiralMajorana,Dua2019_halffluxon}. The tradeoff is that the operations must be performed sufficiently slowly to avoid exciting the gapless Goldstone (phase) mode of the superconductor. However, we show that in the presence of unscreened long-range Coulomb interactions, the resulting critical velocity of vortices does not decrease with increasing system size, paralleling the situation in fully gapped topologically ordered phases.

More broadly, our results point to a novel strategy for engineering ``ultra-quantum" matter, i.e. matter with long-range entanglement, where electron aggregation acts in concert with fractionalization to yield topological defects with enhanced computational functionalities.

\section{Setup: two-component \texorpdfstring{$p+ip$}{} superconductor}

Consider an interacting system of two complex fermion species $c_{\sigma}$ in two spatial dimensions with a global $U(1) \times U(1)$ symmetry associated with the conservation of charge for each species (pseudospin). We assume that the two species are equally populated and both experience zero external magnetic field. This abstract setup can be realized by a spinful electronic system with $S_z$ conservation or a bilayer spin-polarized electron system with suppressed interlayer tunneling. 

Initially, we choose parameters in the Hamiltonian such that each fermion species forms a $p+ip$ $2e$ TSC with broken time-reversal symmetry (generalizations to other TSCs with odd Majorana Chern numbers are straightforward). We refer to this phase as $(p+ip)^2$. This state spontaneously breaks the $U(1) \times U(1)$ symmetry down to a $\mathbb{Z}_2^f \times \mathbb{Z}_2^f$ subgroup, where each $\mathbb{Z}_2^f$ factor corresponds to the fermion parity of each species. 

Fundamental to each symmetry-broken phase are its low-energy excitations and symmetry defects. For a thin-film $2e$ SC, the low energy excitations are gapped Bogoliubov quasiparticles, while the symmetry defects are vortices of its order parameter. In the $(p+ip)^2$ phase, we have two species of vortices $v_{\sigma}$, each with $2\pi$ winding of the SC order parameter $\Delta_{\sigma} \sim e^{i\theta_{\sigma}}$. We denote the composite vortex with $2\pi n$ winding of $\theta_{\sigma}$ by $v_{\sigma}^n$, where $n = -1$ maps to the elementary antivortex $\bar v_{\sigma}$. 

\section{Vortex-antivortex condensation and charge-\texorpdfstring{$4e$}{} superconductivity}

\begin{figure}
    \centering
    \includegraphics[width=0.8\linewidth]{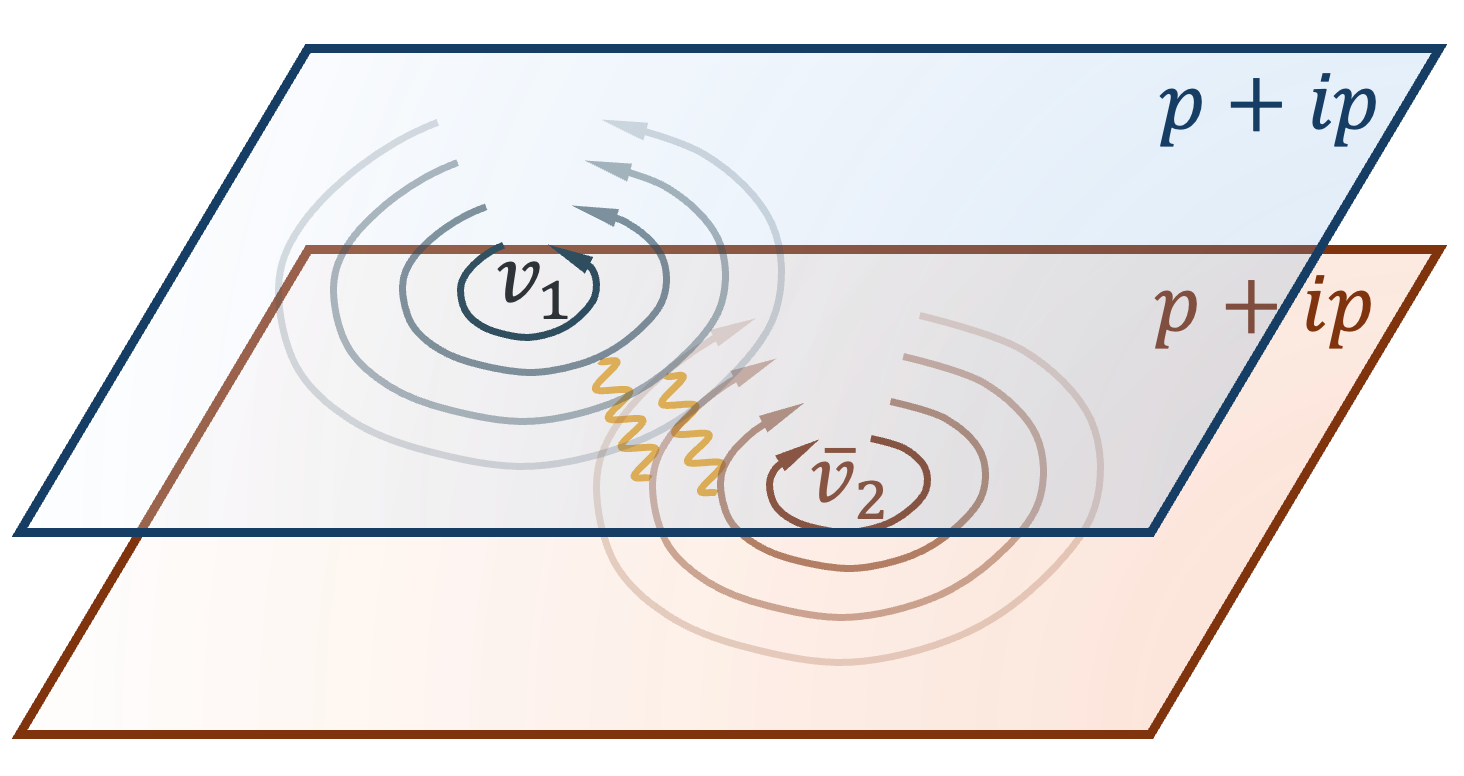}
    \caption{An illustration of vortex-antivortex binding mediated by inter-component current-current coupling between two copies of $p+ip$ TSCs. The condensation of this composite object leads to the $4e$ TSC in \eqref{eq:L_pattern_1}, in which superconductivity coexists with a bosonic chiral $\mathbb{Z}_3$ topological order enriched by the remnant $\mathbb{Z}_4$ charge symmetry. }
    \label{fig:vortex}
\end{figure}

Starting from a general symmetry-broken phase, we can access neighboring phases by condensing different types of symmetry defects. In the $(p+ip)^2$ state, the most general symmetry defects are composite vortices of the form $v_1^n v_2^m$, where $n$ and $m$ are integers. By the dictionary of particle-vortex duality~\cite{Peskin1978_latticePV,Dasgupta1981_latticePV,Fisher1989_continuumPV}, a single vortex $v_{\sigma}$ sees the average particle density $\ev{\rho_{\sigma}}$ as a background magnetic field and gets trapped in localized cyclotron orbits. On the other hand, balanced vortex-antivortex composites $(v_1 \bar v_2)^n$ see zero magnetic field on average and can condense when $v_1$ and $\bar v_2$ experience a sufficiently strong effective attraction. The origin of this attraction can be understood in terms of the charge/pseudospin stiffness $\rho_c$/$\rho_s$ that enters the Hamiltonian for phase fluctuations:
\begin{equation*}
    H_{\rm phase} = \frac{\rho_c}{2} (\nabla \theta_1 + \nabla \theta_2)^2 + \frac{\rho_s}{2} (\nabla \theta_1 - \nabla \theta_2)^2 + \ldots \,.
\end{equation*}
Generically, the inter-species local current-current interaction, $\bm{J}_1\cdot \bm{J}_2\sim (\nabla \theta_1)\cdot (\nabla \theta_2)$ (Andreev-Bashkin coupling~\cite{Andreev1975_threevelocity_hydro}), results in a larger stiffness for charge than for spin $\rho_c > \rho_s$, rendering the relative phase coherence more prone to quantum disordering than the total phase. In the dual perspective of vortices, this term indeed tends to bind $v_1$ and $\bar v_2$, since it is energetically favorable for supercurrents induced by $v_1$ and $\bar v_2$ to minimize their spatial separation (see Fig.~\ref{fig:vortex} for an illustration). This type of current-current interaction could in principle be enhanced through e.g. cavity mode resonance~\cite{sillanpaa2007_cavityQED} especially in the bilayer setup. 

When the current-current interaction is strong enough, the pseudospin $U(1)$ degree of freedom can undergo a symmetry-restoring Mott transition described by the condensation of $v_1 \bar v_2$. Since the phase fields $\theta_1, \theta_2$ wind around $v_1 \bar v_2$ by $2\pi$ and $-2\pi$, the charge-$4e$ Cooper quartet $\sim e^{i(\theta_1 + \theta_2)}$ does not see the phase winding and remains condensed, leading to a $4e$ SC. 

While the simple argument above fixes the symmetry-breaking pattern of the $4e$ SC phase, a more powerful field-theoretic framework is needed to determine its topological properties.

Towards that end, we invoke an elegant effective Lagrangian for a single topological $p+ip$ SC that has been proposed in the vortex basis~\cite{Ma2020_FCIQCD,Shi2025_dopeMR}. The Lagrangian takes the form of a $U(2)$ non-Abelian Maxwell-Chern-Simons theory (see App.~\ref{app:U(2)_charge2e} for a review)
\begin{equation}\label{eq:L_p+ip}
    \begin{aligned}
    &L_{p+ip}[\phi, \alpha, A] \\
    &= - \frac{2}{4\pi} \Tr \left(\alpha \, d \alpha + \frac{2}{3} \alpha^3\right) + \frac{1}{4\pi} \Tr \alpha \, d \Tr \alpha \\
    &+L[\phi, \alpha] + \frac{1}{2\pi} \Tr \alpha \, d A - \mathrm{CS}[A,g] + \frac{1}{2\rho} \Tr F_{\alpha}^2 \,,
    \end{aligned}
\end{equation}
where $\alpha$ is a dynamical $U(2)$ gauge field, $F_{\alpha} = d \alpha + \alpha^2$ is the non-Abelian field strength, $A$ is a background $U(1)$ electromagnetic gauge field, and $\phi$ is minimally coupled to $\alpha$ in the fundamental representation of $U(2)$. The background term $\mathrm{CS}[A,g] = \frac{1}{4\pi} A \, d A + \Omega_g$ is the response theory of a fermionic integer quantum Hall (IQH) state on a curved manifold with metric $g$. The quantized coefficient in front of the gravitational CS term $\Omega_g$ tracks the unit chiral central charge of the IQH state. 

The universal properties of \eqref{eq:L_p+ip} can be deduced by decomposing the $U(2)=SU(2)\times U(1)/\mathbb{Z}_2$ gauge field as $\alpha = \alpha^0 T^0 + \sum_{a=1}^3 \alpha^a T^a$, where $T^0 = I/2, T^a \equiv \sigma^a/2$ are generators of the $\mathfrak{u}(2)$ Lie algebra. The Chern-Simons (CS) term for the $SU(2)$ components $\alpha^a$ endows the elementary vortex with non-Abelian statistics. On the other hand, the $U(1)$ component $\alpha^0$ has no CS term and its dominant low energy coupling comes from the Maxwell term appearing in $\Tr F_{\alpha}^2$. The physical current of electrons $J^\mu =\frac{1}{2\pi} \epsilon^{\mu\nu\eta} \partial_\nu \alpha^0_{\eta}$ is determined by the $U(1)$ part of the gauge field much like in the conventional particle-vortex duality. The coefficient of this Maxwell term is thus related to the superfluid stiffness and determines the vortex energetics.

Now let us consider a pair of $p+ip$ SCs coupled to a pair of external $U(1)$ gauge fields $A_1, A_2$
\begin{equation*}
    L_{(p+ip)^2} = L_{p+ip}[\phi_1, \alpha_1, A_1] + L_{p+ip}[\phi_2, \alpha_2, A_2] + L_{\rm int}  \,,
\end{equation*}
where $\alpha_1, \alpha_2$ are dynamical $U(2)$ gauge fields and $L_{\rm int}$ encodes interactions between the two species. Since $\phi_{\sigma=1,2}$ represents the elementary vortex $v_\sigma$, the vortex-antivortex pair $v_1 \bar v_2$ maps to a matter field $\Phi \sim \phi_1 \otimes \phi_2^{\dagger}$, which transforms in the bi-fundamental representation of $U(2) \times U(2)$. Near the quantum phase transition associated with the condensation of $v_1 \bar v_2$, $L_{\rm int}$ can be described by an effective Lagrangian for $\Phi$
\begin{equation}\label{eq:L_matter}
    \begin{aligned}
    &L_\text{int}[\Phi, \alpha_1 \otimes I_2 - I_1 \otimes \alpha_2] = \Tr \left[D_{\mu} \Phi\right]^{\dagger} \left[D_{\mu} \Phi \right] \\
    &+ m \Tr \,\Phi^{\dagger}\Phi + \lambda_1 \Tr \,\Phi^{\dagger} \Phi \Phi^{\dagger} \Phi + \frac{\lambda_2}{2} \left(\Tr \,\Phi^{\dagger}\Phi\right)^2  \,,
    \end{aligned}
\end{equation}
where the covariant derivative $D_{\mu}$ is defined as
\begin{equation}
    D_{\mu} \Phi = \partial_{\mu} \Phi - i \sum_{a=0}^3 \alpha^a_{1,\mu} T^a \Phi + i \sum_{a=0}^3 \Phi \alpha^a_{2,\mu} T^a \,. 
\end{equation}
In general, the condensation of $\Phi$ in this non-Abelian theory does not lead to a unique outcome. This ambiguity is very physical: since $\Phi$ is a 2 by 2 matrix, different choices of interactions could lead to different internal orientations of the $\Phi$ condensate which preserve distinct subgroups of the $U(2) \times U(2)$ gauge group.

The key parameter that determines the preferred condensate is the coupling $\lambda_1$. While the precise value of $\lambda_1$ depends on microscopic details, we argue in App.~\ref{app:fusion_splitting} that it is positive whenever the interaction between a pair of spin-$1/2$ fields $\phi_{\sigma}$ favors the singlet spin-$0$ channel over the triplet spin-$1$ channel. Remarkably, this energetic condition is generically satisfied when the effective field theory \eqref{eq:L_p+ip} is in a weak-coupling regime. From now on, we will proceed with the assumption $\lambda_1 > 0$ and leave a thorough discussion of the alternative scenario $\lambda_1 < 0$ to App.~\ref{app:Higgs_classification}. 

With $\lambda_1 > 0$, a straightforward analysis in App.~\ref{app:Higgs_classification} shows that the $\Phi$ condensate prefers to be proportional to the identity matrix, thereby preserving a diagonal $U(2)$ subgroup of $U(2) \times U(2)$. In terms of the remaining $U(2)$ gauge field $\alpha \equiv \alpha_1 = \alpha_2$, the final Lagrangian reads
\begin{equation}\label{eq:L_pattern_1}
    \begin{aligned}
    L_{4e} &= -\frac{4}{4\pi} \Tr \left(\alpha \, d \alpha + \frac{2}{3} \alpha^3\right) + \frac{2}{4\pi} \Tr \,\alpha \, d \,\Tr \, \alpha \\
    &+ \frac{1}{2\pi} \Tr\, \alpha \, d (A_1 + A_2) - \mathrm{CS}[A_1,g] - \mathrm{CS}[A_2,g] \,. 
    \end{aligned}
\end{equation}
The $U(1)$ part of $\alpha$ has no self-CS term but has a level-$4$ mutual CS term with the external charge gauge field $A_c = (A_1 + A_2)/2$, indicating that \eqref{eq:L_pattern_1} describes a non-Abelian $4e$ SC (see App.~\ref{app:higher_charge} for a more precise argument). The additional background CS terms for $A_1$ and $A_2$ encode a stacked invertible fermionic phase exhibiting a spin quantum Hall effect with coefficient $-2$~\cite{Senthil1999_SQHE}. Across the transition, the total chiral central charge $c_-$ changes from $1/2+1/2=1$ to $0$. This change in chiral central charge is allowed since $\Phi$ does not source a bosonic quasiparticle in the parent $(p+ip)^2$ phase.

We note that a parton approach can bridge the above field theory to a more microscopic, wavefunction-based description. The idea is to decompose the electron operator of each species as
\begin{align}\label{eq:main_parton}
    c_\sigma =  b_{\sigma,\alpha} \epsilon_{\alpha\alpha'} f_{\sigma,\alpha'}
\end{align}
where $b_{\sigma, \alpha=1,2}$ and $f_{\sigma, \alpha=1,2}$ are a bosonic and a fermionic two-component operator. This decomposition has a $U(2)$ gauge redundancy, under which the $b,f$ fields transform as fundamentals (spin-$1/2$) of the $SU(2)$ part and carry opposite unit charge under the $U(1)$ part. The $(p+ip)^2$ SC state can be projectively constructed~\cite{Wen1990_Projective_construction} through an ansatz where $b,f$ experience opposite internal magnetic fields, which put them into (bosonic and fermionic) IQH states at opposite fillings. In this description, the $\Phi$ field describing the vortex-antivortex pair corresponds to an inter-species exciton of $f$ (or $b$)
\begin{align}
    \Phi_{\alpha\alpha'} \sim f^{\dagger}_{1,\alpha}  f_{2,\alpha'}.
\end{align}
Putting $b,f$ in IQH states for the exciton-hybridized bands yields a wavefunction of the $4e$ SC in \eqref{eq:L_pattern_1} (see App.~\ref{app:parton} for details). We note that tuning the parton dispersions and the exciton pairing wavefunctions does not affect the topological properties of the resulting state. The above projective construction therefore provides a family of trial wavefunctions which may be used in future numerical studies.

\section{Topological order of the \texorpdfstring{$4e$}{} TSC}

Let us now derive the symmetry-enriched topological order (SET) of the $4e$ TSC described by \eqref{eq:L_pattern_1}, treating $A_1, A_2$ as background fields for the $U(1) \times U(1)$ global symmetry. Since the invertible sector described by $-\sum_{i=1}^2 \mathrm{CS}[A_i,g]$ does not have any intrinsic topological order, we will focus on the non-invertible $\alpha$-sector.

As a CS theory, the $\alpha$-dependent part of \eqref{eq:L_pattern_1} describes $U(2)_{4,0}$, which factorizes as $U(2)_{4,0} = SU(2)_{4} \times U(1)_0/\mathbb{Z}_2$~\cite{Seiberg2016_gapped_bdry}. The gapless $U(1)_0$ sector is a topologically trivial Maxwell theory corresponding to a dual description of SC phase fluctuations. The gapped $SU(2)_{4}$ sector is a bosonic topological order with five anyons labeled by their $SU(2)$ spin $j = \{0,1/2,1,3/2,2\}$~\cite{Witten1989_CSjones}. The $\mathbb{Z}_2$ quotient is implemented by gauging the $\mathbb{Z}_2$ 1-form symmetry generated by a Wilson line in the $j = 2$ representation of $SU(2)$. This gauging procedure identifies the $j = 2$ anyon in $SU(2)_4$ with a local excitation, which is the microscopic Cooper pair in the $4e$ SC. Since $j = 2$ braids non-trivially with $j=1/2$ and $j = 3/2$, these half-integer spin lines get confined~\cite{Bais2009_anyon_condensation,Eliens2013_anyon_condensation,Kong2013_anyon_condensation,Burnell2017_anyon_condensation}. Moreover, since $1 \times 1 = 0 + 1 + 2$ in $SU(2)_{4}$, the identification of $2$ with $0$ (up to local operators) leads to a two-fold vacuum multiplicity in the pair fusion of $j = 1$ lines. This multiplicity implies that the $j=1$ line must split into a pair of lines $1_+$ and $1_-$ in the quotient theory $SU(2)_4/\mathbb{Z}_2$ with identical topological spin $\theta = e^{2\pi i/3}$ and fusion rule $1_+^2 = 1_-, 1_-^2 = 1_+$~\cite{Bais2009_anyon_condensation}. The complete topological order $\{0, 1_+, 1_-\}$ is precisely the chiral $\mathbb{Z}_3$ bosonic topological order, equivalent to the Jain state of bosons at Landau level filling $\nu = 2/3$. 

\begin{figure}
    \centering
\includegraphics[width=0.8\linewidth]{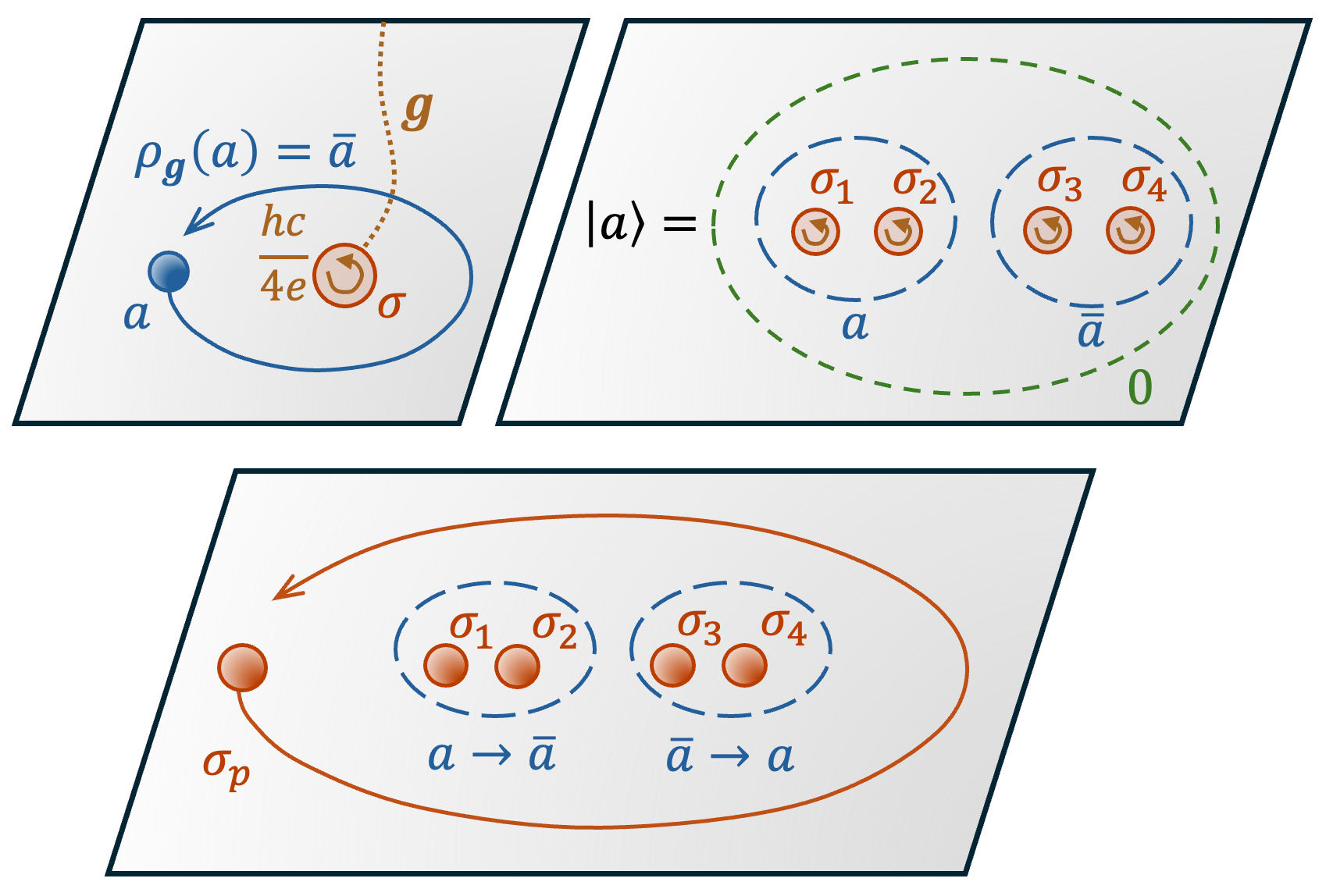}
    \caption{(Upper Left) An illustration of the anyon transmutation effect, in which 
    an Abelian anyon $a$ winding around an elementary $\mathbb{Z}_4$ defect $\sigma$ induced by $hc/(4e)$ flux gets acted on by $\rho_{\bs{g}}$ at the branch cut of phase field, and transmutes into its conjugate partner $\bar a$.  (Upper Right) The qutrit encoding using the fusion space of four $hc/(4e)$ defects split from vacuum, where $a=0,1,2$ represents the topological charge of the Abelian anyon fused from the first two $\sigma$'s.  (Lower) A path of the probe defect $\sigma_p$, which induces a transformation $|a\rangle\leftrightarrow |\bar{a}\rangle$ of the qutrit. \label{fig:illustration}}
\end{figure}

We now ask how symmetries of the $4e$ SC act on these anyons. Since $U(2)_{4,0}$ only couples to the charge gauge field $A_c = (A_1 + A_2)/2$, the spin $U(1)_s$ symmetry acts trivially on $U(2)_{4,0}$ and can be neglected. Focusing on the charge sector, we see that the $4e$-condensate spontaneously breaks $U(1)_c$ down to a $\mathbb{Z}_4$ subgroup generated by $\bs{g}$, with $\bs{g}^2$ acting as the fermion parity $(-1)^F$. In the low energy TQFT, every group element $\bs{g}$ embeds as a map $\rho_{\bs{g}}$ on the set of anyons. Since $SU(2)_4/\mathbb{Z}_2$ has no local fermions, $\rho_{\bs{g}^2}$ must be the identity map and it suffices to understand the action of $\rho_{\bs{g}}$. In App.~\ref{app:U(2)40_TO}, we show that the Wilson lines corresponding to $1_+, 1_-$ in the quotient theory can be decomposed as $1_{\pm} = 1 \pm 1'$ where $1, 1'$ are nonsimple Wilson lines of the parent $SU(2)_4$ theory that differ by a microscopic Cooper pair. As $1_+$ and $1_-$ are invariant under Cooper pair dressing (up to a phase), they define simple line operators in the quotient TQFT $SU(2)_4/\mathbb{Z}_2$. This field-theoretic discussion is consistent with the algebraic viewpoint that anyon types cannot be modified by local operations. Since the Cooper pair operator picks up a $-1$ factor under $\bs{g}$-action, we immediately see that $\rho_{\bs{g}}$ permutes $1_+$ and $1_-$. We therefore conclude that $U(2)_{4,0}$ is a $\mathbb{Z}_3$ chiral bosonic topological order enriched with an anyon permutation symmetry that squares to $(-1)^F$. For notational convenience, we will relabel the anyons as $1_+ \rightarrow 1, 1_- \rightarrow 2$ so that the fusion rule for $a, b \in \{0,1,2\}$ reduces to addition mod 3. 

Given a $G$-symmetric topological order, it is natural to consider its $G$-crossed extension which incorporates the interplay between extrinsic $G$-symmetry defects and intrinsic anyons of the SET~\cite{Barkeshli2014_SETbible} (see App.~\ref{app:SETreview} for a concise review of the $G$-crossed formalism). For the $4e$ TSC, defects of the unbroken $\mathbb{Z}_4$ symmetry are precisely SC vortices that can be trapped by externally applied localized magnetic fluxes. The distinct defect sectors are labeled by group elements $\bs{g}^n \in \mathbb{Z}_4$, with quantized flux $n hc/(4e)$. Since every localized $\bs{g}$-defect is attached to a semi-infinite branch cut of the SC phase, an Abelian anyon $a$ winding around a $\bs{g}$-defect necessarily crosses the branch cut and gets transmuted to $\rho_{\bs{g}}(a)$ (see Fig.~\ref{fig:illustration}). This braiding-induced anyon transmutation is an important feature of the $G$-crossed extension.

Although the $\mathbb{Z}_3$ chiral bosonic topological order is itself Abelian, its symmetry defects (i.e. SC vortices) can have non-invertible fusion rules and correspond to non-Abelian defects~\cite{Barkeshli2012_genon,Barkeshli2013_abelian_defect,Barkeshli2014_SETbible}. To elucidate the non-Abelian nature of these defects, we first restrict the theory to the bosonic sector on which $\bs{g}^2 = (-1)^F$ acts trivially. The unbroken symmetry group thus reduces to $\mathbb{Z}_2$ with a single nontrivial defect sector labeled by $\bs{g}$. In App.~\ref{app:U(2)40_TO}, we show that the elementary vortex $\sigma$ in the nontrivial defect sector has quantum dimension $\sqrt{3}$ and satisfies the fusion rule of a $\mathbb{Z}_3$ parafermion zero mode
\begin{equation}
\label{eq:fusion}
    \sigma \times \sigma = 0 + 1 + 2 \,.
\end{equation}
This parafermion mode is the simplest generalization of the celebrated Majorana zero mode trapped by elementary vortices in a single-component $p+ip$ SC, which is separated from other ($1,2,3$-particle) excitations by finite gaps inversely proportional to the corresponding correlation lengths. In contrast to deconfined anyons, vortices are confined excitations whose braiding induces not only a universal topological phase, but also a non-universal dynamical phase~\cite{Barkeshli2012_genon,Barkeshli2013_abelian_defect}. This means that the total braiding phase $R^{\sigma\sigma}_a$ between a pair of $\sigma$'s in a fixed Abelian fusion channel $a \in \{0,1,2\}$ is not well-defined. Nevertheless, since the non-universal phase does not depend on the fusion channel $a$, the relative ratio between $R^{\sigma\sigma}_a$ and $R^{\sigma\sigma}_b$ is well-defined and takes the form
\begin{equation}
    R^{\sigma\sigma}_a/R^{\sigma\sigma}_b = (-1)^{a-b} e^{i\pi(a^2-b^2)/3} \,. 
\end{equation}
These braiding invariants and the fusion rules, together with other data specified in App.~\ref{app:U(2)40_TO} such as the $F$ symbol, define a $\mathbb{Z}_2$-crossed braided tensor category which characterizes the interplay between $\mathbb{Z}_3$ anyons and vortices in $U(2)_{4,0}$. While the discussion thus far assumes $(-1)^F = 1$, all excitations or defects in the fermionic sector can be obtained by attaching an electron to those in the bosonic sector.

Finally, let us briefly comment on the implication of our theory for bilayer quantum Hall states at Landau level filling $\nu = 1/2 + 1/2$. A widely known  candidate phase at this filling is a bilayer Pfaffian state, which can be mapped to a $p+ip$ TSC of composite fermions in both layers~\cite{Moore1991_nonabelian,Read1999_pairfermion}. Applying the vortex-antivortex condensation to composite fermions, we find a gapped non-Abelian quantum Hall state with topological order $U(2)_{4,-16}$ (see Refs.~\cite{Fradkin1998_LG_nonabelian,Goldman2019_LG_nonabelian,Goldman2020_nonAbelian_fermionization} for related constructions).  This state is closely related, though not identical, to the topological order obtained from the $4e$ TSC state by gauging the unbroken $\mathbb{Z}_4$ global symmetry (as shown in App.~\ref{app:U(2)40_TO}), and it also supports universal TQC through braiding, fusion and measurement~\cite{Hastings2012_metaplectic_power,Cui2014_SU(2)4_gate,Levaillant2015_SU(2)4_gate,Bocharov2015_SU(2)4_compilation}.

\section{Universal quantum computation from vortex braiding and measurement}

The bulk realization of $\mathbb{Z}_3$ parafermions through $hc/(4e)$ fluxes promises enhanced computational power relative to the celebrated Majorana zero modes in $2e$ TSCs. Specifically, the fusion rule \eqref{eq:fusion} implies that every four $\sigma$ vortices can encode a single logical qutrit. The qutrit Hilbert space is spanned by three distinct intermediate channels ($a=0,1,2$) that appear when a full $hc/e$ flux splits into four isolated $hc/(4e)$ fluxes via the splitting tree $0\rightarrow a\bar{a}\rightarrow \sigma\sigma\sigma\sigma$ (Fig.~\ref{fig:illustration}). Physically, a single $hc/(4e)$ flux can be induced by bringing a fluxonium--i.e., a $4e$ SC ring which need not be topological--adjacent to the $4e$ TSC thin film. In App.~\ref{app:Clifford_gates}, following the protocol described in Ref.~\cite{Hutter2015_parafermion_QC}, we demonstrate that successive braiding of $\sigma$ vortices along with the preparation of a tensor product of $|0\rangle$ states can indeed generate all multi-qutrit Clifford gates.

To go beyond Clifford operations and achieve computational universality, the key ingredient is to prepare external fluxes in a quantum superposition of trajectories,
which allows us to design a topologically protected protocol for magic state preparation (i.e. ancilla qutrits in a non-stabilizer state~\cite{Bravyi2005_Cliffordmagic}). 
Specifically, in App.~\ref{app:magicstate}, we describe an interferometry measurement in which a probe $\sigma$ defect winds around four $\sigma$'s that encode the ancilla qutrit. Physically, the interference between the probing and reference arms for the probe $\sigma$ defect can be achieved by preparing a pair of fluxoniums that share a single $hc/(4e)$ flux in a Bell state, sending them through the two paths, and then measuring the relative phase between the two flux states. Note that this recipe is specific to our realization and does not generalize to true $SU(2)_4$ anyons
~\cite{Hastings2012_metaplectic_power,Cui2014_SU(2)4_gate,Levaillant2015_SU(2)4_gate,Bocharov2015_SU(2)4_compilation} (see App.~\ref{app:comparison_SU(2)4}). An advantage of our protocol relative to $SU(2)_4$ is that our qutrit encoding involves four fluxes which fuse into the vacuum. This is in contrast to the standard qutrit encoding in $SU(2)_4$~\cite{Cui2014_SU(2)4_gate} where four $j = 1/2$ anyons fuse into a non-Abelian $j = 1$ anyon, which is more difficult to prepare and manipulate.

As illustrated in Fig.~\ref{fig:illustration}, the key physics underlying this design is that the winding of $\sigma$ around an Abelian anyon induces a transmutation $a \leftrightarrow \bar{a}$, thereby rotating the $|1\rangle,|2\rangle$ states of the ancilla qutrit. The corresponding unitary operation has a peculiar feature that it can be diagonalized in the basis $|0\rangle,|\pm\rangle\equiv ( |1\rangle\pm |2\rangle) /\sqrt{2}$ with distinct eigenvalues for $|\pm\rangle$. This feature allows one to collapse any initial state within the span$\{|1\rangle, |2\rangle\}$ subspace onto $\ket{+}$ or $\ket{-}$ through repeated interferometry measurements with post-selection. A slightly more complex probing path further allows arbitrary qutrit initial states outside of span$\{|1\rangle, |2\rangle\}$ (App.~\ref{app:magicstate}). Since $\ket{+}$ and $\ket{-}$ are both non-stabilizer qutrit states, this protocol provides an efficient and topological way of preparing ancilla qutrits for magic state injection~\footnote{A somewhat similar gate set was recently realized on a quantum processor for $D(S_3)$, the quantum double of the group $S_3$. There is a deep relation between these theories. For instance, $D(S_3)=\left [ SU(2)_4\times SU(2)_{-4}\right ]/\mathbb{Z}_2$~\cite{LeoExpt,LeoTheory}}. We therefore conclude that our non-Abelian $4e$ TSC is sufficient for universal quantum computation in a topologically robust way. 

The preceding protocol works in an ideal limit where the vortices move adiabatically in a noiseless environment at zero temperature. In a realistic setup, all of these assumptions need to be relaxed. While a complete treatment of all decoherence channels is beyond the scope of this work, we analyze two of the most prominent channels in TSC-based platforms.

The first channel is quasiparticle poisoning due to gapped excitations outside the logical subspace. Such poisoning events can originate from excited quasiparticles intrinsic to the system or injection of electrons from the environment. Excitations intrinsic to the system can be exponentially suppressed so long as the temperature is much smaller than the excitation gap. However, particle injection from the environment (e.g. the lead) is less controlled. Notably, in the Majorana platform, a local electron from the environment can flip the parity sector of the Majorana qubit and cause logical errors. In contrast, our $4e$ TSC is robust against this kind of poisoning. Specifically, since $1$ and $2$ are deconfined anyons, the injection of external electrons/Cooper pairs does not change the outcome of various braiding/fusion operations in our protocol. Thus, unlike Majorana based platforms, this system is not susceptible to environment-induced quasiparticle poisoning.


The second channel involves the inevitable gapless Goldstone modes that exist in both the Majorana $p+ip$ platform and our $4e$ TSC. This is the downside of allowing vortices to be freely mobile. Such excitations can mediate long-range communication between spatially separated information-encoding regions and thereby compromise topological protection. Consequently, operations must be performed adiabatically even at zero temperature, with velocities of moving vortices kept below a scale $v_0$ to avoid populating the Goldstone modes. In App.~\ref{app:goldstone} we derive upper bounds on how $v_0$ scales with the system’s linear size $L$ for two experimentally relevant setups. We find that $v_0$ remains $\mathcal{O}(1)$ in an unscreened environment with long-range Coulomb interactions. However,  $v_0$ scales as $\mathcal{O}(1/\sqrt{L})$ and vanishes for large system sizes when Coulomb interactions are screened. Thus, in the former case the protection can remain essentially as robust as in fully gapped topologically ordered phases.

\section{Alternative route towards \texorpdfstring{$4e$}{} and higher-charge SC}

The condensation of vortex-antivortex pairs in a two-component $p+ip$ SC is not the only mechanism for realizing the $U(2)_{4,0}$ $4e$ SC. Here, we briefly describe another path towards the same SC (up to stacking invertible phases) through a quantum phase transition out of the lattice Jain state at band filling $\nu = 2/3$ (this route was proposed independently in Ref.~\cite{Shi2025_FCISC}). The lattice Jain-SC transition we describe occurs in a single-component fermionic system with $U(1)$ symmetry, and there is no separate pseudospin $U(1)$ symmetry.

To describe this alternative route, we begin with the parton construction $c = b_{\alpha} \epsilon_{\alpha\alpha'} f_{\alpha'}$ introduced previously in \eqref{eq:main_parton}. The Jain state at $\nu = 2/3$ can be regarded as two copies of the fermionic integer quantum Hall state stacked with a fractional quantum Hall state at $\nu = -4/3$. The state at $\nu = -4/3$ can be described through a mean-field ansatz in which $b_{\alpha}/f_{\alpha}$ forms a bosonic/fermionic Chern insulator with Hall conductance $\sigma^{xy}_b = -2$/$\sigma^{xy}_f = -4$ (see App.~\ref{app:jain}). A large class of transitions out of the Jain state occur through a jump in the Hall conductance of $b_{\alpha}/f_{\alpha}$. Importantly, since each parton sees $\pm 2\pi /3$ flux per unit cell, lattice translations act projectively on the parton bands, leading to a three-fold internal degeneracy in the band structure of each $b_{\alpha}/f_{\alpha}$~\cite{Cheng2015_translationSET,Shi2024_dopeFQAH}. This means that across any band inversion transition that preserves the $U(2)$ gauge symmetry, the Hall conductance of each parton must jump by $\pm 6$. 

An interesting transition that satisfies these constraints involves a jump in $\sigma^{xy}_b$ from $-2$ to $4$. We show in App.~\ref{app:jain} that the resulting state is a $4e$ SC described by the Chern-Simons theory $U(2)_{4,0} \times U(1)_{-1}^2$, whose non-invertible sector is identical to the $4e$ SC arising from vortex-antivortex proliferation in \eqref{eq:L_pattern_1}.

Despite their apparent differences, there is a remarkable similarity between the $\nu = 2/3$ fermionic Jain state and the $4e$ SC $U(2)_{4,0}$, once the global $U(1)$ symmetry is explicitly broken down to $\mathbb{Z}_4$. Combining the analysis in this work and in Ref.~\cite{Shi2025_anyondeloc}, we know that the intrinsic topological order realized by both theories is the $\mathbb{Z}_3$ chiral bosonic topological order stacked with two copies of fermionic IQH states. The only difference between them is in the symmetry enrichment: the generator of $\mathbb{Z}_4$ acts trivially on anyons of the Jain state, but permutes the nontrivial anyons in $U(2)_{4,0}$. The Jain-SC transition is therefore a transition between two distinct symmetry enrichments of the same underlying topological order, which is in some sense less exotic than transitions from the Jain state to a topologically trivial BCS superconductor~\cite{Shi2025_anyondeloc,Nosov2025_anyonSC,Wang2025_FCISC,Pichler2025_anyonSC_criticality}, across which the intrinsic topological order changes. 

Finally, let us remark that the construction at $\nu = 2/3$ generalizes naturally to arbitrary fillings $\nu = p/(2p+1)$, where the corresponding Jain state transitions into a $2|p|e$ TSC described by $U(2)_{\pm 2p, 0}$. For $|p| \geq 3$, braiding of intrinsic anyons in $U(2)_{\pm 2p,0}$ already generates a universal gate set for quantum computation. 

The simplest example occurs at $\nu = 3/5$, where a $6e$ SC coexists with a non-Abelian intrinsic topological order,  $SU(2)_6/\mathbb{Z}_2 = SO(3)_3$,  represented as $\{1,s\}\times\{1,f\}$, where the second factor corresponds to the local fermion and $s$ is a semion with quantum dimension equal to the ``silver ratio",  $1+\sqrt 2$. This state is identical to the simplest topological order realized at the surface of a 3+1D TSC in class DIII~\cite{Fidkowski2013_surfaceTO}. 

\section{Conclusion}

In this paper, we identified a class of $4e$ topological superconductors which vortices bind $\mathbb{Z}_3$ parafermion zero modes. We highlighted two key and somewhat surprising conclusions. First,  we showed that this phase can emerge from relatively conventional ingredients: either by quantum disordering a bilayer of $p+ip$ SCs, or by melting a $\nu=2/3$ fractional quantum Hall state. In both cases, the transitions into the $4e$ TSC are described by QCD$_3$ (quantum chromodynamics) theories coupled to scalar fields or fermions. In this realization, parafermions are bound to \textit{bulk vortices} in contrast to  previous proposals in which parafermions exist as edge states or on twist defects of discrete lattice symmetries/layer-exchange symmetries. Second, we demonstrated that access to this phase enables a route to computational universality which can be achieved by supplementing parafermion braiding with a topologically protected interferometric measurement.

Our work highlights how the lens of computational power can shed new light on the underlying physics of topological phases, and vice versa. As a case in point, the universal gate set and qurit encoding identified here can be traced back to two key physical features of our topological phase.
The first is that we work with symmetry defects rather than the fully gauged theory. This is what allows a pair of vortices to encode a qurit. In contrast in the standard fusion-tree encoding of the fully gauged theory, $a$ and $\bar{a}$ are identified into a single non-Abelian anyon, which would yield only a qubit. Similar 'non-fusion-tree' encodings have appeared previously in the context of computing with quantum doubles~\cite{Preskill,LeoTheory,LeoExpt}. The second is that our system has a mixed-dimensionality character: the charge-$4e$ TSC occupies a 2D submanifold of 3D space, while the $U(1)$ symmetry is gauged in the ambient 3D space. This allows vortices to be defined via flux trapped in 3D superconducting rings and, crucially, to be placed in superposition, which  ultimately enables magic state preparation.

Our theory opens several exciting avenues for future research. From the perspective of condensed matter physics, a central goal is to identify  microscopic models that realize the transition from the $(p+ip)^2$ bilayer SC/$\nu=2/3$ Jain state into the $4e$ TSC (possibly building on recent progress in Refs.~\cite{Divic2024_anyonSC_criticality,Pichler2025_anyonSC_criticality,Kuhlenkamp2025_dopeCSL,Han2025_anyonexcitonSF_criticality}). Such constructions would provide deeper insight into the range of phenomena that can arise from hierarchical electron aggregation. From the perspective of quantum computation, the $4e$ TSC provides a novel setting for exploring the noise-robustness of topological quantum information processing. It would be fruitful to adapt and generalize some of the techniques developed for noisy Majorana platforms~\cite{Nayak2008_TQC_review} and gapped topological orders~\cite{Bao2023_decohere_transition,Fan2023_mixedstate,Wang2025_FQH_decohere} to clarify the effect of thermal fluctuations, local noise, and gapless Goldstone modes on our braiding/interferometric operations. Addressing these decoherence mechanisms is a crucial step toward translating our theoretical proposal into experimental reality. Finally, characterizing higher-charge topological condensates and their computational functionalities represents a promising direction for future work.

\begin{acknowledgments}
    We thank Meng Cheng, Ho Tat Lam, Zhi-Qiang Gao, Clemens Kuhlenkamp, Hart Goldman and T. Senthil for discussions. ZDS is supported by a Leinweber Institute for Theoretical Physics postdoctoral fellowship at Stanford University and in part by the Gordon and Betty Moore Foundation EPiQS initiative, Grant GBMF8686.01. Z.~H. and A.~V. are supported by the Simons Investigator award, the Simons Collaboration on Ultra-Quantum Matter, which is a grant from the Simons Foundation (651440, A.~V.). S.R. is supported by the US Department of Energy, Office of Basic Energy Sciences, Division of Materials Sciences and Engineering, under Contract No.DE-AC02-76SF00515. 

    {\it Note added. } Near the completion of this work, we became aware of another study on higher-charge TSC~\cite{Gao2025_2ne}. While both works recognize the possibility of such states, they study distinct phases with different approaches and perspectives.
\end{acknowledgments}

\bibliography{TQC4e}

\pagestyle{plain}
\newpage
\appendix
\onecolumngrid

\section{Gauge theory of topological charge-\texorpdfstring{$2e$}{} superconductors}\label{app:U(2)_charge2e}

In this section, we review the construction of bulk effective field theories for arbitrary topological charge-$2e$ superconductors in 2+1 dimensions.

\subsection{Universal properties of topological charge-\texorpdfstring{$2e$}{} superconductors}

Topological charge-$2e$ superconductors in 2+1 dimensions are classified by a single index $\nu \in \mathbb{Z}$~\cite{Schnyder2008_TSC,Kitaev2009_periodic}. For $\nu = 1$, this state is famously realized by the weak-pairing phase of a $p+ip$ superconductor of spinless fermions~\cite{Read1999_pairfermion}. Phases with higher values of $\nu$ can be obtained by stacking multiple copies of $p+ip$ superconductors. The universal low-energy physics of a topological superconductor $\mathrm{TSC}_{\nu}$ can be uniquely specified by three properties:
\begin{enumerate}
    \item $\mathrm{TSC}_{\nu}$ is an invertible phase of matter.
    \item $\mathrm{TSC}_{\nu}$ spontaneously breaks the external $U(1)$ symmetry down to the fermion parity symmetry $\mathbb{Z}_2^f$. 
    \item $\mathrm{TSC}_{\nu}$ has chiral central charge $c_- = \nu/2$. On a manifold with boundary, this value of $c_-$ implies the existence of $\nu$ gapless Majorana edge modes with thermal Hall conductance $\kappa_{xy} = \nu \kappa_0$, where $\kappa_0 = \frac{\pi^2 k_B^2 T}{6}$ 
\end{enumerate}
When $\nu$ is an even integer, the chiral central charge $c_-$ is an integer. It is therefore natural to realize $\mathrm{TSC}_{\nu}$ as an Abelian Chern-Simons theory. When $\nu$ is an odd integer, the chiral central charge $c_-$ is a half-integer, implying that $\mathrm{TSC}_{\nu}$ cannot be realized by any Abelian Chern-Simons theory and the minimal description must involve non-Abelian gauge fields. In what follows, we will explicitly construct the minimal Abelian/non-Abelian Chern-Simons theory for even/odd integer $\nu$. 

\subsection{\texorpdfstring{$U(1)$}{} Abelian Chern-Simons theory for even \texorpdfstring{$\nu$}{}}

For every even $\nu$, we claim that the following Abelian Chern-Simons theory works
\begin{equation}
    L_{\nu \in 2 \mathbb{Z}} = \frac{2}{2\pi} \beta d A + \frac{\nu}{2} \mathrm{CS}[A,g] \,, \quad \mathrm{CS}[A,g] = \frac{1}{4\pi} A d A + \Omega_g \,, \quad \Omega_g = 2 \mathrm{CS}_g \,. 
\end{equation}
Here, $\beta$ is a dynamical $U(1)$ gauge field, $A$ is the external $U(1)$ gauge field (technically a $\mathrm{spin}_{\mathbb{C}}$ connection), and $g$ is a background metric introduced to keep track of the chiral central charge. Since $\beta$ has no self Chern-Simons term, it is clear that this Lagrangian describes an invertible phase. The mutual Chern-Simons term between $\beta$ and $A$ sets $2 A = 0$, implying that the $U(1)$ global symmetry is spontaneously broken down to $\mathbb{Z}_2^f$. Finally, since the chiral central charge of each $\mathrm{CS}[A,g]$ is 1, the total chiral central charge is precisely $c_- = \nu/2$. 

\subsection{\texorpdfstring{$U(2)$}{} non-Abelian Chern-Simons theory for odd \texorpdfstring{$\nu$}{}}

For odd $\nu$, we claim that the following $U(2)$ Chern-Simons theory works
\begin{equation}
    L_{\nu \in 2 \mathbb{Z} + 1} = - \frac{2}{4\pi} \Tr \left(\alpha \, d \alpha + \frac{2}{3} \alpha^3\right) + \frac{1}{4\pi} \Tr \alpha \, d \Tr \alpha + \frac{1}{2\pi} \Tr \alpha \, d A + \frac{\nu - 3}{2} \mathrm{CS}[A,g] \,. 
\end{equation}
By decomposing the $U(2)$ gauge field $\alpha$ as $\alpha = \alpha_0 T^0 + \sum_{a=1}^3 \alpha_a T^a$ where $T^0 = I/2$ and $T^a = \sigma^a/2$, one can check that there is no Chern-Simons level for the $U(1)$ part $\alpha_0$ and there is a Chern-Simons term for the $SU(2)$ part $\alpha_a$ with level $2$. Following standard conventions~\cite{Seiberg2016_gapped_bdry}, the part of the Lagrangian involving $\alpha$ therefore describes $U(2)_{k_1, k_2} = SU(2)_{k_1} \times U(1)_{k_2}/\mathbb{Z}_2$ with $k_1 = 2, k_2 = 0$. The additional term describes $(\nu-3)/2$ copies of the integer quantum Hall state $U(1)_1$. Combining these terms, the full Lagrangian describes $U(2)_{2,0} \times U(1)_1^{(\nu - 3)/2}$. For $\nu = 1$, this Lagrangian reduces to the Lagrangian constructed for the weak-pairing $p+ip$ superconductor of spinless fermions~\cite{Ma2020_FCIQCD,Shi2025_dopeMR}. 

Let us now explain why this Lagrangian produces the three defining properties of $\mathrm{TSC}_{\nu}$. The part of the Lagrangian involving $\alpha$ is a $U(2)$ Chern-Simons theory. The topological order of $U(2)_{2,0}$ is equivalent to $SU(2)_2/\mathbb{Z}_2$, where the $\mathbb{Z}_2$ quotient is implemented by stacking $SU(2)_2$ with a trivial fermion theory $\{0, c\}$ and condensing the bound state of $c$ with the $j = 1$ anyon in $SU(2)_2$. Since $SU(2)_2$ only has three anyons $\{0, 1/2, 1\}$, condensing $j = 1$ confines the only other nontrivial anyon $j = 1/2$ and trivializes the topological order. Therefore, we conclude that $U(2)_{2,0}$ describes an invertible phase. Since the integer quantum Hall state $U(1)_1$ described by the response theory $\mathrm{CS}[A,g]$ is invertible, stacking with copies of $U(1)_1$ preserves the invertible property of the superconductor.

The symmetry-breaking pattern of this superconductor is more subtle. For a general $U(2)$ gauge field defined on a spacetime manifold $M_3$, there is a flux quantization condition
\begin{equation}
    \int_{\Sigma_2} \frac{d \Tr \alpha}{4\pi} + \int_{\Sigma_2} \frac{w_2[PSU(2)]}{2} \in \mathbb{Z} \,.
\end{equation}
In this formula, $\Sigma_2$ is any closed 2-manifold in $M_3$ and $w_2[PSU(2)]$ is the second Stiefel-Whitney class of the $PSU(2)$-bundle on $M_3$. The implication of this condition is that a $2\pi$ flux of $\Tr \alpha$ can only exist in the presence of a $\pi$ flux in the $SU(2)$ center (i.e. $\int_{\Sigma_2} w_2[PSU(2)] = 1$), which carries non-Abelian gauge charge due to the non-Abelian Chern-Simons term. As a result, the minimal gauge-invariant monopole operator that condenses must insert $4\pi$ flux of $\Tr \alpha$. The mutual Chern-Simons term between $A$ and $\Tr \alpha$ implies that this monopole carries physical charge 2e, and its condensation gives rise to a charge-$2e$ superconductor that spontaneously breaks $U(1)$ down to $\mathbb{Z}_2^f$. 

Finally, we turn to the chiral central charge. Since anyon condensation does not modify the chiral central charge, the $U(2)_{2,0}$ theory inherits the chiral central charge of $SU(2)_2$, which is $c_- = 3/2$. Stacking with $(\nu-3)/2$ copies of $U(1)_1$ then gives the correct chiral central charge $c_- = (\nu-3)/2 + 3/2 = \nu/2$ for $\mathrm{TSC}_{\nu}$.

\section{Higher-charge topological superconductors described by \texorpdfstring{$U(2)$}{} gauge theories}\label{app:higher_charge}

In this section, we briefly review higher-charge SC* states that also admit a $U(2)$ gauge theory description. These states are natural generalizations of the $U(2)_{2,0}$ state considered in App.~\ref{app:U(2)_charge2e}.

The most general $U(2)_{k_1,k_2}$ Chern-Simons theory can be factorized as $SU(2)_{k_1} \times U(1)_{k_2}/\mathbb{Z}_2$ and is described by the Lagrangian 
\begin{equation}
    L_{U(2)_{k_1,k_2}} = - \frac{k_1}{4\pi} \Tr \left[\alpha \, d \alpha + \frac{2}{3} \alpha^3\right] + \frac{2k_1 + k_2}{4} \frac{1}{4\pi} \Tr \alpha \, d \Tr \alpha + \frac{n}{2\pi} \Tr \alpha \, d A + \frac{1}{4g^2} \Tr F_{\alpha}^2 \,,
\end{equation}
where $\alpha$ is a $U(2)$ gauge field and $A$ is the external electromagnetic gauge field. We will restrict to effective field theories that are emergeable from a microscopic electronic lattice model. Thus, $A$ is technically a $\mathrm{spin}_{\mathbb{C}}$ connection rather than a $U(1)$ gauge field~\cite{Seiberg2016_gapped_bdry}. In order for this Lagrangian to describe a superconductor, we must set $k_2 = 0$ so that the Maxwell term dominates in the $U(1)$ sector and gives rise to a gapless photon, which is dual to the $U(1)$ Goldstone mode. Moreover, since $2k_1 + k_2$ also needs to be an integer multiple of 4, we can further restrict to cases with $k_1 = 2k$ where $k$ is an arbitrary integer. The subclass of theories of interest to us are therefore
\begin{equation}
    L_{U(2)_{2k,0}} = - \frac{2k}{4\pi} \Tr \left[\alpha \, d \alpha + \frac{2}{3} \alpha^3\right] + \frac{k}{4\pi} \Tr \alpha \, d \Tr \alpha + \frac{n}{2\pi} \Tr \alpha \, d A + \frac{1}{4g^2} \Tr F_{\alpha}^2  \,. 
\end{equation}
As shown in Ref.~\cite{Shi2025_FCISC}, the spin/charge relation for microscopic lattice electronic systems restricts $n$ and $k$ to have the same parity. 

Let us now work out the flux quantization condition of this superconductor. Naively restricting to the Abelian part and fixing $\alpha = \alpha_0 I/2$, it appears that $\alpha_0$ couples to $A$ with a coefficient $n$. This coefficient suggests a charge $n$ superconductor. However, this conclusion is incorrect. Following the argument in Sec.~\ref{app:U(2)_charge2e}, we know that on a general spacetime manifold $M_3$, a $2\pi$ monopole of $\Tr \alpha$ can only exist in the presence of $\pi$ flux in the $SU(2)$ center. When $k = 0$, there is no non-Abelian CS term and the $\pi$ flux of $SU(2)$ does not carry a $SU(2)$ charge. As a result, the $2\pi$ monopole of $\Tr \alpha$ can directly condense and give a charge-$n$ superconductor. Since $k=0$, $n$ must be even, making this superconductor emergeable from an electronic system. On the other hand, for any $k \neq 0$, the $\pi$ flux in the $SU(2)$ sector carries a non-Abelian gauge charge under $\alpha$ and cannot directly condense. The minimal gauge-invariant monopole operator that condenses is a $4\pi$-flux monopole of $\Tr \alpha$, which carries electric charge $2n$. Condensing this monopole then gives a charge-$2n$ superconductor instead. Therefore, vortices in $U(2)_{2k,0}$ have flux quantization in units of $h/ne$ for $k = 0$ and $h/2ne$ for $k \neq 0$. 

As shown in Ref.~\cite{Shi2025_FCISC}, the topological order realized by the $U(2)_{2k,0}$ SC can be obtained by condensing the $j = k$ anyon in $SU(2)_{2k}$. The resulting TQFT is $SU(2)_{2k}/\mathbb{Z}_2 = SO(3)_{k}$, whose general properties can be obtained following the recipe of anyon condensation~\cite{Moore1989_anyon_condensation_CS,Bais2009_anyon_condensation,Eliens2013_anyon_condensation,Kong2013_anyon_condensation,Burnell2017_anyon_condensation}. For all $k > 2$, braiding operations in $SU(2)_{2k}$ can already generate a complete gate set for universal quantum computation~\cite{Freedman2000_CSuniversal}. This property is preserved in the quotient $SU(2)_{2k}/\mathbb{Z}_2$, so long as $d_j^2 \notin \mathbb{Q}$ for some integer spin $j$ Wilson line in $SU(2)_{2k}$. We briefly remark that for $k = 3$, the charge-6 superconductor $U(2)_{6,0}$ realizes a residual fermionic topological order $SO(3)_3$, which is the simplest non-Abelian topological order at the surface of a 3+1D topological superconductor in class DIII~\cite{Fidkowski2013_surfaceTO}. 

\section{Classification of Higgsing patterns for the vortex field theory}\label{app:Higgs_classification}

In this section, we classify different Higgs condensates of the effective matter Lagrangian \eqref{eq:L_matter}. We begin by writing the full Lagrangian for the two-component fermionic system as
\begin{equation}\label{eq:app_Lfull_matter}
    L = \sum_{i=1}^2 \left[-\frac{2}{4\pi} \Tr \left(\alpha_i \, d \alpha_i + \frac{2}{3} \alpha_i^3\right) + \frac{1}{4\pi} \Tr \alpha_i \, d \Tr \alpha_i + \frac{1}{2\pi} \Tr \alpha_i \, d A_i - \mathrm{CS}[A_i,g] \right] + L_{\rm matter}[\Phi, \alpha_1 \otimes I_2 - I_1 \otimes \alpha_2] \,,
\end{equation}
where $\alpha_1, \alpha_2$ are dynamical $U(2)$ gauge fields, $A_1, A_2$ are background fields for the two global $U(1)$ symmetries, and $L_{\rm matter}$ describes the dynamics of the vortex-antivortex pair $\Phi \sim v_1 \bar v_2$. Near the vortex-antivortex condensation transition, the matter Lagrangian admits an expansion in powers of $\Phi$ which we truncate to quartic order
\begin{equation}
\label{eq:LPhi}
    \begin{aligned}
    L_{\rm matter}[\Phi, \alpha_1 \otimes I_2 - I_1 \otimes \alpha_2] = \Tr \left[D_{\mu} \Phi\right]^{\dagger} \left[D_{\mu} \Phi \right] + m \Tr \,\Phi^{\dagger}\Phi + \lambda_1 \Tr \,\Phi^{\dagger} \Phi \Phi^{\dagger} \Phi + \frac{\lambda_2}{2} \left(\Tr \,\Phi^{\dagger}\Phi\right)^2 + \ldots  \,.
    \end{aligned}
\end{equation}
Since the vortex $v_1$ couples to $\alpha_1$ in the fundamental representation and the anti-vortex $\bar v_2$ couples to $\alpha_2$ in the anti-fundamental representation, the pair-field $\Phi$ is a $2 \times 2$ complex matrix in the bifundamental representation. The bifundamental covariant derivative $D_{\mu}$ is defined as
\begin{equation}
    D_{\mu} \Phi = \partial_{\mu} \Phi - i \sum_{a=0}^3 \alpha^a_{1,\mu} T^a \Phi + i \Phi \alpha^a_{2,\mu} T^a \,,
\end{equation}
where $T^0 = I/2$ and $T^a = \sigma^a/2$ are generators of the $\mathfrak{u}(2)$ Lie algebra. 

Let us now consider a state in which $\Phi$ forms a translation-invariant condensate $\ev{\Phi}$. Under the most general gauge rotation $\ev{\Phi} \rightarrow U_1 \ev{\Phi} U_2^{\dagger}$, we can always reduce $\ev{\Phi}$ to the canonical form $\ev{\Phi} = \mathrm{diag}(s_1, s_2)$ where $s_1, s_2 \geq 0$ are the singular values of $\ev{\Phi}$. Without loss of generality, we can rescale the parameters so that $m = -1$ and choose $s_1 \geq s_2$. In terms of $A = s_1^2 + s_2^2$ and $B = s_1^2 - s_2^2$, the interaction potential then takes the simple form
\begin{equation}
    V(A,B) = - A + \frac{\lambda_1 + \lambda_2}{2} A^2 + \frac{\lambda_1}{2} B^2 \,,
\end{equation}
where stability requires $\lambda_1 + \lambda_2 > 0$. We can now classify distinct condensates by the value of $\lambda_1$ and $\lambda_2$. 

\begin{enumerate}
    \item If $\lambda_1 > 0$, then the minimum of $V$ is achieved at $B = 0$. This corresponds to the case $s_1 = s_2 > 0$, which preserves a diagonal $U(2)$ gauge symmetry generated by $\alpha_1 = \alpha_2 = \alpha$. This identification of gauge fields produces a single $U(2)$ gauge theory for $\alpha$ in which all the Chern-Simons levels double. As a result, the final Lagrangian is
    \begin{equation}
        \begin{aligned}
        L &= - \frac{4}{4\pi} \Tr \left(\alpha \, d \alpha + \frac{2}{3} \alpha^3\right) + \frac{2}{4\pi} \Tr \,\alpha \, d \Tr \alpha + \frac{1}{2\pi} \Tr \alpha \, d (A_1 + A_2) - \mathrm{CS}[A_1,g] - \mathrm{CS}[A_2,g] \,. 
        \end{aligned}
    \end{equation}
    which is the $U(2)_{4,0} \times U(1)_{-1}^2$ theory in \eqref{eq:L_pattern_1}. 
    \item If $\lambda_1 < 0$, then it is optimal to maximize $B^2$ by setting $s_2 = 0$. In this case, the potential $V(A,B)$ can be written entirely in terms of $s_1$ as
    \begin{equation}
        V(s_1) = - s_1^2 + \frac{2 \lambda_1 + \lambda_2}{2} s_1^4 \,. 
    \end{equation}
    The stability criterion is now $2 \lambda_1 + \lambda_2 > 0$. When this criterion is satisfied, we find $s_1 > s_2 = 0$. The subgroup of $U(2) \times U(2)$ that preserves this condensate is an Abelian $U(1)^3$ group generated by diagonal gauge fields of the form $\alpha_i = \mathrm{diag}(\alpha_{i, \uparrow}, \alpha_{i,\downarrow})$ subject to the constraint $\alpha_{1,\uparrow} = \alpha_{1,\downarrow} = \alpha$. The resulting Lagrangian takes the form
    \begin{equation}
        \begin{aligned}
        L &= - \frac{1}{4\pi} (\alpha d \alpha + \alpha_{1, \downarrow} d \alpha_{1, \downarrow}) - \frac{1}{4\pi} (\alpha d \alpha + \alpha_{2,\downarrow} d \alpha_{2, \downarrow}) + \frac{1}{2\pi} \alpha_{1,\downarrow} d \alpha + \frac{1}{2\pi} \alpha_{2, \downarrow} d \alpha \\
        &\hspace{0.6cm} + \frac{1}{2\pi} (\alpha + \alpha_{1,\downarrow}) d A_1 + \frac{1}{2\pi} (\alpha + \alpha_{2,\downarrow}) d A_2 - \mathrm{CS}[A_1,g] - \mathrm{CS}[A_2,g] \,. 
        \end{aligned}
    \end{equation}
    Integrating out $\alpha_{1,\downarrow}$ and $\alpha_{2,\downarrow}$ simplifies the Lagrangian to
    \begin{equation}
        \begin{aligned}
            L &= - \frac{2}{4\pi} \alpha d \alpha + \frac{1}{4\pi} (\alpha + A_1) d (\alpha + A_1) + \frac{1}{4\pi} (\alpha + A_2) d (\alpha + A_2) + 2 \Omega_g + \frac{1}{2\pi} \alpha d (A_1+A_2) \\
            &= \frac{2}{2\pi} \alpha d (A_1 + A_2) \,. 
        \end{aligned}
    \end{equation}
    This is a topologically trivial charge-$4e$ SC. 
\end{enumerate}
We remark that the intermediate scenario $s_1 > s_2 > 0$, though possible in principle, is not realized for any choice of phenomenological parameters $m, \lambda_1, \lambda_2$. This conclusion could change when higher order terms in the potential $V(\Phi)$ become sufficiently strong. We leave a thorough exploration of these higher-order effects to future work. 

\section{Higgsing patterns and fusion channel splitting}\label{app:fusion_splitting}

In the main text and in App.~\ref{app:Higgs_classification}, we formulated the quantum condensation of vortex-antivortex pairs as the Higgs transition associated with a matrix field $\Phi_{ab}$. While the field-theoretic analysis after \eqref{eq:app_Lfull_matter} is clear-cut, it does not provide a transparent physical interpretation of the different Higgsing patterns. The goal of this Appendix is to fill this gap by establishing a connection between the choice of Higgsing pattern and the energetic hierarchy of excitations in the parent $p+ip$ superconductor, using ideas from Ref.~\cite{Shi2025_dopeMR}. 

To set the stage, let us first recall some basic properties of the single-component topological $p+ip$ superconductor. The most unusual feature of this superconductor is the non-Abelian statistics of $h/2e$ vortices. Specifically, each $h/2e$ vortex traps a topologically protected Majorana zero mode, and every pair of Majorana zero modes becomes a complex fermion mode that can be either filled or empty. The energy splitting between the empty and filled states is exponentially small in the distance $R$ between the vortices and vanishes as $R \rightarrow \infty$. In the field theory formulation, we can describe each non-Abelian vortex as the endpoint of a Wilson line in the fundamental representation of $U(2)$ carrying spin $j = 1/2$ in the $SU(2)$ factor. The fusion of Majorana zero modes then corresponds to the fusion of Wilson lines in $SU(2)_2$, which reduces to the addition of spin angular momentum $1/2 \times 1/2 = 0 + 1$. Because of this analogy, we will also refer to the 0 channel as the singlet and the 1 channel as the triplet. 

Now let us imagine a phase transition across which the vortex $v_1$ from the first species binds with the antivortex $\bar v_2$ from the second species and condense. In this condensate, there is a nonzero density of vortex-antivortex pairs and the average distance between vortices of each species becomes finite. This finite separation inevitably splits the degeneracy between the singlet and triplet fusion channels of vortex pairs. A striking consequence of this observation is that, unlike Abelian superconductors, the quantum phase of matter induced by vortex-antivortex condensation in a two-component non-Abelian $p+ip$ superconductor is not unique and must depend on the pattern of fusion channel splitting. These distinct patterns give rise to the distinct Higgsing pathways considered in App.~\ref{app:Higgs_classification}.

Based on the general intuition above, we will derive two concrete results: (1) When the interaction between a pair of finitely separated $v_{\sigma}$'s favors the singlet/triplet channel for both choices of $\sigma$, the effective parameters in the Lagrangian \eqref{eq:app_Lfull_matter} satisfy $\lambda_1 > 0$/$\lambda_1 < 0$. (2) In the weak coupling regime of the $U(2)$ Maxwell-CS theory (where gauge fluctuations can be treated within a Gaussian approximation), the singlet channel is always favored over the triplet channel. Combining these two results, we conclude that the weak coupling regime of \eqref{eq:app_Lfull_matter} always selects $\lambda_1 > 0$, as claimed in the main text. In what follows, we will first review fusion channel splitting in a single-component $p+ip$ superconductor. Then we will turn to the two-component $p+ip$ superconductor and derive the aforementioned results. 

\subsection{Fusion channel splitting in a single-component \texorpdfstring{$p+ip$}{} superconductor}

Before tackling the full problem, let us warm up by reviewing fusion channel splitting in a single-component $p+ip$ superconductor. As described in Refs.~\cite{Ma2020_FCIQCD,Shi2025_dopeMR}, an effective Lagrangian that captures the universal low energy physics of this superconductor is
\begin{equation}
    \begin{aligned}
    L_{p+ip} &= - \frac{2}{4\pi} \Tr \left(\alpha d \alpha + \frac{2}{3} \alpha^3\right) + \frac{1}{4\pi} \Tr \alpha \, d \Tr \alpha + \frac{1}{2\pi} \Tr \alpha \, d A - \mathrm{CS}[A,g] + L_{(1/2,1)}[\phi, \alpha] \,,
    \end{aligned}
\end{equation}
where $\phi$ transforms in the fundamental representation $(1/2,1)$ of $U(2)$ and sources the $h/2e$ vortex. A single $h/2e$ vortex traps a Majorana zero mode. A pair of $h/2e$ vortices therefore traps a complex fermion mode, which can be either filled or empty. When the pair of vortices are infinitely far from each other, the filled and empty states of the complex fermion mode are exactly degenerate. However, when the distance $R$ between them is finite, intervortex interactions generically split the energy levels of the filled and empty states. At length scales much larger than $R$, the two-vortex state therefore behaves like a single $h/e$ vortex with a two-level internal structure. 

How does this energy splitting manifest itself in a field-theoretic description? If we represent different types of quasiparticles by irreducible representations $(j, n)$ of $U(2)$, then the vortex fusion rule can be rephrased as
\begin{equation}
    (1/2, 1) \times (1/2, 1) = (0, 2) + (1,2) \,, 
\end{equation}
where the $SU(2)$ singlet $(0,2)$/$SU(2)$ triplet $(1,2)$ corresponds to the empty/filled state of the complex fermion mode. As shown in Ref.~\cite{Shi2025_dopeMR}, adding the simplest Yang-Mills term for the $SU(2)$ gauge fields generates attraction in the singlet channel but repulsion in the triplet channel. In a real system, there will be more complicated interactions in the effective field theory that may flip the relative energies of the two channels, but a nonzero splitting is generic. 

\subsection{From fusion channel splitting to Higgsing patterns in the two-component \texorpdfstring{$p+ip$}{} superconductor} 

Now let us move beyond the single-component case and discuss Higgs transitions out of the two-component $p+ip$ superconductor considered in the main text. Recall the full Lagrangian
\begin{equation}\label{eq:app_Lfull}
    L = L_{\rm gauge} + L_{\rm matter} \,, 
\end{equation}
where 
\begin{equation}
    \begin{aligned}
    L_{\rm gauge} &= - \frac{2}{4\pi} \Tr \left(\alpha_1 \,d\alpha_1 + \frac{2}{3} \alpha_1^3\right) + \frac{1}{4\pi} \Tr \alpha_1 \, d \Tr \alpha_1 + \frac{1}{2\pi} \Tr \alpha_1 \, d A_1 - \mathrm{CS}[A_1, g] \\
    &- \frac{2}{4\pi} \Tr \left(\alpha_2 \,d\alpha_2 + \frac{2}{3} \alpha_2^3\right) + \frac{1}{4\pi} \Tr \alpha_2 \, d \Tr \alpha_2 + \frac{1}{2\pi} \Tr \alpha_2 \, d A_2 - \mathrm{CS}[A_2, g] + L_{\rm Maxwell}[F_{\alpha_1}, F_{\alpha_2}] \,, \\
    L_{\rm matter} &= \Tr D_\mu \Phi (D_\mu \Phi)^{\dagger} + m \Tr \Phi^{\dagger} \Phi + \lambda_1 \Tr \Phi \Phi^{\dagger} \Phi \Phi^{\dagger} + \frac{\lambda_2}{2} (\Tr \Phi^{\dagger} \Phi)^2 \,.
    \end{aligned}
\end{equation}
We choose a normalization in which the $\mathfrak{u}(2)$ generators are given by $T^0 = I/2$ and $T^{1,2,3} = \sigma^{1,2,3}/2$. The equations of motion for the temporal gauge field components $\alpha^a_{1, t}$ and $\alpha^a_{2,t}$ locks the non-Abelian fluxes $F^a_1, F^a_2$ to the non-Abelian matter densities 
\begin{equation}
    \frac{2}{2\pi} F^a_1 = \Tr \, \Phi^{\dagger} T^a \Phi \,, \quad - \frac{2}{2\pi} F^a_2 = \Tr \, \Phi^{\dagger} \Phi T^a \,, \quad F_i = d \alpha_i + \alpha_i^2 \,. 
\end{equation}
This condition gives us a very natural interpretation of the different Higgsing pathways:
\begin{enumerate}
    \item Singlet wins over triplet: if the singlet fusion channel for each species is favored over the triplet fusion channel, then we should demand a flux configuration with vanishing non-Abelian flux $F^a_i$ with $a = \{1,2,3\}$. This assumption puts strong constraints on the form of $\Phi$. Up to a $U(2) \times U(2)$ gauge rotation, we can always rotate the condensate $\Phi$ to a canonical form $U_1 \Phi U_2^{\dagger} = \mathrm{diag}(s_1, s_2)$ where $s_1, s_2 \geq 0$ are the singular values of $\Phi$. After this rotation, we see that the condition of vanishing non-Abelian flux reduces to 
    \begin{equation}
        \Tr \, \mathrm{diag}(s_1, s_2) U_1 \sigma^a U_1^{\dagger} \mathrm{diag}(s_1, s_2) = \Tr \, \mathrm{diag}(s_1, s_2) U_2 \sigma^a U_2^{\dagger} \mathrm{diag}(s_1, s_2) = 0 \,. 
    \end{equation}
    Under a unitary rotation $U$, $\sigma^a$ transforms to $U \sigma^a U^{
    \dagger} = \sum_b X_b(U) \sigma^b$ for some coefficients $X_b(U)$ that depend on $U$. Writing $\mathrm{diag}(s_1^2, s_2^2)$ as $\frac{s_1^2 + s_2^2}{2} I + \frac{s_1^2 - s_2^2}{2} \sigma^3$, we see that the above condition is satisfied iff 
    \begin{equation}
        \Tr \left(\left[\frac{s_1^2 + s_2^2}{2} I + \frac{s_1^2 - s_2^2}{2} \sigma^3\right] \sigma^a\right) = 0 \,, \quad \forall a \in \{1,2,3\} \,. 
    \end{equation}
    This forces $s_1 = s_2$, which corresponds to the Higgsing pattern $U(2) \times U(2) \rightarrow U(2)_{\rm diag}$. 
    \item Triplet wins over singlet: if the triplet fusion channel for each species is favored over the singlet fusion channel, then we should demand a flux configuration with polarized non-Abelian flux. Without loss of generality, we can choose a gauge in which $\Phi = \mathrm{diag}(s_1, s_2)$ and maximally polarize the fluxes in the $z$ direction so that $F^3_1, F^3_2$ are maximized while $F^1_i, F^2_i = 0$. Clearly, the maximum of 
    \begin{equation}
        |F^a_1| = |F^a_2| = \pi \left|\Tr \left(\left[\frac{s_1^2 + s_2^2}{2} I + \frac{s_1^2 - s_2^2}{2} \sigma^3\right] \sigma^a\right) \right|
    \end{equation}
    for fixed $s_1^2 + s_2^2$, is achieved at $s_2 = 0$. The polarized ansatz $\Phi = \mathrm{diag}(s_1, 0)$ also makes sense if we regard $\Phi_{ab}$ as $\phi_{1,a}^{\dagger} \phi_{2,b}$, where $\phi_1$/$\phi_2$ is a scalar field in the fundamental representation of the $U(2)$ gauge groups generated by $\alpha_1$/$\alpha_2$. Favoring the triplet channel for each species leads to the polarized solution $\phi_1, \phi_2 \propto (1,0)$. Taking the outer product of $\phi_1$ and $\phi_2$ then gives $\Phi \propto \mathrm{diag}(1,0)$. 

    Once we favor the triplet channel, each $U(2)$ gauge field $\alpha_i$ gets Higgsed down to $U(1) \times U(1)$ subgroup generated by the diagonal components $\alpha_i = \mathrm{diag}(\alpha_{i,\uparrow}, \alpha_{i,\downarrow})$. Condensing $v_1 \bar v_2$ in the presence of such a polarization further Higgses $\alpha_{1,\uparrow} - \alpha_{2,\uparrow}$. Following the steps in App.~\ref{app:Higgs_classification}, we can simplify the remaining $U(1)^3$ gauge theory down to a topologically trivial charge-$4e$ superconductor. 
\end{enumerate}

Now let us recall from App.~\ref{app:Higgs_classification} that the Higgsing pattern $\Phi \sim I_{2 \times 2}$ is favored for $\lambda_1 > 0$, while $\Phi \sim \mathrm{diag}(1,0)$ is favored for $\lambda_1 < 0$. Combining this observation with the casework above, we establish our result (1): When the interaction between a pair of finitely separated $v_{\sigma}$'s favors the singlet/triplet channel for both choices of $\sigma$, we have $\lambda_1 > 0$/$\lambda_1 < 0$. 

To derive result (2), we need to show that in the weak-coupling regime of the effective field theory $L_{\rm gauge}$, pair fusion of $v_1 \bar v_2$ always prefers the singlet channel over the triplet channel for both species. The argument proceeds as follows. In the weak-coupling regime of $L_{\rm gauge}$, we can treat the gauge fluctuations of $\alpha_1, \alpha_2$ within a Gaussian approximation. To preserve $U(2) \times U(2)$ gauge invariance, the most general quadratic Lagrangian for $F_{\alpha_{\sigma}}$ takes the form
\begin{equation}
    L_{\rm Maxwell} = \frac{1}{4 \rho_1} \Tr \left[F_{\alpha_1}^2\right] + \frac{1}{4 \rho_2} \Tr \left[F_{\alpha_2}^2\right] + \frac{1}{4 g^2} \Tr F_{\alpha_1} \cdot \Tr F_{\alpha_2} \,. 
\end{equation}
Note that there is no interspecies coupling between the $SU(2)$ components of $F_{\alpha_1}$ and $F_{\alpha_2}$, as such a coupling would violate gauge invariance. 

When $g = \infty$, the calculations in Ref.~\cite{Shi2025_dopeMR} imply that pair fusion of $v_{\sigma}$ indeed favors the singlet channel over the triplet channel for each $\sigma$. When $g < \infty$, the additional Abelian interspecies coupling affects the total energy of a pair of $v_1 \bar v_2$, but does not affect the splitting of distinct fusion channels. Therefore, at the Gaussian level, the fusion between a pair of spatially separated $v_1 \bar v_2$ in the $(p+ip)^2$ state always favors the singlet channel in both species, leading to $\lambda_1 > 0$. Again, we emphasize that in a real system, non-Gaussian corrections are generally not suppressed by any small parameter and could subvert this conclusion. However, our analysis at least establishes one pattern of fusion channel splitting in some open region of the phase diagram.

\section{Parton descriptions and representative many-body wavefunctions}
\label{app:parton}

In this section use a parton approach to describe the states analyzed in the main text, which yields a family of representative wavefunctions for each case.

\subsection{\texorpdfstring{$p+ip$}{} charge \texorpdfstring{$2e$}{} SC}
\label{app: parton p+ip}

We consider the following parton decomposition in which a single-component electron operator is written as the product of one spinful boson $b$ and one spinful fermion $f$
\begin{align} \label{eq: parton}
    \hat{c}(\bm{r}) = \hat{b}_a(\bm{r}) \epsilon_{aa'} \hat{f}_{a'}(\bm{r}) \,.
\end{align}
The gauge redundancy of this theory is $U(2) = SU(2)\times U(1)/\mathbb{Z}_2$. Both $b$ and $f$ are spin-$1/2$ under the $SU(2)$, but they carry opposite unit charge under $U(1)$. The $\mathbb{Z}_2$ quotient can be recognized by the fact that the $\mathbb{Z}_2$ centor of $SU(2)$ (i.e.  $e^{\mathrm{i} \frac{2\pi}{2} \vec{n}\cdot\vec{\sigma}}$ around any axis $\vec{n}$) is trivial when combined with the $\pi$ rotation of $U(1)$ $e^{\mathrm{i}\pi}$. The total number of $b$ and $f$ particles should be equal to that of the electrons:
\begin{align}
    \rho_c = \rho_b = \rho_f \,.
\end{align}

The effective field theory description then reads:
\begin{align}
    L = &L[f;\alpha + A] +L[b;\alpha -\Tr\alpha] \,,
\end{align}
where $\alpha$ is $U(2)$ dynamical gauge field (note that this charge assignment ensures that $\alpha$ is not \spinc). We now assume that the mean-field flux configuration of the gauge field is
\begin{align}
    \frac{1}{2\pi}\langle\nabla\times \bm{\alpha}\rangle = \frac{\rho_c}{2} \sigma^0 \,.
\end{align}
This will put $f$ to effective filling fraction $\nu_f = 2$ and $b$ to $\nu_b=-2$. The natural states for the two partons are thus the fermionic and bosonic integer quantum Hall states, respectively. These two states are both spin-singlet, topologically invertible, have different $U(1)$ Hall response $\pm 2$ but the {\it same} spin response $2$. These considerations lead to the following effective topological field theory after integrating out the matter field
\begin{align} \label{eq:L-}
    L_{p+ip} =& \frac{1}{4\pi} \Tr \left[(\alpha +A + \frac{1}{2}\omega)\mathrm{d} (\alpha+A+ \frac{1}{2}\omega) + \frac{2}{3}\alpha^3\right] + 2\Omega_g +\frac{1}{4\pi} \Tr \left[\alpha\mathrm{d} \alpha + \frac{2}{3}\alpha^3\right]-\frac{1}{4\pi}\Tr\alpha \mathrm{d} \Tr \alpha \\
    =&   \frac{2}{4\pi} \Tr \left[\alpha\mathrm{d} \alpha + \frac{2}{3}\alpha^3\right] -\frac{1}{4\pi}\Tr\alpha \mathrm{d} \Tr \alpha+ \frac{1}{2\pi} \Tr\alpha \mathrm{d} (A+\frac{1}{2}\omega) + 2\Omega_g +\frac{2}{4\pi} (A+\frac{1}{2}\omega) \mathrm{d} (A+\frac{1}{2}\omega) \,,
\end{align}
where $2\Omega_g$ is gravitational CS term originating from the framing anomaly, whose coefficient tracks the chiral central charge of the system. $\omega$ is the spatial rotation $SO(2)$ spin connection. The resulting theory is $U(2)_{k_1=-2,k_2= 0}$, meaning that there is no CS term for the $U(1)$ part and thus the theory is a superconductor. The fact that the $U(1)$ part of the gauge field, $\frac{1}{2}\Tr\alpha$, has a mutual CS term with $(A+ \frac{1}{2}\omega)$ with level $2$ indicates that this is a charge-$2$ condensate with angular momentum $1$. The $SU(2)$ part has self-CS with level $-2$, providing no anyon content but a chiral central charge $-3/2$, such that the total $c_-= 2-3/2=1/2$. Moreover, an external $\pi$ flux of $A$ serve as a charge $n=1$ object under the $U(1)$ part of $\alpha$, which must bind with a $j=1/2$ object under the $SU(2)$ part due to selection rule $j+n/2\in Z$. This composite has Ising anyonic statistics due to the level-$2$ $SU(2)$ self-CS term. All of these features match the expectations of a $p+\mathrm{i}p$ superconductor. 

Careful readers may find that the TQFT we obtained is different from the one used in the main text--the $U(2)$ self-CS term has an opposite sign, and they differ by three stacked IQH states. This is not an inconsistency. The two TQFTs truly describe the same phase and are related by a duality transformation: starting from Eq.~\ref{eq:L-}, perform a time reversal and a charge conjugation within a $C=1$ band will generate the TQFT adopted in the main text. Physically, this is a $p-ip$ state of {\it holes} doping from a $C=1$ band, which is indeed topologically equivalent to a $p+ip$ state of {\it electrons} in all aspects. For simplicity, we will stick to this TQFT in this appendix.

The equivalence between the parton wavefunction and the standard BCS wavefunction for a $p+ip$ SC can be directly established without the field theory machinery. Using the projective construction, we can explicitly write down the electronic wavefunction (assuming there are $2N$ electrons) from the parton mean-field ansatz~\cite{Wen1990_Projective_construction} ($z\equiv x-\mathrm{i}y$ and $a=1,2=\updownarrow$ in the below equations) :
\begin{align}
    \Psi(\bm{r}_1 , \dots, \bm{r}_{2N}) = \langle \text{vac}|\prod_i\hat{c}(\bm{r}_{i})\int_{\{\bm{\xi}\}} \Psi_f\left(\left\{\bm{\xi}_\uparrow,\bm{\xi}_\downarrow\right\}\right) \prod_i \hat{f}^\dagger_{\uparrow}(\bm{\xi}_{i,\uparrow})\hat{f}^\dagger_{\downarrow} (\bm{\xi}_{i,\downarrow})\int_{\{\bm{\zeta}\}} \Psi_b\left(\left\{\bm{\zeta}_\uparrow,\bm{\zeta}_\downarrow\right\}\right) \prod_i \hat{f}^\dagger_{\uparrow}(\bm{\zeta}_{i,\uparrow})\hat{f}^\dagger_{\downarrow} (\bm{\zeta}_{i,\downarrow})|\text{vac}\rangle \,.
\end{align}
For example, according to our assumption, we have the following representative wavefunctions for $b$ and $f$:
\begin{align}
    \Psi_f\left(\left\{\bm{r}_\uparrow,\bm{r}_\downarrow\right\}\right) &= \prod_{1\le i < j \le N}(z_{i,\uparrow} - z_{j,\uparrow}) (z_{i,\downarrow} - z_{j,\downarrow}) \mathrm{e}^{-|z_{i,\uparrow}|^2/4\ell^2-|z_{i,\downarrow}|^2/4\ell^2} \,, \\
    \Psi_b\left(\left\{\bm{r}_\uparrow,\bm{r}_\downarrow\right\}\right) &= \prod_{1\le i, j \le N}(z_{i,\uparrow} - z_{j,\downarrow})^{-1} \mathrm{e}^{-|z_{i,\uparrow}|^2/4\ell^2-|z_{i,\downarrow}|^2/4\ell^2} \,.
\end{align}
For later convenience, in the above expression we have adjusted the non-topological factors ($|z-z|$ terms) in $\Psi_b$ such that it is a holomorphic function up to the Gaussian factor while remaining in the same phase. Then the projective construction yields the final electronic wavefunction (omitting Gaussian factors for simplicity)
\begin{align}
\Psi_c(\{\bm{r}\}) = \mathcal{A} \left[\Psi_f\left(\left\{\bm{r}_\uparrow,\bm{r}_\downarrow\right\}\right)\Psi_b\left(\left\{\bm{r}_\uparrow,\bm{r}_\downarrow\right\}\right)\right]= \prod_{1\le i<j\le 2N}(z_i-z_j)^{-1} \mathcal{S}\left[\prod_{1\le i<j<N} (z_{i,\uparrow}-z_{j,\uparrow})^2 \prod_{1\le i<j<N}(z_{i,\downarrow}-z_{j,\downarrow})^2 \right] \,,
\end{align}
where $\mathcal{A},\mathcal{S}$ stand for anti-symmetrization and symmetrization over all coordinates. It was shown in Ref.~\cite{CAPPELLI2001_parafermion_wavefunction} that, to leading order in the inter-particle distance, 
\begin{align}    \mathcal{S}\left[\prod_{1\le i<j<N} (z_{i,\uparrow}-z_{j,\uparrow})^2 \prod_{1\le i<j<N}(z_{i,\downarrow}-z_{j,\downarrow})^2 \right]\sim \Pf\left(\frac{1}{z_i-z_j}\right) \prod_{1\le i<j\le 2N}(z_i-z_j) \,.
\end{align}
Therefore, we see that $\Psi_c \sim \Pf\left(\frac{1}{z_i-z_j}\right)$ is indeed topologically equivalent to the familiar Pfaffian wavefunction of the $p+ip$ state.

\subsection{Charge-\texorpdfstring{$4e$}{} SC from \texorpdfstring{$(p+ip)^2$}{}}

Here we discuss the descendent states of two stacked of $p+ip$ $2e$-SCs, which we call $(p+ip)^2$. We will be using two copies of parton ansatzes constructed in the previous section for the two flavors (labeld by superscript $(i=1,2)$):
\begin{align} \label{eq: flavorful parton}
\hat{c}_\sigma(\bm{r}) = \hat{b}_{\sigma,a}(\bm{r}) \epsilon_{aa'} \hat{f}_{\sigma,a'}(\bm{r}) \,.
\end{align}

If the two $p+ip$ SC are simply stacked without inter-flavor interaction, we will simply end up with a direct product of the state obtained in the previous section. However, interesting topological $4e$ states arises when the exciton order spontaneously occurs due to inter-flavor interaction between the fermionic partons (or bosonic parton) across the two flavors. This can be approached effectively through a Hubbard-Stratonovich transformation to the interaction:
\begin{align}
    L_\text{int} = &\int_{\bm{R},\bm{r}} V_{12}(\bm{r}) f^{\dagger}_{1,a}(\bm{R}+\bm{r}) f_{1,a} (\bm{R}+\bm{r})f^{\dagger}_{2,a'}(\bm{R}-\bm{r}) f_{2,a'} (\bm{R}-\bm{r})\\
     \rightarrow  &\left[\int_{\bm{R},\bm{r}} \Phi_{aa'}(\bm{R},\bm{r})  f^{\dagger}_{1,a} (\bm{R}+\bm{r})f_{2,a'} (\bm{R}-\bm{r}) + \text{h.c.} \right] + L[\Phi ; \alpha_1 \otimes I_2 -I_1 \otimes \alpha_2] \,,
\end{align}
where $\Phi$ transforms as (dropping time dependence in the below equations for simplicity) 
\begin{align}
    \Phi(\bm{R},\bm{r}) \rightarrow U_1 (\bm{R}+\bm{r}) \Phi(\bm{R},\bm{r}) U^{\dagger}_2(\bm{R}-\bm{r}) 
\end{align}
under gauge transformation
\begin{align}
    \alpha_{\sigma,\mu} (\bm{r}) \rightarrow \alpha_{\sigma,\mu}  (\bm{r}) + U_{\sigma}^{\dagger}(\bm{r}) \partial_\mu U_{\sigma}(\bm{r})  \,.
\end{align}
In the long-wavelength limit, we can factorize $\Phi(\bm{R},\bm{r})\approx  \Phi(\bm{R})F(\bm{r})$ and assume that the pairing form factor $F(\bm{r})$ is local in $\bm{r}$. Then $\Phi(\bs{R})$ transforms in the same way as the Higgs field $\Phi$ we discussed in Eq.~\ref{eq:LPhi} of App.~\ref{app:Higgs_classification}. Hence we use the same symbol for both fields, with the understanding that this exciton pair actually represents the vortex-antivortex pair across the two copies of $2e$ SCs as we intepreted in the above sections. 

Several important remarks are in order: (1) It is crucial that this bosonic field doesn't see net magnetic flux, so that it can indeed condense without breaking translation symmetry. (2) The form factor $F(\bm{r})$ is in principle determined by the interaction. At the level of trial wavefunctions, $F(\bs{r})$ can be treated as a set of variational parameters that generate a class of wavefunctions representing the same phase. The physics discussed below should not depend on the specific form of it except in the pathological limit $F(\bm{r})\propto  \delta(\bm{r})$. 

Given these considerations, now consider a state with a uniform condensate $\langle\Phi(\bm{R})\rangle = \bar{\Phi} \neq 0$. As discussed in App.~\ref{app:Higgs_classification}, different condensation patterns lead to different Higgsing pathways. 
In the pathway of interest to us (\eqref{eq:L_pattern_1} in the main text), the two dynamical gauge fields of the two flavors $\alpha_\sigma$ are Higgsed to be the same one, which we call $\alpha$. A simple wavefunction can then be written down according to the mean field action for the partons:
\begin{align}
    L_\text{MF}  = &\sum_\sigma L[f_\sigma;\alpha + A_\sigma] +L[b_{\sigma};\alpha -\Tr\alpha] + \left[\int_{\bm{R},\bm{r}} \bar{\Phi}_{aa'}F(\bm{r})  f^{\dagger}_{1, a}(\bm{R}+\bm{r})f_{2,a'} (\bm{R}-\bm{r}) + \text{h.c.} \right] \,.
\end{align}
This procedure gives rise to a family of electron wavefunctions after applying the projective construction.

\section{Review of symmetry-enriched topological order in 2+1 dimensions}\label{app:SETreview}

In this Appendix, we provide a brief review of symmetry-enriched topological orders in 2+1D. The goal of this Appendix is to provide the minimal background necessary for the analysis of $U(2)_{4,0}$ presented in App.~\ref{app:U(2)40_TO}. We refer readers who are interested in a more complete and general treatment to Ref.~\cite{Barkeshli2014_SETbible}. 

\subsection{Algebraic theory of topological orders}

In 2+1 dimensions, a general topological order is described by a unitary modular tensor category (UMTC). To define this notion, we begin with a set $\mathcal{C}$ of anyons with an associative fusion rule 
\begin{equation}
    a \times b = \sum_c N^c_{ab} c \,, \quad a, b, c \in \mathcal{C} \,,
\end{equation}
where the fusion multiplicities $N^c_{ab}$ are non-negative integers encoding the number of different ways in which $a$ and $b$ can fuse to produce $c$. In general, each anyon $a$ has an associated quantum dimension $d_a$, such that the Hilbert space dimension of $N$ anyons of type $a$ grows asymptotically as $d_a^N$ as $N \rightarrow \infty$. The $d_a$'s satisfy a similar equation 
\begin{equation}
    d_a d_b = \sum_{c \in \mathcal{C}} N^c_{ab} d_c \,, \quad \mathcal{D} = \sqrt{\sum_{a \in \mathcal{C}} d_a^2} \,,
\end{equation}
where $\mathcal{D}$ is defined to be the total quantum dimension of the topological order.

For each triplet of anyons with $N^c_{ab} > 0$, we can draw a trivalent vertex for anyon fusion and splitting:
\begin{equation}\label{eq:fuse_split_vertex}
    \left(\frac{d_c}{d_a d_b}\right)^{1/4} \includegraphics[height=1.4cm,valign=c]{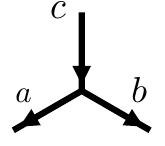} = \bra{a,b;c} \in V^c_{ab} \,, \quad \left(\frac{d_c}{d_a d_b}\right)^{1/4} \includegraphics[height=1.4cm,valign=c]{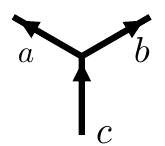} = \ket{a,b;c} \in V^{ab}_c \,,
\end{equation}
where $V^c_{ab}$ and $V^{ab}_c$ are fusion and splitting vector spaces respectively. In what follows, we will assume that $N^c_{ab} = 0/1$ for all $a,b,c \in \mathcal{C}$ so that $\dim V^{ab}_c = \dim V^{c}_{ab} = 1$ if and only if $N^c_{ab} = 1$. This assumption holds for a large class of topological orders of physical interest, including the ones we encounter in this work.

The contraction and connection of fusion and splitting vertices lead to two elementary identities 
\begin{equation}
    \includegraphics[height=2cm,valign=c]{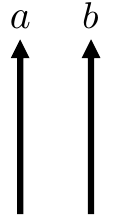} = \sum_c  \sqrt{\frac{d_c}{d_a d_b}} \includegraphics[height=2cm,valign=c]{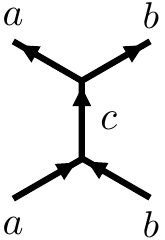} \,, \quad \quad \includegraphics[height=2cm,valign=c]{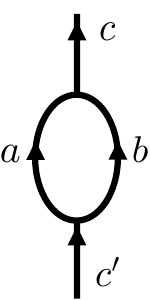} = \delta_{cc'} \sqrt{\frac{d_a d_b}{d_c}} \, \includegraphics[height=2cm,valign=c]{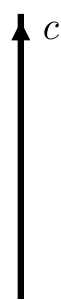} \,.
\end{equation}

Complicated fusion diagrams can be generated from these basic trivalent vertices. The associativity of fusion means that different multi-anyon fusion spaces are related by isomorphisms known as ``$F$-moves". The basic $F$-move is defined by the following mapping
\begin{equation}\label{eq:F_move_def}
    \includegraphics[height=1.4cm,valign=c]{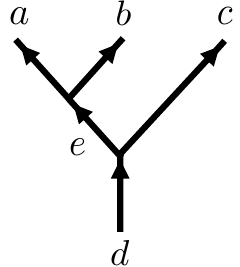} = \left[F^{abc}_d\right]_{ef} \includegraphics[height=1.4cm,valign=c]{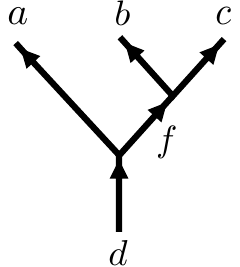} \,, \quad F^{abc}_d: \oplus_e V^{ab}_c \otimes V^{ec}_d \rightarrow \oplus_f V^{af}_d \otimes V^{bc}_f \,.  
\end{equation}
Consistency of fusion demands that two sequences of $F$-moves with the same initial and final configuration must be equivalent. Mac Lane's coherence theorem~\cite{MacLane1998_category} guarantees that this consistency condition holds whenever the $F$-symbols satisfy the pentagon equation
\begin{equation}\label{eq:pentagon}
    \includegraphics[height=5cm,valign=c]{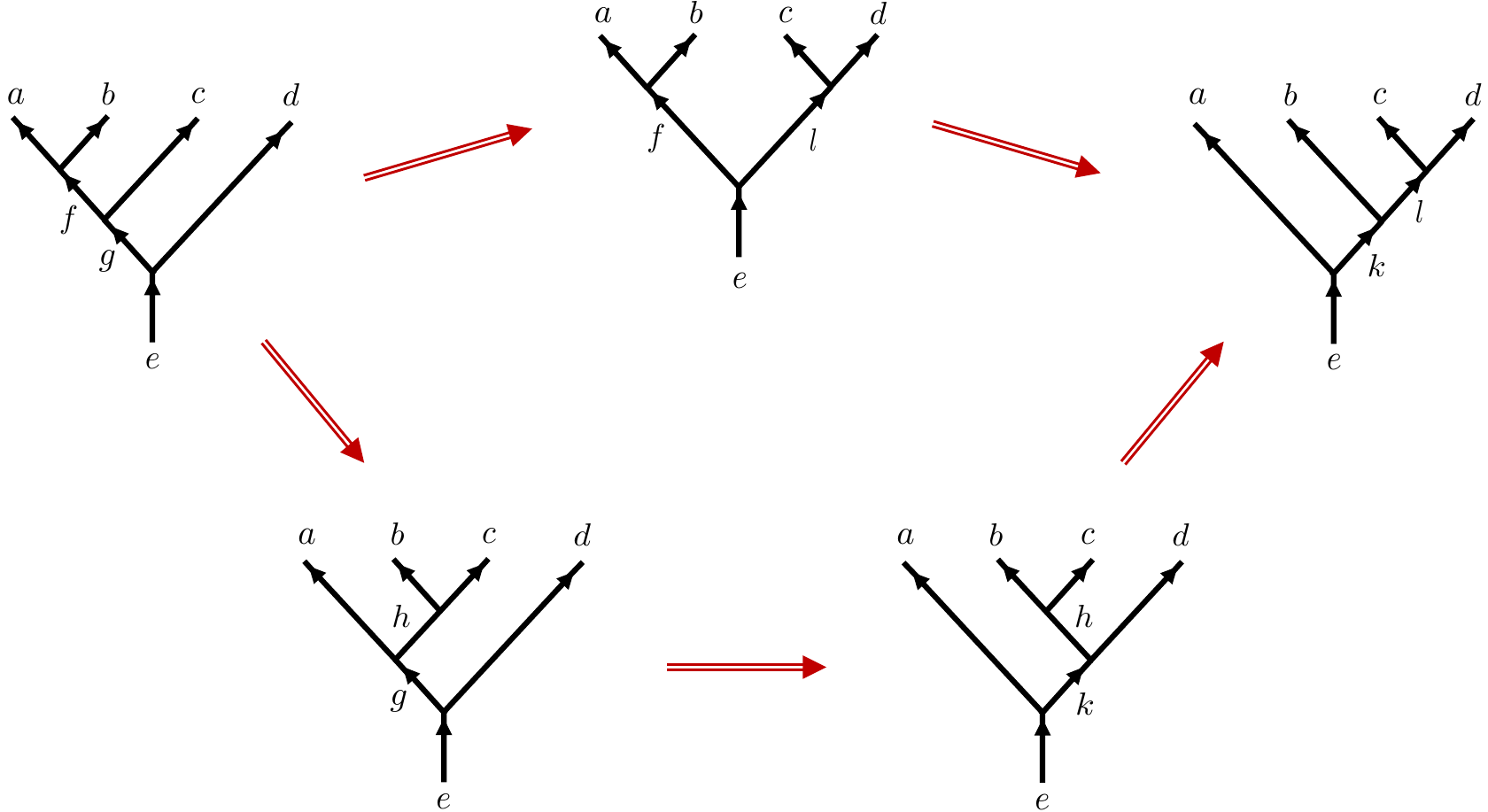} \quad \sum_h F^{abc}_{gfh} F^{ahd}_{egk} F^{bcd}_{khl} = \sum F^{fcd}_{egl} F^{abl}_{efk} \,. 
\end{equation}
Finally, we demand that there is a unique vacuum $0 \in \mathcal{C}$ and a unique anti-particle $\bar a$ for each $a \in \mathcal{C}$ satisfying $N^0_{ab} = \delta_{b \bar a}$. We can construct the Frobenius-Schur indicator $\varkappa_a$ via the equation $F^{a \bar a a}_{a 00} = \mathcal{\varkappa}_a/d_a$. For $a = \bar a$, $\varkappa_a$ is an invariant of the topological order and is always equal to $\pm 1$. There are more constraints on various coefficients arising from the existence of antiparticles (see Section II.A of Ref.~\cite{Barkeshli2014_SETbible}) but we will not enumerate them as they are not essential for our analysis in App.~\ref{app:U(2)40_TO}.

The set $\mathcal{C}$, equipped with the fusion rules and the $F$-symbols, defines a unitary fusion tensor category. When fusion is commutative, we can further define the structure of anyon braiding which is essential for topological order. The basic braiding operator is the $R$-symbol $R^{ab}_c$ for pairwise exchange, from which one can derive the topological twist $\theta_a$, the topological $S$-matrix, and the monodromy scalar component $M_{ab}$
\begin{equation}\label{eq:R_def}
    \includegraphics[height=1.8cm,valign=c]{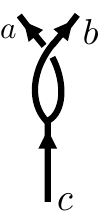} = R^{ab}_c \includegraphics[height=1.8cm,valign=c]{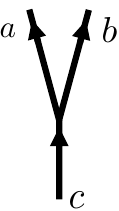} \,, \quad \theta_a = \theta_{\bar a} = \sum_c \frac{d_c}{d_a} R^{aa}_c \,, \quad S_{ab} = \mathcal{D}^{-1} \sum_c N^c_{\bar a b}\frac{\theta_c}{\theta_a \theta_b} d_c \,, \quad M_{ab} = \frac{S_{ab}^* S_{00}}{S_{0a} S_{0b}} \,.
\end{equation}
Consistency of braiding with fusion gives an additional set of constraints called the hexagon equations. The explicit form of the hexagon equations is given in Fig.~2 of Ref.~\cite{Barkeshli2014_SETbible} and will not be repeated here. Every set of $R$ and $F$ symbols that solves the pentagon and hexagon equations (modulo basis transformations of $V^{ab}_c$ and $V^c_{ab}$) defines a braided tensor category.

Finally, in any topological order, we must demand that each anyon braids nontrivially with at least one other anyon. This non-degeneracy condition ensures the unitarity of the $S$ matrix. A braided tensor category with a unitary $S$ matrix is a unitary modular tensor category (UMTC). It is conjectured that topological orders in 2+1 dimensions are in one-to-one correspondence with UMTCs, though a rigorous proof remains to be found~\cite{Kitaev2006_honeycomb}. 

\subsection{Symmetry enrichment of topological orders}

The mathematical structure of topological orders is further enriched when the microscopic Hamiltonian realizing the topological order enjoys a global symmetry. For the purpose of this brief review, let us restrict to the case where the global symmetry of the microscopic Hamiltonian is a 0-form internal symmetry described by a group $G$. If a topological order described by the UMTC $\mathcal{C}$ emerges in the low energy limit, then there must exist a map $\rho: G \rightarrow \mathrm{Aut}(\mathcal{C})$, where $\mathrm{Aut}(\mathcal{C})$ is the group of automorphisms of $\mathcal{C}$. Since different maps related by natural isomorphisms of $\mathcal{C}$ should be identified, we denote the equivalence class of maps under natural isomorphisms by $[\rho]$. Simple examples of such automorphisms include the e-m duality symmetry in $\mathbb{Z}_2$ gauge theory and the layer-permutation symmetry in multi-layer quantum Hall states. 

For every group element $\bs{g} \in G$, a representative $\rho_{\bs{g}}$ acts not only on the anyon labels $a \in \mathcal{C}$ but also on all the topological data that define $\mathcal{C}$, including fusion multiplicities $N^c_{ab}$, state vectors in fusion spaces $V^c_{ab}/V^{ab}_c$, as well as $F$-symbols and $R$-symbols. The precise action is described in Section III.B of Ref.~\cite{Barkeshli2014_SETbible}. Importantly, since anyons cannot be locally created, the microscopic symmetry $G$ can act projectively on isolated localized anyon excitations. This projective action is encoded in a set of projective phase factors $\eta_a(\bs{g}, \bs{h})$. When these phases are nontrivial, we say that the topological order exhibits ``fractionalization" of the symmetry $G$. 

Given a UMTC $\mathcal{C}$ and a group action $[\rho]$, whether symmetry fractionalization is possible at all is determined by a specific obstruction class $[\mathfrak{O}] \in H^3_{[\rho]}(G,\mathcal{A})$, where $\mathcal{A}$ is the group of Abelian anyons in $\mathcal{C}$~\cite{Barkeshli2014_SETbible}. When the obstruction class vanishes, symmetry fractionalization of $G$ is possible but generally not unique. 
The distinct fractionalization patterns are classified by $H^2_{[\rho]}(G, \mathcal{A})$. 

We will not describe the full mathematical machinery that is needed to understand this classification, as it is not necessary for the relatively simple topological order $SU(2)_4/\mathbb{Z}_2$ appearing in our work, enriched with an internal symmetry $G = \mathbb{Z}_2$. As we will see, for $\mathcal{C} = SU(2)_4/\mathbb{Z}_2$, $\mathrm{Aut}(\mathcal{C}) = \mathbb{Z}_2$ is generated by a permutation that exchanges the only two nontrivial anyons in $\mathcal{C}$ and the group of Abelian anyons in $\mathcal{C}$ is $\mathcal{A} = \mathbb{Z}_3$. The only nontrivial element $\bs{g}$ of the microscopic symmetry $G = \mathbb{Z}_2$ embeds precisely as the anyon permutation in $\mathrm{Aut}(\mathcal{C}) = \mathbb{Z}_2$. Moreover, since $H^3_{[\rho]}(\mathbb{Z}_2, \mathbb{Z}_3) = 0$ and $H^2_{[\rho]}(\mathbb{Z}_2, \mathbb{Z}_3) = \mathbb{Z}_1$, $G$ has a single fractionalization class, which acts trivially on the anyons. 

\subsection{Defectification of a symmetry-enriched topological order}

Given a topological order $\mathcal{C}$ enriched with a discrete symmetry $G$, we can consider extrinsic point-like $G$-defects that live on the endpoint of one-dimensional defect lines. Concretely, in order to create a pair of defects labeled by $\bs{g}$ and $\bs{g}^{-1}$ at locations $\bs{x}$ and $\bs{x}'$, we draw an oriented path from $\bs{x}'$ to $\bs{x}$ and modify every term in the Hamiltonian that intersects the path by acting with the symmetry element $\bs{g}$ locally on one side of the path. This path itself is often referred to as the ``branch cut" connecting the defect endpoints. From this definition, we see that when a deconfined anyon excitation $a$ winds around the defect $\bs{g}$, it crosses the branch cut and gets acted on by the symmetry transformation $\rho_{\bs{g}} \in \mathrm{Aut}(\mathcal{C})$. The presence of this branch cut is one of the key properties that distinguish confined defects from deconfined anyons. 

The full mathematical structure that describes the interplay between anyons and symmetry defects will be a $G$-crossed braided tensor category. The process of extending the original UMTC $\mathcal{C}$ to the associated $G$-crossed braided tensor category is referred to as defectification. Let us now outline the essential ingredients that go into this extension.

For each symmetry element $\bs{g}$, we can define the set of topologically distinct defects in the defect sector $\bs{g}$ by $\mathcal{C}_{\bs{g}}$, with $\mathcal{C}_0 \equiv \mathcal{C}$. Elements in $\mathcal{C}_{\bs{g}}$ will be labeled as $a_{\bs{g}}$ and can be understood as an elementary $\bs{g}$-defect bound to distinct anyons in $\mathcal{C}_0$. The set of all anyons and defects therefore has a graded structure and obeys a generalized fusion rule
\begin{equation}
    \mathcal{C}_G = \oplus_{\bs{g} \in G} \, \mathcal{C}_{\bs{g}} \,, \quad a_{\bs{g}} \times b_{\bs{h}} = \sum_{c_{\bs{gh}}} N^c_{ab} \, c_{\bs{gh}} \,. 
\end{equation}
This basic fusion rule, along with the axioms satisfied by the fusion multiplicities, guarantees that
\begin{equation}\label{eq:Qdim_equivalence}
    \mathcal{D}_0^2 = \sum_{a_{\bs{0}}} d_{a_{\bs{0}}}^2 = \sum_{a_{\bs{g}}} d_{a_{\bs{g}}}^2 = \mathcal{D}_{\bs{g}}^2 \,, 
\end{equation}
for every nontrivial $\mathcal{C}_{\bs{g}}$. This means that the defectified theory $\mathcal{C}_G$ has a total quantum dimension 
\begin{equation}
    \mathcal{D}_{\mathcal{C}_G} = |G|^{1/2} \mathcal{D}_{\bs{0}} \,, 
\end{equation}
where $|G|$ is the order of the group $G$. As in the case of UMTCs, we can use the fusion rules to define trivalent vertices for fusion and splitting (analogous to \eqref{eq:fuse_split_vertex}) as well as $F$-symbols $F^{abc}_d$ that map between isomorphic fusion spaces in which $a_{\bs{g}}, b_{\bs{h}}, c_{\bs{k}}$ fuse into $d_{\bs{ghk}}$ (analogous to \eqref{eq:F_move_def}). The $F$ symbols satisfy the same pentagon equation as in \eqref{eq:pentagon}. The fusion structure specified by $N^c_{ab}$ and $F^{abc}_d$ endows the set $\mathcal{C}_G$ with the structure of a $G$-crossed fusion category. This $G$-crossed fusion category will be referred to as $\mathcal{C}^{\times}_G$.

As in the construction of UMTCs, braiding can be introduced as an additional structure on top of fusion. The main new ingredient is that nontrivial defects $a_{\bs{g}}$ with $\bs{g} \neq 0$ carry one-dimensional branch cuts extending to spatial infinity. The process of exchanging the locations of two defects $a_{\bs{g}}, b_{\bs{h}}$ inevitably induces a crossing between the branch cut of $a_{\bs{g}}$ with the defect $b_{\bs{h}}$ and its associated branch cut. These crossings lead to an action of $\rho_{\bs{g}}$ on $b_{\bs{h}}$ such that $\rho_{\bs{g}}(b_{\bs{h}}) \in \mathcal{C}_{\bs{g} \bs{h} \bs{g}^{-1}}$. At the level of spacetime diagrams, we can account for this symmetry action by imagining that each defect line $x_{\bs{k}}$ with $\bs{k} \neq 0$ has a branch sheet attached to it which goes into the page. As a result, the counterclockwise exchange operator $R^{a_{\bs{g}} b_{\bs{h}}}$ sends $b_{\bs{h}}$ to itself but $a_{\bs{g}}$ to $\rho_{\bar{\bs{h}}}(a_{\bs{g}})$ since $a_{\bs{g}}$ crosses the branch sheet of $b_{\bs{h}}$ in spacetime (and similarly for the clockwise exchange operator). These considerations lead to a modified diagrammatic representation of the $G$-crossed $R$-symbol (analogous to \eqref{eq:R_def})
\begin{equation}
    \includegraphics[height=2cm, valign=c]{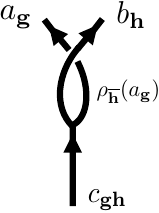} \, = R^{a_{\bs{g}} b_{\bs{h}}}_{c_{\bs{gh}}} \includegraphics[height=2cm, valign=c]{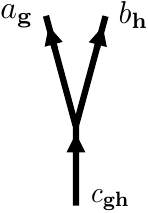} \,, \quad R^{a_{\bs{g}} b_{\bs{h}}} = \includegraphics[height=2cm, valign=c]{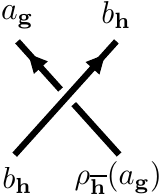} = \sum_c \sqrt{\frac{d_c}{d_a d_b}} R^{a_{\bs{g}} b_{\bs{h}}}_{c_{\bs{gh}}}  \,\, \includegraphics[height=2cm, valign=c]{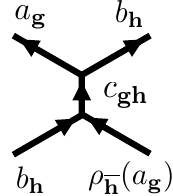} \,. 
\end{equation}

Moreover, unlike in a UMTC, where anyon worldlines can slide freely both above and below any trivalent vertex, sliding a defect worldline of $x_{\bs{k}}$ above a trivalent vertex induces a unitary transformation $U_{\bs{k}}(a,b,c)$ from the state space $V^{ab}_c$ to $V^{\rho_{\bs{k}}(a) \rho_{\bs{k}}(b)}_{\rho_{\bs{k}}(c)}$ due to the crossing between the $\bs{k}$-branch cut with $a,b,c \in \mathcal{C}^{\times}_G$. On the other hand, since the trivalent vertex indicates where the $\bs{gh}$ branch cut splits into $\bs{g}$ and $\bs{h}$, sliding a defect worldline of $x_{\bs{k}}$ below the trivalent vertex picks up the difference between acting on $x_{\bs{k}}$ with $\bs{gh}$ and acting with $\bs{g}$ and $\bs{h}$ in succession. This difference is precisely the projective phase $\eta_{x_{\bs{k}}}(\bs{g}, \bs{h})$ that naturally generalizes the projective phase $\eta_a(\bs{g}, \bs{h})$ (with $a \in \mathcal{C}_0$) to nontrivial defect sectors. These new constraints on sliding moves are summarized as
\begin{equation}\label{eq:Gcross_slide}
    \includegraphics[height=2cm, valign=c]{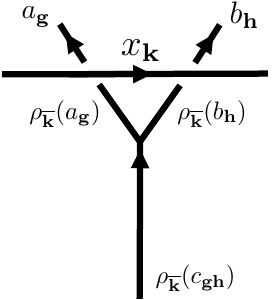} = U_{\bs{k}}(a_{\bs{g}}, b_{\bs{h}}, c_{\bs{gh}}) \,\includegraphics[height=2cm, valign=c]{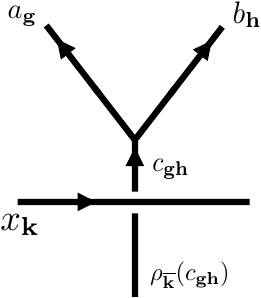} \,, \quad\quad \quad \includegraphics[height=2cm, valign=c]{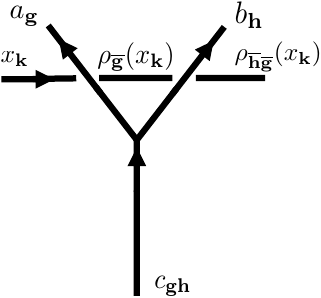} = \eta_{x_{\bs{k}}}(\bs{g}, \bs{h})\, \includegraphics[height=2cm, valign=c]{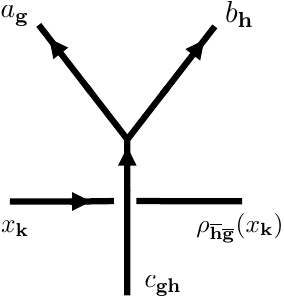} \,.
\end{equation}
Sliding defect lines in different orders gives a new consistency relation for $(U,\eta)$ alone
\begin{equation}
    \eta_{\rho_{\overline{\bs{g}}}(x)}(\bs{h},\bs{k}) \eta_x(\bs{g}, \bs{hk}) = \eta_x(\bs{g},\bs{h}) \eta_x(\bs{gh},\bs{k}) \,, \quad  \eta_b(\bs{k}, \bs{l}) \eta_a(\bs{k},\bs{l}) U_{\bs{l}}(\rho_{\overline{\bs{k}}}(a), \rho_{\overline{\bs{k}}}(b) ; \rho_{\overline{\bs{k}}}(c)) U_{\bs{k}}(a,b;c) = U_{\bs{kl}}(a,b;c) \eta_c(\bs{k},\bs{l}) \,. 
\end{equation}
Furthermore, consistency between $G$-crossed braiding, fusion, and the sliding moves in \eqref{eq:Gcross_slide} can be guaranteed by a more complicated heptagon equation in Fig.~10 of Ref.~\cite{Barkeshli2014_SETbible}, which generalizes the hexagon equation in a UMTC. A solution of these equations (modulo redundancies to be described next) defines a $G$-crossed braided tensor category, which is the complete mathematical structure of the $G$-defectification of $\mathcal{C}$.

As in the definition of UMTCs, it is important to remember that the $F$-symbols, $R$-symbols, and the symmetry action data $U, \eta$ are not physical observables as they depend on the arbitrary choice of basis for each trivalent vertex as well as the arbitrary choice of representative $\rho \in [\rho]$. These arbitrary choices are often referred to as gauge redundancies in the topological data. 

It turns out that, like in a UMTC, the quantum dimensions $d_{a_{\bs{g}}}$, the fusion multiplicities $N^c_{ab}$, and the Frobenius-Schur indicator $\varkappa_a$ (when $a = \bar a$) in the $G$-crossed theory remain gauge-invariant. However, unlike in a UMTC, the topological spin $\theta_{a_{\bs{g}}}$ and the braiding $S$-matrix of the $G$-crossed theory are not gauge-invariant. The failure of braiding to be gauge-invariant has a very physical origin: because the defects are introduced externally, they are not deconfined excitations of the original Hamiltonian. As a result, braiding of defects introduces non-topological Berry phases in addition to universal statistical phases. However, when $a$ and $b$ are braided in the fusion channel $c$, the non-topological Berry phase does not depend on the choice of $c$. Therefore, while the absolute phases accumulated in the fusion channels $c$ and $c'$ are not physical, the relative ratio between them is physical, so long as the branch cut of $b$ does not act on $a$ and vice versa. The existence of these relative invariants implies a projective notion of braiding for any $G$-crossed theory, as first introduced in Ref.~\cite{You2012_projective_stats,Barkeshli2012_nematic_disloc,Barkeshli2012_genon}. Using the full mathematical machinery developed in Ref.~\cite{Barkeshli2014_SETbible,Delaney2018_defectTQC}, one can formulate the projective braiding invariants as
\begin{equation}\label{eq:proj_braiding}
    \left\{\frac{R^{a_{\bs{g}} a_{\bs{g}}}_{c_{\bs{g}^2}}}{R^{a_{\bs{g}} a_{\bs{g}}}_{c'_{\bs{g}^2}}} \,, \quad \forall \bs{g}\right\}  \,, \quad \left\{\frac{(R^{2n})^{a_{\bs{g}} b_{\bs{h}}}_{c_{\bs{gh}}}}{(R^{2n})^{a_{\bs{g}} b_{\bs{h}}}_{c'_{\bs{gh}}}} \,, \quad \forall \,  \rho_{\bs{k}}(a_{\bs{g}}) = a_{\bs{g}} \,, \quad \rho_{\bs{k}}(b_{\bs{h}}) = b_{\bs{h}} \,, \quad \bs{k} = (\bs{gh})^n \right\} \,.  
\end{equation}

In principle, starting from the complete data for a $G$-crossed braided tensor category, one can construct arbitrary combinations of $F, R, U, \eta$ and check whether they are invariant under all the gauge redundancies. The invariants described in \eqref{eq:proj_braiding} are but a small subset of all possible invariants. A more complete (but still not exhaustive) list of invariants can be found in Table I of Ref.~\cite{Barkeshli2014_SETbible}. 

Finally, let us briefly comment on the gauging operation, which promotes the global symmetry group $G$ into a gauge symmetry. Algebraically, this process is complicated in general and can be carried out following the recipe in Section VIII of Ref.~\cite{Barkeshli2014_SETbible}. However, the recipe becomes vastly simplified when the parent topological order $\mathcal{C}$ admits a Lagrangian description. In particular, let us suppose that the partition function of the parent topological order $\mathcal{C}$ is given by a path integral
\begin{equation}
    Z[\mathcal{C}] = \int \prod_i D a_i \exp{- S[a_i]} \,, 
\end{equation}
where $a_i$ is a set of dynamical fields (in the case of Chern-Simons theories, these are precisely the Chern-Simons gauge fields). Starting from this theory, one can introduce extrinsic $G$-defects by coupling $S$ to static background $G$-gauge fields $B_G$. Gauging can then be achieved by promoting the background gauge field $B_G$ into a fluctuating dynamical gauge field $b_G$. The partition function of the new theory is therefore related to the original action $S$ via 
\begin{equation}
    Z[(\mathcal{C}^{\times}_G)^G] = \int \prod_i D a_i \int D b_G \exp{- S[a_i, b_G]} \,. 
\end{equation}
By examining the structure of $S[a_i, b_G]$, it is often possible to directly deduce all the topological data of the gauged theory $(\mathcal{C}^{\times}_G)^G$ without going through the algebraic construction. This shortcut will suffice for our case study in App.~\ref{app:U(2)40_TO}.

\section{Topological order of a charge-\texorpdfstring{$4e$}{} superconductor described by \texorpdfstring{$U(2)_{4,0}$}{}}\label{app:U(2)40_TO}

In this Appendix, we work out the topological order of the $U(2)_{4,0}$ $4e$ SC state in detail, following the general theoretical framework reviewed in App.~\ref{app:SETreview}. The background gauge fields can be decomposed into the charge gauge field $A = (A_1 + A_2)/2$ and the spin gauge field $A_s = A_1 - A_2$. Since the spin gauge field does not couple to the $U(2)_{4,0}$ theory, we will neglect it for the moment and keep track of only the charge gauge field $A$. 

Following the order of presentation in the main text, we will first keep $A$ static and work out the intrinsic topological order of $U(2)_{4,0}$ as well as its enrichment by the unbroken $\mathbb{Z}_4$ symmetry of the $4e$ SC. Then we will introduce vortices as extrinsic $\mathbb{Z}_4$ defects and work out the structure of the resulting $G$-crossed extension. Finally, we will gauge the $\mathbb{Z}_4$ global symmetry and deduce the resulting topological order in which vortices are promoted to dynamical deconfined excitations. 

\subsection{Topological order with non-dynamical electromagnetic gauge field}
\label{app:TO}

Let us first deduce the topological order realized by $U(2)_{4,0}$ directly from the Chern-Simons description
\begin{equation}
    L = - \frac{4}{4\pi} \Tr \left(\alpha \, d \alpha + \frac{2}{3} \alpha^3\right) + \frac{2}{4\pi} \Tr \alpha \, d \Tr \alpha + \frac{2}{2\pi} \Tr \alpha\, d A \,,
\end{equation}
where $\alpha$ is a $U(2)$ gauge field. By definition, the $U(2)$ gauge theory can be obtained from the $SU(2)_4 \times U(1)_0$ theory by gauging a $\mathbb{Z}_2$ one-form symmetry. In terms of the $SU(2)$ gauge field $\hat \alpha$ and the $U(1)$ gauge field $\alpha_0$, we can write down the line operator that generates the $\mathbb{Z}_2$ 1-form symmetry
\begin{equation}
    U = W_{j = 2} \cdot H_{\pi} \,, 
\end{equation}
where $W_{j=2}$ is the $j = 2$ Wilson line of the $SU(2)$ gauge field $\hat \alpha$ and $H_{\pi}$ is the $\pi$-flux 't Hooft line of $\alpha_0$. The $\pi$-flux of $\alpha_0$ carries electric charge 2 and can be identified with the local Cooper pair operator. Upon gauging the 1-form symmetry generated by $U$, the local Cooper pair gets identified with the endpoint of the Wilson line $W_{j=2}$. This fact will soon be important when we study symmetry enrichment of the $U(2)_{4,0}$ topological order. In the $U(2)_{4,0}$ theory, since the $U(1)$ sector is a gapless Maxwell theory, magnetic monopoles of $\alpha_0$ are condensed. As a result, gauge charges of $\alpha_0$ are not part of the low energy Hilbert space. Since worldlines of these gauge charges are precisely Wilson lines of $\alpha_0$ charged under $H_{\pi}$, we conclude that $H_{\pi}$ acts trivially on the low energy Hilbert space and can be replaced with the identity. 

\begin{figure}
    \centering
    \includegraphics[width=0.8\linewidth]{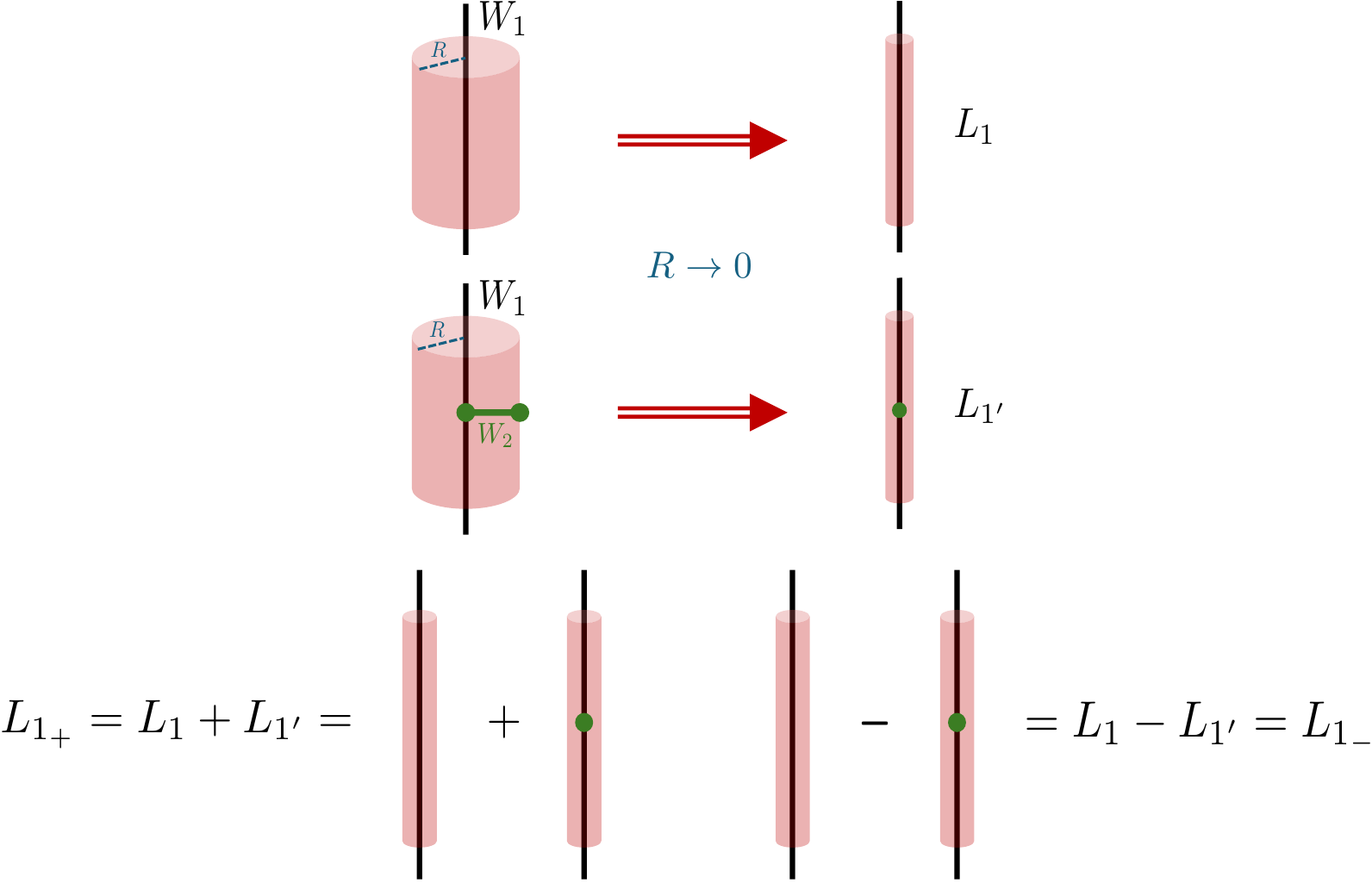}
    \caption{In non-Abelian topological orders, anyon condensation can lead to the splitting of non-Abelian anyons into descendent Abelian anyons. Here we illustrate this phenomenon through the example of $SU(2)_4/\mathbb{Z}_2$. For each cylindrical surface with radius $R$, the interior is $SU(2)_4$ while the exterior is $SU(2)_4/\mathbb{Z}_2$ (with the $j =2$ anyon condensed). $W_j$ labels the Wilson line with spin $j$ in $SU(2)_4$.}
    \label{fig:Anyon_splitting}
\end{figure}

Given the trivial action of $H_{\pi}$, the $\mathbb{Z}_2$ 1-form symmetry generator reduces to $U = W_{j=2}$, and the topological order of the quotient theory can be deduced by summing over $W_{j=2}$ insertions in the path integral for the $SU(2)_4$ Chern-Simons theory. The irreducible anyons in $SU(2)_4$ correspond to line operators with spin $j = \{0, 1/2, 1, 3/2, 2\}$. Following the rules first described in Ref.~\cite{Moore1989_anyon_condensation_CS}, the summation over $W_{j=2}$ confines all other lines with half-integer spin that braid nontrivially with $W_{j=2}$. The only remaining nontrivial line is $W_{j = 1}$. We now argue that after gauging, $W_{j = 1}$ becomes nonsimple and must split into a pair of simple lines.\footnote{We credit the general understanding of anyon splitting in terms of Wilson lines to the authors of Ref.~\cite{Lam2026_splitting}.} To see that, let us divide spacetime into two parts by a cylindrical surface with radius $R$. We refer to the region bounded by the surface as $\mathcal{M}$ and the complement of this region $\mathcal{M}^c$. Consider a Lagrangian which describes $SU(2)_4$ in $\mathcal{M}$ and $SU(2)_4/\mathbb{Z}_2$ in $\mathcal{M}^c$. In other words, the $\mathbb{Z}_2$ 1-form symmetry is gauged on the $\mathcal{M}^c$ region and the cylindrical surface is the gauging interface. Now consider two possible operator insertions: (a) line $L_1$ is a vertical $W_{j=1}$ insertion in $\mathcal{M}$; (b) line $L_{1'}$ is a vertical $W_{j=1}$ in $\mathcal{M}$ dressed by a $W_{j=2}$ insertion with one endpoint on $W_{j=1}$ and the other endpoint on the gauging interface (see Fig.~\ref{fig:Anyon_splitting}). Since $1 \times 2 = 1$ in $SU(2)_4$, the trivalent junction between $W_{j=1}$ and $W_{j=2}$ and the intersection between $W_{j=2}$ and the gauging interface are both topological. 

Now let us shrink the gauging interface. In the $R \rightarrow 0$ limit, the $\mathcal{M}$ region disappears and the two distinct operator insertions become two genuine line operators in the gauged theory $SU(2)_4/\mathbb{Z}_2$. While $1$ is a bare Wilson line, $1'$ is dressed by a topological local operator $\mathcal{O}$ (that can freely slide along the Wilson line) descending from $W_{j=2}$. The existence of this topological local operator implies that the $j = 1$ line becomes nonsimple in the new theory $SU(2)_4/\mathbb{Z}_2$ and must split into simple lines that are eigenvectors of the topological local operator $\mathcal{O}$. Since $2 \times 2 = 0$, $\mathcal{O} L_1 = L_{1'}$ and $\mathcal{O}L_{1'} = L_1$. Therefore, the eigenvectors of $\mathcal{O}$ are precisely given by the symmetric and antisymmetric combinations of $L_1$ and $L_{1'}$, which we label as
\begin{equation}
    L_{1_{\pm}} = L_1 \pm L_{1'} \,, \quad \mathcal{O} L_{1_+} = L_{1_+} \,, \quad \mathcal{O} L_{1_-} = - L_{1_-} \,.
\end{equation}
As $L_1$ and $L_{1'}$ are related by a topological local operator, the corresponding anyons $1_+$ and $1_-$ have identical topological spin $e^{2\pi i/3}$ inherited from the parent theory. Moreover, they satisfy a simple fusion relation $1_+^2 = 1_-$, $1_-^2 = 1_+$. These are precisely the fusion rules of the bosonic $\mathbb{Z}_3$ topological order, also known as $SU(3)_1$~\cite{Bais2009_anyon_condensation}.

Using the same Chern-Simons framework, we can also work out the enrichment of this bosonic $\mathbb{Z}_3$ topological order by the residual $\mathbb{Z}_4$ symmetry of the $4e$ SC. Let $\bs{g}$ be the generator of $\mathbb{Z}_4$ such that $\bs{g}^2 = (-1)^F$ acts as the fermion parity. Since $U(2)_{4,0}$ is a bosonic theory, the fermionic nature of the microscopic system is simply incorporated by stacking $U(2)_{4,0}$ with an invertible fermion sector $\{1, c\}$. Given this trivial grading, $\bs{g}^2$ acts trivially on all the nontrivial anyons of $U(2)_{4,0}$ and it suffices to work in the parity-even sector where the $\mathbb{Z}_4$ symmetry is reduced to $\mathbb{Z}_2$. The operator that carries unit charge under the $\mathbb{Z}_2$ generator $\bs{g}$ is precisely the Cooper pair operator. In the Chern-Simons description, the Cooper pair lives on the endpoint of the $W_{j=2}$ Wilson line, as we previously showed. 

Now let us recall that $L_{1'}$ is related to $L_1$ by dressing with a localized Cooper pair. Since $\bs{g}$ acts on the Cooper pair with a $-1$ factor, we immediately deduce that $\bs{g}$ sends $L_1 + L_{1'}$ to $L_1 - L_{1'}$ and vice versa. Physically, this means that when an anyon $1_+$ ($1_-$) winds around a localized $hc/(4e)$ vortex, it transmutes into $1_-$ ($1_+$). Therefore, the internal $\mathbb{Z}_2$ symmetry in the UV theory acts as an anyon permutation symmetry in the IR that exchanges the two nontrivial anyons $1_+, 1_-$ in $SU(2)_4/\mathbb{Z}_2$. This is the result quoted in the main text. 

Given the simple $\mathbb{Z}_3$ fusion rule of $SU(2)_4/\mathbb{Z}_2$, we will henceforth relabel its anyons by a single integer $a = \{0, 1, 2\}$ with $a = 0$ denoting the vacuum and $1, 2$ denoting the nontrivial Abelian anyons $1_+, 1_-$. This notation will make it easier to write down formulae for more complicated topological data in the next section.

\subsection{Defectification and parafermion zero modes}

In a real system, we have the freedom to introduce and manipulate vortices of the superconductor carrying quantized vorticity. These vortices can be viewed as extrinsic defects of the residual $\mathbb{Z}_2$ symmetry. Including these extrinsic defects promotes the UMTC $\mathcal{C} = SU(2)_4/\mathbb{Z}_2$ to a $G$-crossed braided tensor category $\mathcal{C}_{\mathbb{Z}_2}^{\times}$. From the general theory reviewed in App.~\ref{app:SETreview}, we know that $\mathcal{C}_{\mathbb{Z}_2}^{\times}$ has a graded fusion structure
\begin{equation}
    \mathcal{C}_{\mathbb{Z}_2}^{\times} = \mathcal{C}_0 \oplus \mathcal{C}_{\bs{g}} \,, \quad \mathcal{D}_0^2 = \mathcal{D}_{\bs{g}}^2 = 3 \,,
\end{equation}
where $\mathcal{D}_{0}, \mathcal{D}_{\bs{g}}$ are the quantum dimensions of the two defect sectors. Moreover, the number of distinct elements in $\mathcal{C}_{\bs{g}}$ is equal to the number of elements in $\mathcal{C}_0$ that are invariant under the action of $\bs{g}$. These conditions imply that $\mathcal{C}_{\bs{g}}$ contains a single element $\sigma$ with $d_{\sigma} = \sqrt{3}$. Since fusion has to be consistent with the anyon permutation symmetry, the only possible fusion rule is
\begin{equation}\label{eq:Z3_fusion}
    \sigma \times \sigma = \oplus_{a = 0}^2 \, a \,, \quad \sigma \times a = a \times \sigma = \sigma \,. 
\end{equation}
This fusion rule matches the fusion of $\mathbb{Z}_3$ parafermions, sometimes also referred to as the $\mathbb{Z}_3$ Tambara-Yamagami (TY) fusion category~\cite{Tambara1998_TYcategory}. 

To further specify the universal data of this theory, we need to pick a gauge. Following Ref.~\cite{Barkeshli2014_SETbible}, we can choose a gauge in which $\eta_a(\bs{g}, \bs{h}) = 1$ and $U_{\bs{g}}(a, b; [a+b]_3) = 1$. This choice is allowed because the $\mathbb{Z}_2$ symmetry fractionalization class is trivial in $\mathcal{C} = SU(2)_4/\mathbb{Z}_2$. With this gauge choice, we can extract the gauge-dependent $F$ and $R$ symbols. 

We begin with the $F$ symbols, which are maps between fusion spaces:
\begin{equation}
    F^{abc}_d: \oplus_e V^{ab}_e \otimes V^{ec}_d \rightarrow \oplus_f V^{af}_d \otimes V^{bc}_f \,, \quad \dim V^{ab}_c = N^c_{ab} \,. 
\end{equation}
Since all the fusion multiplicities are 0 or 1 in \eqref{eq:Z3_fusion}, the $F$-symbol does not carry additional indices and we can write it in explicit component form $F^{abc}_{def} = [F^{abc}_d]_{ef}$, where each entry is a complex number. The $F$-symbols satisfy the pentagon identity 
\begin{equation}
    \sum_{h} F^{abc}_{gfh} F^{ahd}_{egk} F^{bcd}_{khl} = F^{fcd}_{egl} F^{abl}_{efk} \,.
\end{equation}
Based on the fusion rules, we see that the only nontrivial $F$-symbols involve four factors of $\sigma$. If $a,b,c,d$ are all $\sigma$, then $e, f$ have to be both Abelian. If $d,e,f$ are all $\sigma$, then $b$ has to be $\sigma$, while $a,c$ can be Abelian. The only other possibility with four factors of $\sigma$ is $a = c = \sigma$ and $e = f = \sigma$. It turns out that in these cases, the most general solution to the pentagon equations takes the form
\begin{equation}\label{eq: F symbol}
    F^{e \sigma f}_{\sigma \sigma \sigma} = F^{\sigma e \sigma}_{f \sigma \sigma} = \chi(e,f) \,, \quad F^{\sigma\sigma\sigma}_{\sigma ef} = \frac{\kappa_{\sigma}}{\sqrt{3}} \chi(e,f)^{-1} \,,
\end{equation}
where $\kappa_{\sigma}$ is the Frobenius-Schur indicator of $\sigma$ and $\chi:\mathbb{Z}_3 \times \mathbb{Z}_3 \rightarrow U(1)$ is a symmetric bicharacter satisfying 
\begin{equation}
    \chi(e, f) = \chi(f, e) \,, \quad \chi(ef, g) = \chi(e, g) \chi(f, g) \,, \quad \chi(e, fg) = \chi(e,f) \chi(e,g) \,, \quad \chi(0, e) = \chi(e,0) = 1 \,. 
\end{equation}
The two choices of the Frobenius-Schur indicator $\kappa_{\sigma} = \pm 1$ correspond to the two nontrivial defectification classes in $H^3(\mathbb{Z}_2, U(1)) = \mathbb{Z}_2$. For $\mathcal{C}_0 = SU(2)_4/\mathbb{Z}_2$, we are in the class with $\kappa_{\sigma} = -1$~\cite{Barkeshli2014_SETbible} and the symmetric bicharacter takes the form
\begin{equation}
    \chi(e, f) = e^{2\pi i ef/3} \,, \quad e, f = \{0,1,2\} \,. 
\end{equation}

Next, we turn to the $R$-symbols that encode braiding. Following Ref.~\cite{Barkeshli2014_SETbible}, the general solution to the consistency equations between $R$ and $F$ symbols takes the form
\begin{equation}\label{eq: R symbol}
    R^{\sigma\sigma}_a = Y (-1)^a e^{i\pi a^2/3} \,, \quad R^{\sigma a}_{\sigma} = (-1)^a e^{-i\pi a^2/3} \,, \quad R^{a \sigma}_{\sigma} = r^a (-1)^a e^{-i\pi a^2/3} \,,
\end{equation}
\begin{equation}
    U_1(a, \sigma; \sigma) = U_1(\sigma, a; \sigma) = r^a \,, \quad U_1(\sigma,\sigma;a) = r^{-a} \,, \quad Y^2 = \frac{\kappa_{\sigma}}{\sqrt{3}} \sum_{a=0}^2 (-1)^a e^{-i \pi a^2/3} = - i \,,
\end{equation}
where $r = e^{2\pi n i/3}$ is an arbitrary choice for some integer $n$. The remaining gauge freedom allows us to set $r = 1$ and fix the sign of the phase factor $Y$ (in any case, this sign will cancel out in all gauge-invariant expressions). 

From the $R$ symbols, we can obtain the topological twist and the S-matrix
\begin{equation}
    \theta_{\sigma} = \sum_a \frac{d_a}{d_{\sigma}} R^{\sigma\sigma}_a = \kappa_{\sigma} Y^{-1} \,, \quad S_{0\sigma} = S_{\sigma0} = 1 \,, \quad S_{\sigma \sigma} = \frac{1}{\sqrt{3}} \sum_a (R^{\sigma\sigma}_a)^2 = \Theta_0 \theta_{\sigma}^{-2} \,.
\end{equation}
We note that although $S$ and $\theta$ are not gauge-invariant, the specific combination $\Theta_0 = e^{i\pi c_-/4}$ is gauge-invariant and defines the chiral central charge $c_-$.

From these data, we can compute braiding invariants of the general form \eqref{eq:proj_braiding}. If $a$ is Abelian, then $c$ must also be Abelian and the braiding matrix $R^{aa}_c$ is itself gauge-invariant because it does not involve any nontrivial defects. On the other hand, if $a = \sigma$, only the ratio $R^{\sigma\sigma}_a/R^{\sigma\sigma}_b$ is well-defined
\begin{equation}\label{eq:Z3_Rinvariant_1}
    \frac{R^{\sigma\sigma}_a}{R^{\sigma\sigma}_b} = (-1)^{a-b} e^{-i\pi (a^2-b^2)/3} \,. 
\end{equation}
A more general class of projective invariants in \eqref{eq:proj_braiding} take the form
\begin{equation}
    \frac{(R^{2n})^{a_{\bs{g}} b_{\bs{h}}}_{c_{\bs{gh}}}}{(R^{2n})^{a_{\bs{g}} b_{\bs{h}}}_{c'_{\bs{gh}}}} \,, \quad \rho_{(\bs{gh})^n}(a_{\bs{g}}) = a_{\bs{g}} \,, \quad \rho_{(\bs{gh})^n}(b_{\bs{h}}) = b_{\bs{h}} \,. 
\end{equation}
If $\bs{gh}$ is a trivial $\mathbb{Z}_2$ element, and $\bs{g}, \bs{h}$ are both trivial, then we recover the gauge-invariant $R$ symbols of the parent Abelian anyon theory. If $\bs{g}, \bs{h}$ are both non-trivial, then we recover the exchange statistics of $\sigma$. Finally, if $\bs{gh}$ is a nontrivial $\mathbb{Z}_2$ element, then $c_{\bs{gh}} = c'_{\bs{gh}} = \sigma$ and the numerator equals the denominator. Thus, we conclude that there is no new invariant that goes beyond \eqref{eq:Z3_Rinvariant_1}. These braiding data will be used to construct the qutrit Clifford gates in App.~\ref{app:Clifford_gates}. 

\subsection{Topological order with dynamical electromagnetic gauge field}

Finally, let us promote the $\mathbb{Z}_2$ global symmetry to a gauge symmetry and obtain the gauged topological order $(\mathcal{C}_{\mathbb{Z}_2}^{\times})^{\mathbb{Z}_2}$. As explained in App.~\ref{app:SETreview}, it is easiest to carry out this procedure in the field-theoretic description. 

Suppose we start with the $U(2)_{4,0}$ theory coupled to a background $\mathbb{Z}_2$ gauge field $B$ through the Lagrangian 
\begin{equation}
    L[\alpha, B] = -\frac{4}{4\pi} \Tr \left(\alpha \, d \alpha + \frac{2}{3} \alpha^3\right) + \frac{2}{4\pi} \Tr \alpha \, d \Tr \alpha + \frac{1}{2\pi} \Tr \alpha \, d B \,. 
\end{equation}
Gauging the $\mathbb{Z}_2$ symmetry means promoting the background field $B$ to a dynamical gauge field $b$. Integrating over $b$ sets the $U(1)$ part of the $U(2)$ gauge field $\alpha$ to zero. As a result, the gauged theory is simply $SU(2)_4$, in agreement with the algebraic answer in Ref.~\cite{Bais2009_anyon_condensation}. Note that a different choice of the Frobenius-Schur indicator $\kappa_{\sigma} = 1$ would produce the same $SU(2)_4$ topological order upon gauging. The phenomenon in which multiple distinct SETs become identical after gauging has been discussed in Refs.~\cite{Lu2013_SETCS,Burnell2015_gapbdry_semion}.

Now let us bring back the electron sector so that the full unbroken symmetry group is actually $\mathbb{Z}_4$. In the UV system, this $\mathbb{Z}_4$ is part of a $U(1)$ symmetry under which the electron carries unit charge. Moreover, we have to remember that the $U(2)_{4,0}$ theory is stacked with two copies of fermionic integer quantum Hall states with total Hall conductance $\sigma_{xy} = -2 e^2/h$. Therefore, the complete Lagrangian takes the form
\begin{equation}
    L[\alpha, A] = -\frac{4}{4\pi} \Tr \left(\alpha \, d \alpha + \frac{2}{3} \alpha^3\right) + \frac{2}{4\pi} \Tr \alpha \, d \Tr \alpha + \frac{2}{2\pi} \Tr \alpha \, d A - \frac{2}{4\pi} A dA - 2 \Omega_g \,,
\end{equation}
where $A$ is the background $\mathrm{spin}_{\mathbb{C}}$ connection corresponding to the $U(1)$ symmetry. Note that the integration over $\Tr \alpha$ Higgses $A$ to a $\mathbb{Z}_4$ gauge field, which is precisely the unbroken symmetry group of the $4e$ superconductor. 

Now we promote $A$ to a dynamical $\mathbb{Z}_4$ gauge field $a$. To simplify this Lagrangian further, we can ``integrate in" an auxiliary $U(1)$ gauge field $\beta$ and rewrite the Lagrangian as 
\begin{equation}
    L[\alpha, \beta, a] = - \frac{4}{4\pi} \Tr \left(\alpha \, d\alpha+ \frac{2}{3} \alpha^3\right) + \frac{2}{4\pi} \Tr \alpha \, d \Tr \alpha + \frac{2}{2\pi} \Tr \alpha d a + \frac{1}{4\pi} \beta d \beta - \frac{1}{2\pi} \beta d a - \frac{1}{4\pi} a da - \Omega_g \,. 
\end{equation}
Since $a$ is a dynamical gauge field, we can integrate it out and reduce the Lagrangian to
\begin{equation}
    L[\alpha, \beta] = - \frac{4}{4\pi} \Tr \left(\alpha \, d\alpha+ \frac{2}{3} \alpha^3\right) + \frac{2}{4\pi} \Tr \alpha \, d \Tr \alpha + \frac{1}{4\pi} \beta d \beta + \frac{1}{4\pi} (\beta - 2 \Tr \alpha) \, d (\beta - 2 \Tr \alpha) \,. 
\end{equation}
Further shifting $\beta \rightarrow \beta + \Tr \alpha$ removes the mutual coupling between $\beta$ and $\Tr \alpha$, giving rise to the final Lagrangian
\begin{equation}
    L[\alpha, \beta] = - \frac{4}{4\pi} \Tr \left(\alpha \, d\alpha+ \frac{2}{3} \alpha^3\right) + \frac{4}{4\pi} \Tr \alpha \, d \Tr \alpha + \frac{2}{4\pi} \beta d \beta \,.
\end{equation}
We recognize this theory as the factorized topological order $(\mathcal{C}_{\mathbb{Z}_4}^{\times})^{\mathbb{Z}_4} = U(2)_{4,-8} \times U(1)_{-2}$.

Note that $(\mathcal{C}_{\mathbb{Z}_4}^{\times})^{\mathbb{Z}_4}$ is very different from $(\mathcal{C}_{\mathbb{Z}_2}^{\times})^{\mathbb{Z}_2}$. This is expected because $\mathcal{C}_{\mathbb{Z}_2}^{\times}$ and $\mathcal{C}_{\mathbb{Z}_4}^{\times}$ are two different defectifications of $\mathcal{C}$ related by stacking fermionic invertible phases. Nevertheless, since the stacking of invertible phases mostly affects the Abelian sector of the gauged theory, the computational power of $(\mathcal{C}_{\mathbb{Z}_4}^{\times})^{\mathbb{Z}_4}$ and $(\mathcal{C}_{\mathbb{Z}_2}^{\times})^{\mathbb{Z}_2}$ are identical. In particular, both are weakly integral topological orders that are not computationally universal through braiding alone but becomes universal when braiding is augmented by fusion and topological charge measurements~\cite{Hastings2012_metaplectic_power,Cui2014_SU(2)4_gate,Levaillant2015_SU(2)4_gate, Bocharov2015_SU(2)4_compilation}.

\subsection{Embedding in bilayer quantum Hall}

Finally, let us comment on an application to bilayer quantum Hall states at $\nu = 1/2 + 1/2$. Using the composite fermion construction, we can describe the bilayer system with an effective Lagrangian
\begin{equation}\label{eq:bilayer_halfLL}
    L = \sum_{\sigma = 1}^2 \left\{ L[f_{\sigma}, a_{\sigma}] - \frac{2}{4\pi} \beta_{\sigma} d \beta_{\sigma} + \frac{1}{2\pi} \beta_{\sigma} d (A_{\sigma} - a_{\sigma}) \right\} \,,
\end{equation}
where $A_{\sigma}$ is the background $U(1)$ gauge field for layer $\sigma$, and $f_{\sigma}$ is the composite fermion in layer $\sigma$ coupled to an emergent $U(1)$ gauge field $a_{\sigma}$ that implements flux attachment. At $\nu = 1/2$, a bilayer composite Fermi liquid is obtained if each $f_{\sigma}$ forms a Fermi surface (this state turns out to be unstable and flows to an exciton superfluid at low temperature). On the other hand, if each $f_{\sigma}$ forms a $p+ip$ topological superconductor, the full system realizes a stable bilayer Moore-Read state with $\mathrm{Pf}^2$ topological order. 

Now let us imagine that the composite fermion bilayer $p+ip$ SC undergoes a vortex-antivortex condensation transition. The effective Lagrangian for the composite fermion sector becomes
\begin{equation}
    \sum_{\sigma} L[f_{\sigma}, a_{\sigma}] \rightarrow - \frac{4}{4\pi} \Tr \left(\alpha \, d \alpha + \frac{2}{3} \alpha^3\right) + \frac{2}{4\pi} \Tr \alpha \, d \Tr \alpha + \frac{1}{2\pi} \Tr \alpha \, d (a_1 + a_2) - \mathrm{CS}[a_1,g] - \mathrm{CS}[a_2,g] \,. 
\end{equation}
Plugging this theory back into \eqref{eq:bilayer_halfLL} gives
\begin{equation}   
    \begin{aligned}
    L &= - \frac{4}{4\pi} \Tr \left(\alpha \, d \alpha + \frac{2}{3} \alpha^3\right) + \frac{2}{4\pi} \Tr \alpha \, d \Tr \alpha - \mathrm{CS}[a_1,g] - \mathrm{CS}[a_2,g] \\
    &- \frac{2}{4\pi} \beta_1 d \beta_1 + \frac{1}{2\pi} \beta_1 d A_1 - \frac{2}{4\pi} \beta_2 d \beta_2 + \frac{1}{2\pi} \beta_2 d A_2 + \frac{1}{2\pi} a_1 d (\Tr \alpha - \beta_1) + \frac{1}{2\pi} a_2 d (\Tr \alpha - \beta_2) \,.
    \end{aligned}
\end{equation}
Integrating out the emergent gauge fields $a_{\sigma}$ simplifies the Lagrangian to
\begin{equation}
    L = - \frac{4}{4\pi} \Tr \left(\alpha \, d \alpha + \frac{2}{3} \alpha^3\right) + \frac{4}{4\pi} \Tr \alpha \, d \Tr \alpha + \sum_{\sigma=1}^2 \left\{- \frac{1}{4\pi} \beta_{\sigma} d \beta_{\sigma} + \frac{1}{2\pi} \beta_{\sigma} d (A_{\sigma} - \Tr \alpha) \right\} \,. 
\end{equation}
Further integrating over $\beta_{\sigma}$ gives 
\begin{equation}
    L = - \frac{4}{4\pi} \Tr \left(\alpha \, d \alpha + \frac{2}{3} \alpha^3\right) + \frac{6}{4\pi} \Tr \alpha \, d \Tr \alpha - \frac{1}{2\pi} \Tr \alpha d (A_1 + A_2) + \mathrm{CS}[A_1,g] + \mathrm{CS}[A_2,g] \,. 
\end{equation}
Treating $A_1, A_2$ as background fields, this final Lagrangian describes a gapped topological phase with non-Abelian topological order $U(2)_{4, -16}$, stacked with two copies of the fermionic IQH state. Like $SU(2)_4$ and $U(2)_{4,-8}$, the theory $U(2)_{4,-16}$ is also weakly integral. This means that computational universality can be achieved when braiding is augmented by fusion and topological charge measurements~\cite{Hastings2012_metaplectic_power,Cui2014_SU(2)4_gate,Levaillant2015_SU(2)4_gate, Bocharov2015_SU(2)4_compilation}.

\section{Computational power of the charge-\texorpdfstring{$4e$}{} superconductor}
\label{app: computational power}

In this section we discuss the computational power of the defects and anyons of the topological charge-\texorpdfstring{$4e$}{} superconductor. For convenience we define $\omega\equiv \mathrm{e}^{\mathrm{i}2\pi/3}$.

We will adopt the standard encoding of a qutrit through the splitting process $0 \rightarrow a\bar{a}\rightarrow\sigma\sigma\sigma\sigma$. In our case each $\sigma$ is a vortex of the $4e$ SC, induced and controlled by an external $\pm hc/(4e)$ flux. We will fix a convention for the flux associated with each anyon as follows:
\begin{equation}
    |a\rangle \equiv \begin{tikzpicture}[baseline=(current bounding box.center), x=1cm,y=1cm, line cap=round, line join=round]
  \def\YA{1.2}
  \def\YB{2.6}

  \TreeOneTwoFour{1}{0}{\YA}{\YB}

\node[xshift=-1pt, yshift=-2pt, fill=white, inner sep=1pt,align=center]  at (0,\YA) {$a$ \\ $(+\frac{hc}{2e})$};

\node[xshift=82pt, yshift=-2pt, fill=white, inner sep=1pt,align=center]  at (0,\YA) {$\bar{a}$ \\ $(+\frac{hc}{2e})$};

  \pgfmathsetmacro{\YTOP}{\YB + \BraidT*\BraidDY}
  \pgfmathsetmacro{\Xone}{\Xof{1}}
  \pgfmathsetmacro{\Xtwo}{\Xof{2}}
  \pgfmathsetmacro{\Xthree}{\Xof{3}}
  \pgfmathsetmacro{\Xfour}{\Xof{4}}
\TopDotsLabels{4}{\YTOP}

\node[above=2pt, xshift=-7pt, yshift=2pt, fill=white, inner sep=1pt,align=center]
  at (\Xone,\YTOP) {$\sigma_1$ \\ $(+\frac{hc}{4e})$};

\node[above=2pt, xshift=-7pt, yshift=2pt, fill=white, inner sep=1pt,align=center]
  at (\Xtwo,\YTOP) {$\sigma_2$ \\ $(+\frac{hc}{4e})$};
  
  \node[above=2pt, xshift=-7pt, yshift=2pt, fill=white, inner sep=1pt,align=center]
  at (\Xthree,\YTOP) {$\sigma_3$ \\ $(+\frac{hc}{4e})$};
  
  \node[above=2pt, xshift=-7pt, yshift=2pt, fill=white, inner sep=1pt,align=center]
  at (\Xfour,\YTOP) {$\sigma_4 $\\ $(+\frac{hc}{4e})$};

  \pgfmathsetmacro{\Xroot}{(\Xof{1}+\Xof{4})/2}
  \node[treelabel, below=2pt,align=center] at (\Xroot,0) {$0$ \\ $
  (+\frac{hc}{e})$};
\end{tikzpicture}
\end{equation}
For simplicity, we will omit the flux assignment in the illustrations below. 

We assume the ability of preparing qutrit states in the  $|0\rangle$ state. Physically, this corresponds to the following process: First, we induce an $hc/e$ flux through the boundary while carefully making sure that there is no topological excitation trapped. Then, split the flux into two $hc/(2e)$ fluxes and separate them far away, and measure the anyon topological charge trapped by each of these fluxes again making sure there is no anyon excitation trapped. 
Eventually, we further split the two $hc/(2e)$ fluxes into four $hc/(4e)$ fluxes, and separate them far away. This realize the initialization of a qutrit in the $|0\rangle$ topologically encoded in the four $\sigma$'s trapped by the four $hc/(4e)$ fluxes. Combined with the Clifford gates we construct below, one can achieve the complete set of Clifford operations in the sense that all the stabilizer states can be reached~\cite{Farinholt2014_Clifford}. 

We note that all the motions of the vortices should be slow enough to avoid non-adiabatic effects (see discussions in the context of $2e$ TSC~\cite{Cheng2011_Nonadiabatic,Cheng2012_topological}) and to avoid exciting phonons (see App.~\ref{app:goldstone} for more detailed discussions).

In all the calculations in this appendix, we fix a gauge $r = 1$ and $\eta_b(\bs{g},\bs{h}) = 1$ (see App.~\ref{app:U(2)40_TO}) and neglect the overall phase factor, as it is not physically meaningful. 

\subsection{Multi-qutrit Clifford gates from braiding vortices}
\label{app:Clifford_gates}

We first illustrate a minimal set of gates that generate the many-qutrit Clifford group: phase, Hadarmard and control-$Z$ gates. Along with the ability of preparing qutrits in the $|0\rangle$ state, they suffice to reach all the stabilizer states. 

The protocol is largely following Ref.~\cite{Hutter2015_parafermion_QC} but we have independently verified the results using the $F$ and $R$ symbols for our topological $4e$ SC in Eqs.~\eqref{eq: F symbol}\&\eqref{eq: R symbol}. The braiding between $\sigma_i$ and $\sigma_{i+1}$ is represented by $B_i$. Each braid implements a unitary transformation $U$ on the qutrit, and we will represent the operation with a matrix in the basis $|a\rangle$: $U_{aa'}\equiv \langle a| U|a'\rangle$. For all the gates below we will drop the possible overall phase factors. 

First we list the unitary operators corresponding to the elementary braids within a single qutrit:
\begin{align}
    (B_1)_{aa'} =&  (B_3)_{aa'}= \begin{bmatrix}
        1 & 0 & 0 \\
        0 & \omega^{-1}  & 0 \\
        0 & 0 & \omega^{-1}
    \end{bmatrix} \,,  \\
(B_2)_{aa'} =& \frac{1}{\sqrt{3}}\begin{bmatrix}
        1 & \omega & \omega \\
        \omega & 1  & \omega \\
        \omega & \omega & 1
    \end{bmatrix} \,. 
\end{align}
These give rise to the following generators of the single-qutrit Clifford gates: Phase gate $B_1$ and Hadamard $B_1B_2B_1$.

\begin{center}
$Q=$\begin{tikzpicture}[ x=1cm,y=1cm,baseline=(current bounding box.center), line cap=round, line join=round]
  \def\YA{1.2}
  \def\YB{2.6}

  \TreeOneTwoFour{1}{0}{\YA}{\YB}

  \BraidFromOffset{\YB}{4}{+1}

  \pgfmathsetmacro{\YTOP}{\YB + \BraidT*\BraidDY}
  \TopDotsLabels{4}{\YTOP}

  \pgfmathsetmacro{\Xroot}{(\Xof{1}+\Xof{4})/2}
  \node[treelabel, below=2pt] at (\Xroot,0) {$0$};
\end{tikzpicture}  \ \ $H=$\begin{tikzpicture}[x=1cm,y=1cm,baseline=(current bounding box.center), line cap=round, line join=round]
  \def\YA{1.2}
  \def\YB{2.6}

  \TreeOneTwoFour{1}{0}{\YA}{\YB}

  \BraidFromOffset{\YB}{4}{+1,+2,+1}

  \pgfmathsetmacro{\YTOP}{\YB + \BraidT*\BraidDY}
  \TopDotsLabels{4}{\YTOP}

  \pgfmathsetmacro{\Xroot}{(\Xof{1}+\Xof{4})/2}
  \node[treelabel, below=2pt] at (\Xroot,0) {$0$};
\end{tikzpicture}
\end{center}

\begin{align}
    Q_{aa'} =& \delta_{aa'} \omega^{-a^2} \,, \\
    H_{aa'} =& \frac{1}{\sqrt{3}}\omega^{aa'} \,. 
\end{align}


For two qutrits encoded in $\sigma_{1,2,3,4}$ and $\sigma_{5,6,7,8}$, there is a simple construction of the entangling control-$Z$ gate involving the mutual braiding of $(\sigma_3\sigma_4)$ and $(\sigma_5\sigma_6)$: $(B_4B_3B_5B_4)^2$:

\begin{center}
$C_Z = $ \begin{tikzpicture}[x=1cm,y=1cm,baseline=(current bounding box.center), line cap=round, line join=round]
  \def\YA{1.2}
  \def\YB{2.6}

  \TreeOneTwoFour{1}{0}{\YA}{\YB}
  \TreeOuterLabels{1}{1.45}{a}{\Bar{a}}

  \TreeOneTwoFour{5}{0}{\YA}{\YB}
  \TreeOuterLabels{5}{1.45}{b}{\Bar{b}}

  \BraidN=8\relax
  \BraidT=0\relax
  \BraidParStep{\YB}{+4}
  \BraidParStep{\YB}{+3,+5}
  \BraidParStep{\YB}{+4}
  \BraidParStep{\YB}{+4}
  \BraidParStep{\YB}{+3,+5}
  \BraidParStep{\YB}{+4}

  \pgfmathsetmacro{\YTOP}{\YB + \BraidT*\BraidDY}
  \TopDotsLabels{8}{\YTOP}

  \pgfmathsetmacro{\XrootL}{(\Xof{1}+\Xof{4})/2}
  \pgfmathsetmacro{\XrootR}{(\Xof{5}+\Xof{8})/2}
  \node[treelabel, below=2pt] at (\XrootL,0) {$0$};
  \node[treelabel, below=2pt] at (\XrootR,0) {$0$};
\end{tikzpicture}
\end{center}
Note that, for two qutrits in the state $|a\rangle\otimes |b\rangle$, this is simply a braiding between $\bar{a}$ and $b$ anyons encoded in the two pairs of  $\sigma$'s. This doesn't change the state but will assign a state-dependent phase. Specifically, it acts as:
\begin{align}
    (C_Z)_{ab;a'b'} = \delta_{aa'}\delta_{bb'} \omega^{ab} \,.
\end{align}

\subsection{Measurement preparation of a qutrit magic state}
\label{app:magicstate}

Finally we show how to use interferometric measurements to prepare a qutrit magic state (non-stabilizer state), which will supplement the Clifford gates through magic state injection to achieve a universal gate set. 

\begin{figure}[t]
    \centering
    \includegraphics[width=0.5\linewidth]{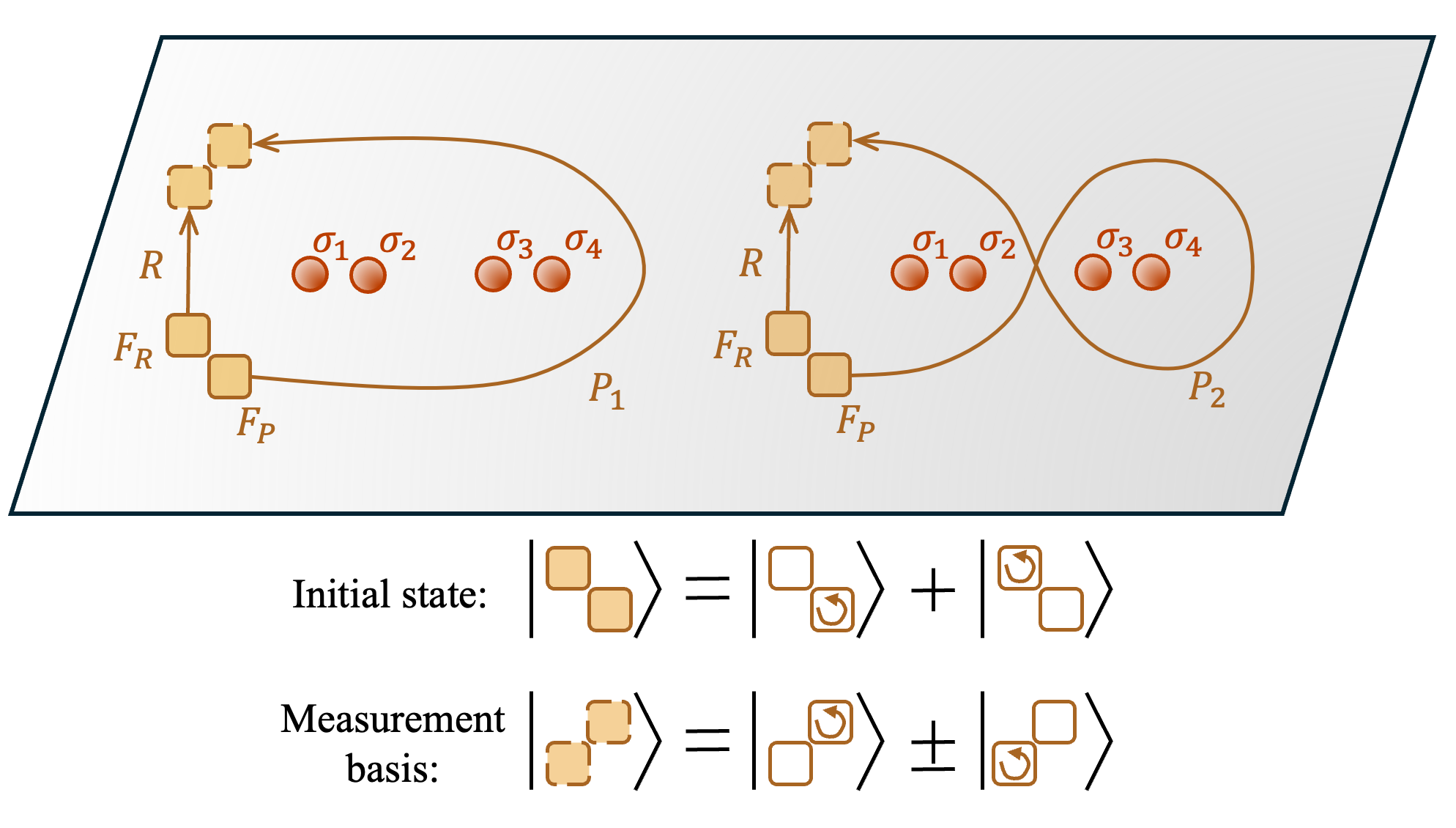}
    \caption{An illustration of the interferometry measurement protocol. Each yellow, rounded square (labeled by $F$) represents a fluxonium (essentially a $4e$ superconducting ring, which doesn't need to be topological) sitting above the plane, which have two meta-stable states: with or without an $hc/(4e)$ magnetic flux. The state with an $hc/(4e)$ magnetic flux with induce a $\sigma$ mode in the adjacent $4e$ TSC thin film. The protocol requires one to prepare a pair of fluxoniums in a Bell state, and then send the two fluxoniums in different paths respectively--reference (R) and probing (P) paths--and eventually measure the outcome in the Bell basis.  Two possible probing paths, $P_{1,2}$, are plotted. 
    }
    \label{fig:interferometry}
\end{figure}

Our protocol uses a single probe $\sigma$ (whose fusion history doesn't matter) and is illustrated in Fig.~\ref{fig:interferometry}. This is a slightly generalized Mach-Zehnder interferometry setup, where we consider two paths $R$ and $P$ for the probe $\sigma$. These two paths are to be coherently superposed for the probe anyon in the interferometry arms (we will discuss how to physically realize this later). We denote the corresponding states in the space of paths $|R\rangle$ and $|P\rangle$. The path $R$ is a completely trivial reference path, where the probe anyon does not braid with the anyons in the target qutrit. The probing path $P$, however, will braid non-trivially with the target qutrit following either of the paths shown below (the red lines represent probe anyons whereas the black lines represent the anyons constituting the target qutrit):
\begin{center}
$P_1$:
\begin{tikzpicture}[x=1cm,y=1cm,baseline=(current bounding box.center), line cap=round, line join=round]

  \def\YA{-0.6}
  \def\YB{1.0}
  \pgfmathsetmacro{\YOuterRoot}{-2.0}

  \begin{scope}[xshift={-\BraidDX cm}]

    \pgfmathsetmacro{\XposOne}{0}
    \pgfmathsetmacro{\XposTwo}{1*\BraidDX}
    \pgfmathsetmacro{\XposFive}{4*\BraidDX}

    \TreeOneTwoFourPositions{2}{3}{4}{5}{\YOuterRoot}{\YA}{\YB}

    \pgfmathsetmacro{\XrootTree}{(\XposTwo+\XposFive)/2}
    \node[treelabel, below=2pt] at (\XrootTree,\YOuterRoot) {$1$};

    \draw[strandline, draw=red] (\XposOne,\YOuterRoot+\JoinEps) -- (\XposOne,\YB+\JoinEps);
    \fill[red] (\XposOne,\YOuterRoot) circle[radius=\BraidDot]; 

    \BraidN=5\relax
    \BraidT=0\relax
    \InitPerm{5}

    \BraidParStepTracked{\YB}{+1}
    \BraidParStepTracked{\YB}{+2}
    \BraidParStepTracked{\YB}{+3}
    \BraidParStepTracked{\YB}{+4}
    \BraidParStepTracked{\YB}{+4}
    \BraidParStepTracked{\YB}{+3}
    \BraidParStepTracked{\YB}{+2}
    \BraidParStepTracked{\YB}{+1}

    \pgfmathsetmacro{\YTOP}{\YB + \BraidT*\BraidDY}

    \foreach \pos in {1,2,3,4,5}{%
      \pgfmathsetmacro{\Xpos}{(\pos-1)*\BraidDX}%
      \ifnum\pos=1\relax
        \fill[red] (\Xpos,\YTOP) circle[radius=\BraidDot]; 
        \node[treelabel, above=2pt] at (\Xpos,\YTOP) {{\color{red}$\sigma_{p}$}};
      \else
        \fill (\Xpos,\YTOP) circle[radius=\BraidDot];
        \pgfmathtruncatemacro{\lab}{\pos-1}%
        \node[treelabel, above=2pt] at (\Xpos,\YTOP) {$\sigma_{\lab}$};
      \fi
    }%
    
      \TreeOuterLabels{2}{-0.4}{a}{\Bar{a}}

  \end{scope}

\end{tikzpicture} 
 \ \ $P_2$:
\begin{tikzpicture}[x=1cm,y=1cm, baseline=(current bounding box.center), line cap=round, line join=round]

  \def\YA{-0.6}
  \def\YB{1.0}
  \pgfmathsetmacro{\YOuterRoot}{-2.0}

  \begin{scope}[xshift={-\BraidDX cm}]

    \pgfmathsetmacro{\XposOne}{0}
    \pgfmathsetmacro{\XposTwo}{1*\BraidDX}
    \pgfmathsetmacro{\XposFive}{4*\BraidDX}

    \TreeOneTwoFourPositions{2}{3}{4}{5}{\YOuterRoot}{\YA}{\YB}

    \pgfmathsetmacro{\XrootTree}{(\XposTwo+\XposFive)/2}
    \node[treelabel, below=2pt] at (\XrootTree,\YOuterRoot) {$1$};

    \draw[strandline, draw=red] (\XposOne,\YOuterRoot+\JoinEps) -- (\XposOne,\YB+\JoinEps);
    \fill[red] (\XposOne,\YOuterRoot) circle[radius=\BraidDot]; 

    \BraidN=5\relax
    \BraidT=0\relax
    \InitPerm{5}

    \BraidParStepTracked{\YB}{+1}
    \BraidParStepTracked{\YB}{+2}
    \BraidParStepTracked{\YB}{-3}
    \BraidParStepTracked{\YB}{-4}
    \BraidParStepTracked{\YB}{-4}
    \BraidParStepTracked{\YB}{-3}
    \BraidParStepTracked{\YB}{+2}
    \BraidParStepTracked{\YB}{+1}

    \pgfmathsetmacro{\YTOP}{\YB + \BraidT*\BraidDY}

    \foreach \pos in {1,2,3,4,5}{%
      \pgfmathsetmacro{\Xpos}{(\pos-1)*\BraidDX}%
      \ifnum\pos=1\relax
        \fill[red] (\Xpos,\YTOP) circle[radius=\BraidDot]; 
        \node[treelabel, above=2pt] at (\Xpos,\YTOP) {{\color{red}$\sigma_{p}$}};
      \else
        \fill (\Xpos,\YTOP) circle[radius=\BraidDot];
        \pgfmathtruncatemacro{\lab}{\pos-1}%
        \node[treelabel, above=2pt] at (\Xpos,\YTOP) {$\sigma_{\lab}$};
      \fi
    }%
    
      \TreeOuterLabels{2}{-0.4}{a}{\Bar{a}}

  \end{scope}

\end{tikzpicture}
\end{center}

This first path simply winds around the qutrit,  whereas the second path does an reverted winding for the second two $\sigma$'s. The first path is simpler but the second one is more powerful as we will see below. The non-trivially designed property of these probing paths is that it does not entangle the qutrit with the probe. Instead, it simply implements a unitary operator on the target qutrit
\begin{align}
    (M_1)_{a,a'} = \begin{bmatrix}
        1 & 0 & 0\\
        0 & 0 & 1 \\
        0& 1 & 0
    \end{bmatrix}_{a,a'} = (H^2)_{a,a'}  \ , \ \ \  (M_2)_{a,a'} = \begin{bmatrix}
        1 & 0 & 0\\
        0 & 0 & \omega^{-1} \\
        0& \omega^{-1} & 0
    \end{bmatrix}_{a,a'} = (H^2 Q^{-1})_{a,a'} . 
\end{align}

As in usual Mach-Zehnder interferometry experiments~\cite{wiseman2009_quantum}, one generates multiple probe anyons and then sends them into a beam splitter to prepare them in the state $(|R\rangle +|P\rangle )/\sqrt{2}$. After the braids, the two paths recombine at the output beam splitter, and then the probe is detected at one of two output ports $s=\pm$, which corresponds to two possible measuring outcomes $(|R\rangle \pm |P\rangle )/\sqrt{2}$ for the path degree of freedom. This procedure realizes a two-outcome generalized measurement on the target with Kraus operators (up to an overall phase)
\begin{align}
    K_\pm = \frac{1}{2} \left(I\pm\mathrm{e}^{\mathrm{i}\phi}M\right) \,,
\end{align}
where $\phi$ is a non-topological relative phase between the two paths, which can be tuned by the details in the paths and the dynamical process.

What is special about the $M$'s (compared to most other unitary operations) is that it can be diagonalized as 
\begin{align}
    M = \sum_{i=0,\pm} \lambda_i P_i
\end{align}
where $|\pm\rangle \equiv (|1\rangle \pm |2\rangle)/\sqrt{2}$ and $P_i$ is the projector onto the corresponding subspace. For $M_1$, $\lambda_{i=0,+,-} = 1,1,-1$, whereas for $M_2$, $\lambda_{i=0,+,-} = 1,\omega^{-1},-\omega^{-1}$. This special structure means that the $K_{\pm} $ can also be decomposed as:
\begin{align}
    K_{s=\pm} = \sum_{i} k_s(\lambda_i) P_i \ \ , \ \ k_s(\lambda)\equiv  \frac{1}{2}(1+s \mathrm{e}^{\mathrm{i}\phi} \lambda)
\end{align}

For an arbitrary initial state of the target qutrit $|\psi\rangle$, the probability of outcome $s$ is $p(s)=\langle \psi|K_s^\dagger K_s|\psi\rangle$ and the corresponding post-measurement state is $|\tilde{\psi}\rangle =K_s|\psi\rangle$. Due to the special decomposition of $K_s$, the outcome probability only depends on the weights $w_i\equiv \langle P_i\rangle$ and measurement does not induce transitions across different $|i\rangle$, meaning that the measurement is quantum non-demolition: it extracts information about the eigenvalue sector of $M$ while preserving that sector space under the measurement.

After each measurement, the weights will undergo a Bayesian reweighting with likelihood $|k_s(\lambda_i)|^2$, which is the standard quantum-trajectory (selective measurement) update rule for Kraus measurements~\cite{wiseman2009_quantum}. Repeating the probe many times multiplies these likelihood factors, so that (whenever the outcome statistics are distinguishable for different $i$) the posterior distribution $w_i$ becomes sharply peaked and the conditioned state is driven toward support on a single eigenspace $P_{i_0}$; this is the operational meaning of ``collapse'' under repeated weak measurements. In the anyon-interferometry setting, this collapse mechanism and its rate are analyzed explicitly in terms of probe-induced distinguishability (e.g., via monodromy for charge measurements)~\cite{BONDERSON2008_interferometry}.

Experimentally, one can record the full measurement record $\vec{s}=(s_1,\dots,s_N)$ of output ports from $N\gg 1$ repeated probes. For each candidate sector $i$, the model predicts different single-shot outcome probabilities $p(s\!\mid\! i)=|k_s(\lambda_i)|^2$ for different $i$, thus providing a way to distinguish them. For our case with $M_2$, this process can fully distinguish the three states $|0\rangle$, $|+\rangle$, and $|-\rangle$. Therefore, for any initial qutrit state, one can achieve projection onto $|+\rangle$ or $|-\rangle$ (both of which are non-stabilizer states) through post-selection. For the slightly weaker case of $M_1$, we can still distinguish between the subspace spanned by $\ket{0}, |+\rangle$ from the subspace spanned by $|-\rangle$. Therefore, as long as one can prepare a qutrit initial state in the $\text{span}\{|1\rangle,|2\rangle\}$ subspace, one can still projectively prepare $|+\rangle$ or $|-\rangle$. 

Practically, the phase $\phi$ should be carefully tuned in order to know exactly the resulting state. One possible trick is to slowly deform the path $P$ to $R$, and record the change of the probability distribution $p(s\!\mid\! i)$ during the process.  A discontinuous jump in this case signals that $i=\pm $. This type of trick has been experimentally realized in e.g. Refs.~\cite{nakamura2020_anyoninterferometry,carrega2021_anyons} for anyons in certain fractional quantum Hall states. 

Compared to the traditional anyon interferometry, here we considered a slightly generalized setup. The key difference is that here the probe anyons $\sigma_p$ are really symmetry defects introduced and controlled by external fluxes. This external flux can be introduced by bringing a superconducting ring device (usually called a fluxonium) adjacent to the thin-film superconductor, which itself is another bulk $4e$ superconductor (which doesn't need to be topological) and thus could trap $hc/(4e)$ flux quanta. Let us denote the state of a fluxonium with a $hc/(4e)$ flux by $|1\rangle$ and the no-flux state by $|0\rangle$. Then the Mach-Zehnder-type interference we imagined really amounts to creating a pair of fluxoniums, preparing them in a Bell state $(|10\rangle +|01\rangle )/\sqrt{2}$, sending them separately through the reference and probing paths $R$ and $P$, and then measuring their flux state in the Bell basis $(|10\rangle +|01\rangle )/\sqrt{2}$ again. Each of these steps is already achieved experimentally (with $2e$ superconductors, see e.g.~\cite{Steffen2006_tomography,Bao2022_fluxonium,Ding2023_fluxonium}). We note, that a perfect initial Bell state  and a perfect measurement in the Bell basis is not necessary for the two fluxoniums hosting one unit flux. The interferometry setup can work as long as these states are prepared and measured in an entangled way.

\subsection{Comparison with TQC protocols for $SU(2)_4$}
\label{app:comparison_SU(2)4}

Let us now briefly contrast our protocol with the protocol for $SU(2)_4$ devised in Refs.~\cite{Hastings2012_metaplectic_power,Cui2014_SU(2)4_gate,Levaillant2015_SU(2)4_gate,Bocharov2015_SU(2)4_compilation}. The topological order that we used is a $\mathbb{Z}_3$ chiral topological order, enriched by a $\mathbb{Z}_2$ anyon permutation symmetry. The computational power of this setup derives from the non-Abelian statistics of the nontrivial $\mathbb{Z}_2$ symmetry defect $\sigma$. The $SU(2)_4$ topological order can be obtained by gauging the $\mathbb{Z}_2$ symmetry in the $\mathbb{Z}_3$ topological order (i.e. promoting the background $\mathbb{Z}_2$ gauge field to a dynamical $\mathbb{Z}_2$ gauge field). Under the gauging map, the $\sigma$ defect evolves into the $j = 1/2$ anyon in $SU(2)_4$, while the Abelian anyons $a, \bar a$ (permuted by the $Z_2$ symmetry action) in the $\mathbb{Z}_3$ chiral topological order merge into the non-Abelian anyon $j = 1$ in $SU(2)_4$.

An immediate consequence of these distinctions is that the qutrit encoding $0 \rightarrow a \bar a \rightarrow \sigma \sigma \sigma \sigma$ used in our protocol does not apply to $SU(2)_4$. The analogous splitting $(j=0) \rightarrow \sigma \sigma \sigma\sigma$ in $SU(2)_4$ only has two intermediate states $(j_1 = 0, j_2 = 0)$ or $(j_1 = 1, j_2 = 1)$ and therefore encodes a qubit rather than a qutrit. Qutrit encoding requires a different splitting $(j = 1) \rightarrow \sigma \sigma \sigma \sigma$ which contains three intermediate states $(j_1 = 1, j_2 = 0), (j_1 = 0, j_2 = 1), (j_1 = 1, j_2 = 1)$. Once we use this encoding, the braiding of $\sigma$ anyons in $SU(2)_4$ becomes just as powerful as the braiding of $\sigma$ defects in the $Z_3$ topological order: both generate the entire multi-qutrit Clifford group.

As for magic state preparation, the protocol we used crucially relies on the fact that braiding with $\sigma$ permutes the Abelian anyons $a$ and $\bar a$ in the $Z_3$ topological order. Once $a$ and $\bar a$ merge into the $j = 1$ anyon in $SU(2)_4$, braiding $j = 1$ with $\sigma$ no longer implements an anyon permutation and the interferometric measurement in App.~\ref{app:magicstate} does not project onto a qutrit magic state. Of course, this statement does not contradict the computational universality of $SU(2)_4$: with more sophisticated protocols in Refs.~\cite{Hastings2012_metaplectic_power,Cui2014_SU(2)4_gate,Levaillant2015_SU(2)4_gate,Bocharov2015_SU(2)4_compilation}, one can still generate non-Clifford gates from fusion + measurements in $SU(2)_4$.

\subsection{Effects of the Goldstone mode}
\label{app:goldstone}


A crucial caveat of our proposal is the presence of gapless Goldstone (phase) modes (phonons), which can mediate unwanted coupling between the internal Hilbert spaces of distinct information-encoding blocks. To suppress this channel, all the operations outlined in our paper must be performed slowly enough to avoid exciting phonons. In this section we estimate the system size dependence of the corresponding critical velocity. We note that these analyses generically apply to vortices in all thin-film SCs, including the ones in traditional $p+ip$ $2e$ SCs.


Since here we focus on the phase mode with the assumption that the parafermionic zero modes are confined to the core of vortices and energetically separated by a finite gap to other excitations, it is justifiable to simply consider a quantum XY model with an applied localized external flux at $\bm{R}$. Generically, the Hamiltonian takes the form:
\begin{align}
    \hat{H}(\bm{R}) = \int \mathrm{d}^2\bm{r}\mathrm{d}^2\bm{r}' ~ \frac{1}{2}\left\{ V(\bm{r},\bm{r}') \hat{n}(\bm{r})\hat{n}(\bm{r}') + G(\bm{r},\bm{r}')   
     \left[\nabla\hat{\theta}(\bm{r}) -Q\bm{A}(\bm{r}) \right]\left[\nabla\hat{\theta}(\bm{r}') -Q\bm{A}(\bm{r}') \right]
    \right\}
\end{align}
where $Q$ is the charge of the condensate ($Q=4$ for our case but this is irrelevant to this part of the analysis), $\hat{\theta}$ is an $U(1)$ phase operator satisfying the canonical commutation relation with the density operator $[\hat{\theta}(\bm{r}),\hat{n}(\bm{r}')]=\mathrm{i}\delta^{2}(\bm{r}-\bm{r}') $. The vector potential $\bm{A}(\bm{r})$ encodes the localized flux:
\begin{align}
    \nabla\times \bm{A}(\bm{r}) = B_F(\bm{r}-\bm{R})
\end{align}
with certain localized profile of a quantum magnetic flux with characteristic length scale $\xi$, e.g. $B_F(\bm{r}) =1 /(Q\xi^2) \mathrm{e}^{-\bm{r}^2/(2\xi^2)}$ such that the total flux is $2\pi/Q$. 
$V$ and $G$ are two interaction kernels which are mostly translationally invariant but only locally affected by the presence of external flux (due to the change of local screening environment and pairing amplitude, etc): 
\begin{align}
    V(\bm{r},\bm{r}') = V_0(\bm{r}-\bm{r}')+ \delta V(\bm{r},\bm{r}';\bm{R}) \\
    G(\bm{r},\bm{r}') = G_0(\bm{r}-\bm{r}')+ \delta G(\bm{r},\bm{r}';\bm{R})
\end{align}
where $\delta V$ and $\delta G$ rapidly decay with both $|\bm{r}-\bm{R}|/\xi$ and $|\bm{r}'-\bm{R}|/\xi$. In summary, the Hamiltonian can be decomposed as:
\begin{align} \label{eq:HR}
    \hat{H}(\bm{R}) =\hat{H}_0+\delta\hat{H}(\bm{R})
\end{align}
where the bare $\hat{H}_0$ is translationally invariant and $\delta\hat{H}(\bm{R})$ is supported in a local regime of radius $\xi$ around position $\bm{R}$. $\hat{H}_0$ can be readily diagonalized in momentum space as:
\begin{align}
    \hat{H}_0 = \sum_{\bm{k}} \frac{1}{2} \left[\frac{|\hat{n}_{\bm{k}}|^2}{\chi_{\bm{k}}} + D_{\bm{k}} |\hat{\theta}_{\bm{k}}|^2\right] = \sum_{\bm{k}} \omega_{\bm{k}} \left[\hat{a}^{\dagger}_{\bm{k}}\hat{a}_{\bm{k}}+\frac{1}{2}\right]
\end{align}
where
\begin{align}
    \omega_{\bm{k}} &\equiv \sqrt{D_{\bm{k}}/\chi_{\bm{k}}} \\
    \hat{a}_{\bm{k}} &\equiv \frac{1}{\sqrt{2}} \left[(D_{\bm{k}}\chi_{\bm{k}})^{1/4}\hat{\theta}_{\bm{k}} + \mathrm{i} (D_{\bm{k}}\chi_{\bm{k}})^{-1/4}\hat{n}_{\bm{k}} \right] .
\end{align}

We first consider the structure of phonon eigenmodes in this setup on a finite $L\times L$ plane, assuming $L\gg \xi$. There are two effects of the precense of a quantum flux, one is topological while the other is dynamical. The topological effect is to re-organize the low-energy states around a classical vortex configuration $\bar{\theta} (\bm{r}) \equiv \arg(\bm{r}-\bm{R})$. Then, the phonon modes corresponding to the deviation $\delta\theta = \theta - \bar{\theta}$ are still formally described by the Hamiltonian Eq.~\ref{eq:HR}, but now with a modified effective magnetic field 
\begin{align}
    B_F'(\bm{r}) = B_F(\bm{r}) - 2\pi \delta^2(\bm{r})/Q
\end{align}
which doesn't have net flux. 
Next, the dynamical effects, described by the local perturbation $\delta\hat{H}(\bm{R})$, serve as a local scatterer for the phonons of $\hat{H}_0$. For small momenta $\bm{k} \ll |\xi|^{-1}$, such perturbations only cause $\mathcal{O}[(\xi/L)^2]$ mixing of the plane-wave states $u_{\bm{k}}(\bm{r})\sim \mathrm{e}^{\mathrm{i}\bm{k}\cdot \bm{r}}$, thus do not alter the low-energy phonon spectrum and eigenmodes significantly compared to that of $\hat{H}_0$.

Now we consider a dynamical process in which $\bm{R}(t)$ evolves with time from $t=0$ to $t=T$ with smooth start and stop. Assuming we start with the instantaneous ground state (no excited phonons), the adiabatic theorem implies that the total leakage out of the ground state is upper bounded by 
\begin{align}
    P_\text{leak} \lesssim \sum_{n>0} \max_{t} \left|\frac{\langle n| \frac{\mathrm{d}\delta\hat{H}}{\mathrm{d}\bm{R}}|0\rangle \dot{\bm{R}}}{(E_n - E_0)^2}\right|^2
\end{align}
Expanding the perturbations in the momentum space, we find
\begin{align}
\delta\hat{H}= \sum_{\bm{k}}\left( \alpha_{\bm{k}} \hat{\delta\theta}_{\bm{k}} +  \beta_{\bm{k}} \hat{n}_{\bm{k}}\right) + \sum_{\bm{k},\bm{k}'} \left(\gamma_{\bm{k},\bm{k'}}\hat{\delta\theta}_{\bm{k}} \hat{\delta\theta}_{\bm{k}'} +\eta_{\bm{k},\bm{k}'} \hat{n}_{\bm{k}}\hat{n}_{\bm{k}'}\right)
\end{align}
For small $|\bm{k}|\ll \xi^{-1}$, 
\begin{align}
    \alpha_{\bm{k}}\sim & \epsilon_p \left(\frac{\xi}{L}\right) |\bm{k}| \\
    \gamma_{\bm{k},\bm{k}'}\sim & \epsilon_p  \left(\frac{\xi}{L}\right)^2 |\bm{k}| |\bm{k}'| \\
    \beta_{\bm{k}}\sim & \epsilon_p \left(\frac{\xi^2}{L}\right) \\
    \eta_{\bm{k},\bm{k}'}\sim & \epsilon_p \left(\frac{\xi^2}{L}\right)^2,
\end{align}
where $\epsilon_p $ is an $\mathcal{O}(1)$ energy scale characterizing the strength of the perturbation, $\left(\frac{\xi}{L}\right)$ factors account for the fact that $\delta\hat{H}$ is supported locally, and 
$\alpha$ and $\gamma$ couplings receive additional suppression from $|\xi\bm{k}|$ factors due to the fact that they should analytically vanish when $|\bm{k}|\rightarrow 0$. When taking derivative with respect to $\bm{R}$, the magnitude of these coupling coefficients takes an additional factor of order $\xi^{-1}$, characterizing the local structure of the perturbation. With these scaling arguments, we can loosely estimate the total probability of exciting phonons during the dynamical process with maximum speed $v_\text{max}$:
\begin{align}\label{eq:Pleak}
    P_\text{leak} \lesssim & \frac{v^2_\text{max}  }{\xi^2}\left\{  L^2\int_{2\pi/L <|\bm{k}|<\xi^{-1}} \mathrm{d}^2\bm{k} \frac{\left[\alpha_{\bm{k}}(D_{\bm{k}}\chi_{\bm{k}})^{-1/4}+\beta_{\bm{k}}(D_{\bm{k}}\chi_{\bm{k}})^{1/4}\right]^2}{\omega_{\bm{k}}^4} \right. \nonumber\\
    &\left. \ \ \ \ \ \ \ \ \ \ \ \ \ \ \ \ \ + L^4\int_{2\pi/L <|\bm{k}|,|\bm{k}'|<\xi^{-1}} \mathrm{d}^2\bm{k} \mathrm{d}^2\bm{k}' \frac{\left[\gamma_{\bm{k},\bm{k}'}(D_{\bm{k}}\chi_{\bm{k}}D_{\bm{k}'}\chi_{\bm{k}'})^{-1/4}+\eta_{\bm{k},\bm{k}'}(D_{\bm{k}}\chi_{\bm{k}}D_{\bm{k}'}\chi_{\bm{k}'})^{1/4}\right]^2}{(\omega_{\bm{k}}+\omega_{\bm{k'}})^4} \right\}
\end{align}

Finally, let's discuss several specific cases of particular relevance to experiments.
\begin{itemize}
    \item For short-range charge-charge interaction $V$ and short-range current-current interaction $G$ corresponding to a metallically screening environment, we have $\chi_{\bm{k}}\sim \kappa_s /v^2_s$, $D_{\bm{k}}\sim \kappa_s |\bm{k}|^2$ and $\omega_{\bm{k}}\sim v_s |\bm{k}|$, where $\kappa_s$ is the phase stiffness and $v_s$ is the speed of sound. In this case, we can estimate the upper bound on the leakage probability as:
    \begin{align}  
        P_\text{leak} \lesssim L \left[\xi/\sqrt{\ell_0} +\sqrt{\ell_0}\right]^2 \epsilon_p^2 v^2_\text{max} / v_s^4
    \end{align}
    where we define a characteristic length scale of the system $\ell_0 \sim v_s/\kappa_s$. 
    Requiring $P_\text{leak} \ll 1$, we obtain a sufficient condition 
    \begin{align}
       v_\text{max}  \ll v_0\equiv  \frac{v_s^2}{\epsilon_p \sqrt{L} \left[\xi/\sqrt{\ell_0} +\sqrt{\ell_0}\right]} .
    \end{align}

    \item For long-range charge-charge interaction $V$ and short-range current-current interaction $G$ corresponding to a magnetically screening environment (since the fine structure constant of our universe is small, this is also a good approximation to unscreened systems), we have $\chi_{\bm{k}}\sim \epsilon |\bm{k}| $ ($\epsilon$ is the permittivity), $D_{\bm{k}}\sim \kappa_s |\bm{k}|^2$ and $\omega_{\bm{k}}\sim \sqrt{(\kappa_s /\epsilon) |\bm{k}|} $. In this case, the integrals in Eq.~\ref{eq:Pleak} do not have infra-red divergence, so the estimate converges to a size-independent, microscopic-detail-dependent quantity:
    \begin{align}
         P_\text{leak} \lesssim \left[(\ell_0'/\xi)+(\ell_0'/\xi)^{-1}\right] \epsilon_p^2 v^2_\text{max} / (\kappa_s^4 \ell_0'^2)
    \end{align}
     where we define a characteristic length scale of the system $\ell_0' \sim 1/(\kappa_s\epsilon)$. Requiring $P_\text{leak} \ll 1$, we obtain a sufficient condition
     \begin{align}
         v_\text{max}  \ll v_0\equiv \frac{\kappa_s^2 \ell_0'}{\left[(\ell_0'/\xi)+(\ell_0'/\xi)^{-1}\right]^{1/2} \epsilon_p }
     \end{align}

\end{itemize}

In conclusion, we find that for systems with long-range charge-charge interaction and short-range current-current interaction, the adiabatic condition can be met by requiring the motion of the vortices to be small compared to a fixed, system-size-independent scale.


\section{Transitions between Jain states and higher-charge topological superconductors}
\label{app:jain}

In this Appendix, we provide some more details on transitions between the Jain states at $\nu = p/(2p+1)$ and $2pe$ topological superconductors. 

Let us begin with the general parton decomposition $c = b_{a} \epsilon_{aa'} f_{a'}$, which introduces a local $U(2)$ gauge redundancy such that $b, f$ carry spin $1/2$ under the $SU(2)$ part of $U(2)$ but charge $\pm 1$ under the $U(1)$ part of $U(2)$. The general low energy Lagrangian that preserves the $U(2)$ gauge symmetry takes the form
\begin{equation}
    L = L[b, \alpha - \Tr \alpha] + L[f, \alpha + A ] \,, 
\end{equation}
where $\alpha$ is a dynamical $U(2)$ gauge field and $A$ is the external electromagnetic $U(1)$ gauge field (technically a $\mathrm{spin}_{\mathbb{C}}$ connection. 

At general $\nu = p/(2p+1)$, the Jain state can be regarded as $-p$ copies of the fermionic integer quantum Hall state stacked with another fractional quantum Hall state at $\nu = p + \frac{p}{2p+1} = \frac{2p(p+1)}{2p+1}$. The $\nu = \frac{2p(p+1)}{2p+1}$ state can be constructed by choosing a mean-field flux assignment in which 
\begin{equation}
    \frac{1}{2\pi} \nabla \times \bs{\alpha_0} = - \frac{p+1}{2p+1} \frac{1}{2\pi} \nabla \times \bs{A} \,. 
\end{equation}
This flux assignment ensures that $b$/$f$ is at Landau level filling $2p$/$2p+2$ and can form bosonic/fermionic integer quantum Hall states with Hall conductance $\sigma^{xy}_{b} = 2p, \sigma^{xy}_f = 2p+2$. The resulting topological Lagrangian then takes the form
\begin{equation}\label{eq:lag_Jain}
    \begin{aligned}
    L_p^{+} &= -p \mathrm{CS}[A,g] + \frac{p+1}{4\pi} \Tr \left[ (\alpha + AI) d (\alpha + AI) + \frac{2}{3} \alpha^3\right] + 2(p+1) \Omega_g - \frac{p}{4\pi} \Tr \left[\alpha d \alpha + \frac{2}{3} \alpha^3\right] + \frac{p}{4\pi} \Tr \alpha \, d \Tr \alpha  \\
    &= \frac{1}{4\pi} \Tr \left[\alpha d \alpha + \frac{2}{3} \alpha^3\right] + \frac{p}{4\pi} \Tr \alpha \, d \Tr \alpha + \frac{p+1}{2\pi} \Tr \alpha d A + (p+2) \mathrm{CS}[A,g] \,,
    \end{aligned}
\end{equation}
which is simply $U(2)_{-1,-2(2p+1)} \times U(1)_1^{p+2}$ where $U(2)_{-1,-2(2p+1)} = SU(2)_{-1} \times U(1)_{-2(2p+1)}/\mathbb{Z}_2$. By the theorems in Ref.~\cite{Cheng2025_orderingFQH}, we can verify that this Lagrangian indeed describes the Jain topological order at $\nu = p/(2p+1)$ by checking that it has Hall conductance $\sigma_{xy} = p/(2p+1)$, chiral central charge $c_- = p + 1 - \mathrm{sgn}(p)$, and total squared quantum dimension $2|2p+1|$. The Hall conductance of $U(2)_{-1,-2(2p+1)}$ is easily computed to be $-\frac{2(p+1)^2}{2(2p+1)}$. Stacking with $U(1)_1^{p+2}$ gives the expected total Hall conductance $\sigma_{xy} = \frac{(2p+1)(p+2) - 2(p+1)^2}{2p+1} = \frac{p}{2p+1}$. The chiral central charge receives a contribution $-1$ from $SU(2)_{-1}$ and $-\mathrm{sgn}(p)$ from $U(1)_{-2(2p+1)}$. Stacking with $U(1)_1^{p+2}$ then gives the expected $c_- = -1 - \mathrm{sgn}(p) + p+2 = p+1 - \mathrm{sgn}(p)$. Finally, since $SU(2)_{-1}$ has squared quantum dimension $2$, and $U(1)_{-2(2p+1)}$ has squared quantum dimension $2|2p+1|$, the fermionic $\mathbb{Z}_2$ quotient has squared quantum dimension $2 \times 2|2p+1|/2 = 2|2p+1|$. This argument completes the identification of \eqref{eq:lag_Jain} with the Jain topological order at $\nu = p/(2p+1)$. 

An alternative description of the Jain state can be obtained by choosing $\sigma^{xy}_{b} = 2p+2, \sigma^{xy}_f = 2p$ instead. The corresponding Lagrangian is now
\begin{equation}
    \begin{aligned}
    L_{p}^{-} &= -p \mathrm{CS}[A,g] + \frac{p}{4\pi} \Tr \left[ (\alpha + AI) d (\alpha + AI) + \frac{2}{3} \alpha^3\right] + 2p \Omega_g - \frac{p+1}{4\pi} \Tr \left[\alpha d \alpha + \frac{2}{3} \alpha^3\right] + \frac{p+1}{4\pi} \Tr \alpha \, d \Tr \alpha  \\
    &= - \frac{1}{4\pi} \Tr \left[\alpha d \alpha + \frac{2}{3} \alpha^3\right] + \frac{p+1}{4\pi} \Tr \alpha \, d \Tr \alpha + \frac{p}{2\pi} \Tr \alpha d A + p\mathrm{CS}[A,g] \,,
    \end{aligned}
\end{equation}
which we identify as $U(2)_{1,-2(2p+1)} \times U(1)_1^p$. We can again verify that the Hall conductance of this new Lagrangian is $\sigma_{xy} = - \frac{2p^2}{2p+1} + p = \frac{p}{2p+1}$, the chiral central charge is $p+ 1 - \mathrm{sgn}(p)$, and the squared quantum dimension is $2 \times 2|2p+1|/2 = 2|2p+1|$. Therefore, $L^+_p, L^-_p$ provide two equivalent Lagrangian descriptions of the same Jain topological order at $\nu = p/(2p+1)$. 

In both cases, $b_{a}$ and $f_{a}$ source an anyon with fractional charge proportional to $e/(2p+1)$. This means that lattice translations act projectively on both $b_{a}, f_{a}$ and the dispersion of $b_{a}, f_{a}$ must each have a $|2p+1|$-fold degeneracy. Across a Chern-number changing transition for $b_{a}$ or $f_{a}$ that preserves the $U(2)$ gauge invariance, the jump in total Hall conductance must therefore be $|\Delta \sigma^{xy}_{b}| = 2|2p+1|, |\Delta \sigma^{xy}_f| = 2|2p+1|$.

Given these constraints, we immediately see two possible transitions from the Jain state to a $2pe$ superconductor:
\begin{enumerate}
    \item If we use the $L^+_{p}$ presentation and start with $\sigma^{xy}_{b} = 2p, \sigma^{xy}_f = 2p+2$, then we can go through a transition across which $\sigma^{xy}_f$ jumps to $-2p$. The resulting theory is described by the Lagrangian
    \begin{equation}
        \begin{aligned}
        L^+_{2pe} &= -p \mathrm{CS}[A,g] - \frac{p}{4\pi} \Tr \left[ (\alpha + AI) d (\alpha + AI) + \frac{2}{3} \alpha^3\right] - 2p \Omega_g - \frac{p}{4\pi} \Tr \left[\alpha d \alpha + \frac{2}{3} \alpha^3\right] + \frac{p}{4\pi} \Tr \alpha \, d \Tr \alpha  \\
        &= - \frac{2p}{4\pi} \Tr \left[\alpha d \alpha + \frac{2}{3} \alpha^3\right] + \frac{p}{4\pi} \Tr \alpha \, d \Tr \alpha - \frac{p}{2\pi} \Tr \alpha d A - 3p\mathrm{CS}[A,g] \,. 
        \end{aligned}
    \end{equation}
    This Lagrangian describes a $U(2)_{2p, 0} \times U(1)_1^{-3p}$ Chern-Simions theory for a topological $2pe$ superconductor. When $p = -2$, it reduces to the $U(2)_{-4,0} \times U(1)_1^6$ $4e$ superconductor quoted in the main text. The intrinsic topological order of $U(2)_{-4,0}$ is related to the Jain topological order at $\nu = 2/3$ by time-reversal. The critical point involves $N_f = |2p+1|$ species of two-component massless Dirac fermions coupled to the $U(2)$ gauge field $\alpha$
    \begin{equation}
        L^+_{\rm crit} = L^+_{p}[\alpha, A] + \sum_{i=1}^{|2p+1|} \bar \Psi_{i, \sigma} i \gamma_{\mu} D^{\mu}_{\alpha+A} \Psi_{i,\sigma} \,.
    \end{equation}
    \item If we use the $L^-_{p}$ presentation and start with $\sigma^{xy}_{b} = 2p+2, \sigma^{xy}_f = 2p$, then we can go through a transition across which $\sigma^{xy}_{b}$ jumps to $-2p$. The resulting theory is described by the Lagrangian
    \begin{equation}
        \begin{aligned}
        L^-_{2pe} &= -p \mathrm{CS}[A,g] + \frac{p}{4\pi} \Tr \left[ (\alpha + AI) d (\alpha + AI) + \frac{2}{3} \alpha^3\right] + 2p \Omega_g +  \frac{p}{4\pi} \Tr \left[\alpha d \alpha + \frac{2}{3} \alpha^3\right] - \frac{p}{4\pi} \Tr \alpha \, d \Tr \alpha  \\
        &= \frac{2p}{4\pi} \Tr \left[\alpha d \alpha + \frac{2}{3} \alpha^3\right] - \frac{p}{4\pi} \Tr \alpha \, d \Tr \alpha + \frac{p}{2\pi} \Tr \alpha d A + p \mathrm{CS}[A,g] \,. 
        \end{aligned}
    \end{equation}
    This new theory is $U(2)_{-2p,0} \times U(1)_1^p$. At $p = -2$, the intrinsic topological order of $U(2)_{-2p,0}$ is equivalent to the Jain topological order at $\nu = 2/3$. The critical theory now corresponds (at the mean-field level) to a $U(2)$-symmetric transition between boson integer quantum Hall states with $\Delta \sigma^{xy}_b = -2(2p+1)$. The critical Lagrangian is more involved and contains a dynamical $U(2)$ gauge field $\alpha$ as well as $|2p+1|$ dynamical $U(1)$ gauge fields $\beta_i$. We refer interested readers to Ref.~\cite{Shi2025_FCISC} for a detailed parton construction of such a $U(2)$-symmetric boson integer quantum Hall transition. 
\end{enumerate}

We emphasize that although $L^+_p, L^-_p$ describe the same Jain state at $\nu = p/(2p+1)$, the $2pe$ superconductors described by $L^+_{2pe}$ and $L^-_{2pe}$ are not identical (though they can be related by time-reversal + stacking invertible phases). 

Finally, let us remark that the transition from $L^+_p$ to $L^+_{2pe}$ can also be obtained through a more familiar parton construction $c = f_1 f_2 \psi$. Choosing a mean-field that preserves a $U(2)$ gauge redundancy under which $(f_1, f_2)$ transforms as a doublet, we have the general effective Lagrangian 
\begin{equation}
    L = L[f_i, \alpha + A] + L[\psi, - A - \Tr \alpha] \,. 
\end{equation}
The Jain state at $\nu = p/(2p+1)$ corresponds to putting $f_i, \psi$ in Chern insulator states with $C_{f_1} = C_{f_2} = 1$ and $C_{\psi} = p$. After integrating out the gapped partons, we obtain an effective $U(2)$ gauge theory
\begin{equation}
    \begin{aligned}
    L &= \frac{1}{4\pi} \Tr \left[(\alpha + AI) d (\alpha + AI) + \frac{2}{3} \alpha^3\right] + 2 \Omega_g + \frac{p}{4\pi} (\Tr \alpha + A) d (\Tr \alpha + A) + p \Omega_g \\
    &= \frac{1}{4\pi} \Tr \left[\alpha d \alpha + \frac{2}{3} \alpha^3\right] + \frac{p}{4\pi} \Tr \alpha \, d \Tr \alpha + \frac{p+1}{2\pi} \Tr \alpha d A + (p+2) \mathrm{CS}[A,g] \,.
    \end{aligned}
\end{equation}
The final line is precisely equal to $L^+_p$, which was previously derived from the $c = b_a \epsilon_{aa'} f_{a'}$ parton. From here, we immediately see that if $C_{f_1}, C_{f_2}$ jump from $1$ to $-2p$, we arrive at a superconducting Lagrangian
\begin{equation}
    \begin{aligned}
    L &= - \frac{2p}{4\pi} \Tr \left[(\alpha + AI) d (\alpha + AI) + \frac{2}{3} \alpha^3\right] - 4p \Omega_g + \frac{p}{4\pi} (\Tr \alpha + A) d (\Tr \alpha + A) + p \Omega_g  \\
    &= - \frac{2p}{4\pi} \Tr \left[\alpha d \alpha + \frac{2}{3} \alpha^3\right] - \frac{p}{2\pi} \Tr \alpha d A + \frac{p}{4\pi} \Tr \alpha d \Tr \alpha - 3 p \mathrm{CS}[A,g] \,. 
    \end{aligned}
\end{equation}
which precisely reproduces $L^+_{2pe}$.

\end{document}